\documentclass[12pt]{article}
\usepackage[utf8]{inputenc}
\usepackage[english]{babel}
\usepackage{amsmath}
\usepackage{mathrsfs}
\usepackage{amssymb}
\usepackage{amsthm}
\usepackage{bm}
\usepackage{amsfonts}
\usepackage{xspace}
\usepackage[normalem]{ulem}
\usepackage{graphicx}
\usepackage{multirow}
\usepackage{booktabs}
\usepackage{appendix}
\usepackage{enumerate}
\usepackage{url} 
\usepackage{authblk}
\usepackage{xcolor}
\usepackage{subcaption}
\usepackage{xr}
\usepackage{adjustbox}
\usepackage[colorlinks=true,linkcolor=red,citecolor=blue]{hyperref}
\usepackage{enumitem}
 \usepackage{natbib}

\newcommand{\blind}{1}

\makeatletter
\newcommand*{\addFileDependency}[1]{
\typeout{(#1)}
\@addtofilelist{#1}
\IfFileExists{#1}{}{\typeout{No file #1.}}
}\makeatother
\newcommand{\balpha}{\boldsymbol{\alpha}}
\newcommand{\bbeta}{\boldsymbol{\beta}}
\newcommand{\btheta}{\boldsymbol{\theta}}
\newcommand{\bgamma}{\boldsymbol{\gamma}}
\newcommand{\bX}{\mathbf{X}}
\newcommand{\mX}{\mathbf{X}}
\newcommand{\mP}{\mathbb{P}}

\newtheorem{assumption}{Assumption}
\newtheorem{theorem}{Theorem}
\newtheorem{prop}{Proposition}
\newtheorem{lemma}{Lemma}
\newtheorem{example}{Example}
\newtheorem{proposition}{Proposition}

\begin{document}

\def\spacingset#1{\renewcommand{\baselinestretch}%
{#1}\small\normalsize} \spacingset{1}


\if1\blind
{
  \title{\bf Doubly robust estimation and sensitivity analysis with outcomes truncated by death in multi-arm clinical trials}
\author{\textbf{Jiaqi Tong$^{1,2}$, Chao Cheng$^{3}$, Guangyu Tong$^{1,2,4}$, Michael O. Harhay$^{5}$ and Fan Li$^{1,2,*}$}

$^{1}$Department of Biostatistics, Yale School of Public Health, New Haven, CT, USA

$^{2}$Center for Methods in Implementation and Prevention Science, Yale School of Public Health, New Haven, CT, USA

$^{3}$Department of Statistics and Data Science, Washington University in St. Louis, St. Louis, CT, USA

$^{4}$Department of Internal Medicine, Section of Cardiovascular Medicine, Yale School of Medicine, New Haven, CT, USA

$^{5}$Department of Biostatistics, Epidemiology and Informatics, Perelman School of Medicine, 
University of Pennsylvania, Philadelphia, PA, USA


\emph{*email}: fan.f.li@yale.edu}
  \maketitle
} \fi

\if0\blind
{
  \bigskip
  \bigskip
  \bigskip
  \begin{center}
    {\LARGE\bf Doubly robust estimation and sensitivity analysis with outcomes truncated by death in multi-arm clinical trials}
\end{center}
  \medskip
} \fi

\bigskip
\begin{abstract}
In clinical trials, the observation of participant outcomes may frequently be hindered by death, leading to ambiguity in defining a scientifically meaningful final outcome for those who die. Principal stratification methods are valuable tools for addressing the average causal effect among always-survivors, i.e., the average treatment effect among a subpopulation defined as those who would survive regardless of treatment assignment. Although robust methods for the truncation-by-death problem in two-arm clinical trials have been previously studied, its expansion to multi-arm clinical trials remains elusive. In this article, we study the identification of a class of survivor average causal effect estimands with multiple treatments under monotonicity and principal ignorability, and first propose simple weighting and regression approaches for point estimation. As a further improvement, we derive the efficient influence function to motivate doubly robust estimators for the survivor average causal effects in multi-arm clinical trials. We also propose sensitivity methods under violations of key causal assumptions. Extensive simulations are conducted to investigate the finite-sample performance of the proposed methods against the existing methods, and a real data example is used to illustrate how to operationalize the proposed estimators and the sensitivity methods in practice. 
\end{abstract}

\noindent%
{\it Keywords:}  Causal inference; Multiple treatments; Principal stratification; Principal ignorability; Sensitivity analysis; Survivor average causal effect.
\vfill

\newpage
\spacingset{1.45} 

\section{Introduction}
\label{s:intro}


Truncation-by-death refers to the occurrence of death as an intermediate outcome (or intercurrent event) in a study that, in effect, precludes complete or partial observation of the outcome of interest \citep{rubin2006causal}. This issue is common in randomized clinical trials and impacts either the estimand definition or the interpretation of non-mortality outcomes. As survival status can be affected by treatment assignment, naive adjustment conditioning on survivors does not ensure a valid causal effect estimate. For example, a direct comparison of the quality of life outcomes between those who survive in the control versus those in the active treatment is prone to selection bias since treated survivors may not have survived had they been assigned to the control arm. Instead, a relevant causal estimand can be defined among those who would have survived regardless of the treatment assigned. Under the potential outcomes framework, \cite{frangakis2002principal} developed the principal stratification approach to define the principal causal effects by treating the joint potential values of the intermediate outcomes as pre-treatment covariates. Using this framework, the survivor average causal effect (SACE) represents the average potential outcome contrasts among a principal strata consisting of those who would have survived irrespective of the treatment assignment, and is causally interpretable. More broadly, the ICH E9(R1) addendum for the analysis of clinical trials \citep{ema2020} now explicitly specified principal stratification as one of the five strategies for dealing with intercurrent events with improved transparency in estimands.


Although principal stratification methods for a binary treatment have been previously developed, many randomized clinical trials include more than two arms. For example, a review of all randomized trials published in one month in 2012 found that 14\% had 3 arms and 7\% had 4 or more arms \citep{juszczak2019reporting}. Nevertheless, relatively fewer efforts have been devoted to principal stratification methods with multiple treatments with a few exceptions \citep{rubin2006causal}. Under monotonicity, \cite{elliott2006potential} proposed a Bayesian Gaussian mixture model to empirically identify SACEs with continuous outcomes, and \cite{wang2017causal} constructed testing procedures for detecting clinically meaningful SACEs in trials with ordinal treatments and binary outcomes. Extending the work in \cite{ding2011identifiability}, {\cite{luo2023causal} established point identification of SACEs by assuming either a scalar instrument variable that affects the final outcome only through the latent principal strata variable or a linear structural model for the outcome mean given the latent principal strata variable, treatment, and covariates, and further derived sharp bounds in the presence of covariates (see Section \ref{subs:sim_study} for details).} A summary of the literature on principal stratification with multiple treatments is provided in Table \ref{tabs:summary_litrevi}. For point identification of SACEs in multi-arm studies, a key limitation of the existing methods is that consistent estimation typically requires fully correctly specified parametric models, whereas estimators more robust to model misspecification are scarce. With a binary treatment, Ding and Lu\cite{DingandLu2016} proposed the principal score weighting estimator under principal ignorability; Jiang et al.\cite{JiangJRSSB2022} and Cheng et al.\cite{cheng2023multiply} studied triply robust estimators that leverage multiple working models to provide more chances to consistently estimate the principal causal effects. These robust methods, while attractive, have not been generalized to accommodate multiple treatments. 


\begin{table}[htbp]
\caption{Summary of literature on estimating SACEs with multiple treatments. We summarize the following features: i) whether applicable to randomized trials or observation studies; ii) number of treatments; iii) structural causal assumptions; iv) type of outcome; v) with or without covariates; vi) statistical methods; vii) whether sensitivity analysis is provided.}     \label{tabs:summary_litrevi}
    \centering
      \begin{adjustbox}{max width=\textwidth} 
    \begin{tabular}{lllll}
    \toprule
         & \cite{elliott2006potential}  & \cite{wang2017causal} &  \cite{luo2023causal} &This article\\ \hline
        \emph{Study design} & Randomized & Randomized  & Randomized \& Observational &Randomized \& Observational \\ 
        \emph{Number of treatments} & $\geq 3$  & $\geq 3$  & $\geq 3$&$\geq 3$\\ 
        \emph{Key assumptions} & Monoconicity  & Monotonicity  &Instrument \& monotonicity &Monotonicity \\ 
        \emph{Outcome type} & Continuous & Binary  & Continuous &Continuous\\ 
        \emph{Covariates} & With  & Without & With &With \\ 
        \emph{Methods} & Mixture model & Hypothesis testing& Model-based \& bounds &Semiparametric doubly robust \\ 
        \emph{Sensitivity analysis} & No  & No  & Partial& General framework\\ \bottomrule
    \end{tabular}
  \end{adjustbox}
\end{table}

In this article, we expand the work of \cite{DingandLu2016} and \cite{JiangJRSSB2022} to derive doubly robust estimators for the SACE estimands with multiple treatments under principal ignorability, with a focus on randomized clinical trials. We first develop the principal score weighting and outcome regression estimators. These two estimators are motivated by the moment conditions and are consistent if the associated working models are correctly specified, and hence only singly robust. To improve the model robustness, we further construct the efficient influence function to motivate doubly robust estimators, which are consistent if one set of working models is correctly specified, but not necessarily both. When all working models are correctly specified, the resulting estimators are semiparametrically efficient and achieve the variance lower bound among the class of regular and asymptotically linear estimators. Additionally, because doubly robust estimators rely on monotonicity and principal ignorability, we propose a sensitivity function approach to evaluate the estimation results when these assumptions are violated. In general, sensitivity methods for principal stratification analysis with multiple treatments are rare, except for \cite{luo2023causal}, who assessed monotonicity. However, their method is restricted to partial deviation from monotonicity between adjacent strata. In contrast, we provide a more general approach that accommodates broader departures from this assumption. {Furthermore, we generalize our developments to handle ignorable treatment assignment in the observational study settings \citep{li2019propensity}.} Our method is then illustrated by a four-arm randomized trial conducted by the National Toxicology Program to evaluate chemical effects in biological systems, where the final outcome - animal body weight - is truncated by death occurring before the conclusion of the study. We apply our proposed methods to estimate the survivor average causal effects and assess the sensitivity of results when key structural assumptions are violated.

The remainder of this manuscript is organized as follows. In Section \ref{s:model}, we introduce the notation, causal estimands, and the necessary causal structural assumptions to facilitate nonparametric identification. In Section \ref{s:iden&est}, we establish the nonparametric identification of the causal estimands and provide statistical inference procedures. In Section \ref{s:SA}, we present a sensitivity analysis framework to assess departures from the causal structural assumptions. {Section \ref{sec:generalization-os} provides a generalization of our methods to the observational studies with ignorable treatment assignments.} In Section \ref{subs:sim_study}, we conduct a thorough simulation study to investigate the performance of our methods against existing methods. In Section \ref{s:DE}, we present a case study to illustrate practical implementation. Section \ref{s:Dis} concludes with a discussion.

\section{Notation, causal estimands, and assumptions}
\label{s:model}
We consider a multi-arm randomized trial with $n$ units. For each unit, we observe a vector of pre-treatment covariates $\mathbf{X}$, an ordinal treatment $Z\in\mathcal{J}=\{1,\ldots,J\}$ with $J\geq 2$ levels, an intermediate survival status $S$ with $S=1$ indicating survival and $S=0$ indicating death, and a non-mortality outcome $Y$. We assume that $Y$ is measured at the end of the study and hence only well-defined among survivors with $S=1$ (the survival status is determined prior to the final outcome measurement). We pursue the potential outcomes framework, and define $S(z)\in\{0,1\}$ and $Y(z)$ as the potential values of the survival status and final outcome that would have been observed under treatment condition $z$. The Stable Unit Treatment Value Assumption allows us to connect $S$ and $Y$ with their potential values through $S=\sum_{z=1}^J \mathbf{1}(Z=z)S(z)$ and $Y=\sum_{z=1}^J \mathbf{1}(Z=z)Y(z)$ where $\mathbf{1}(\bullet)$ is the indicator function. 

{Under the principal stratification framework \citep{frangakis2002principal}, the joint potential survival status can be considered as a pre-treatment covariate that defines subgroup causal effects. Specifically, we define the basic principal stratum as 
$$G\in{\mathcal{G}}=\{(S(1),S(2),\ldots,S(J)):S(z)\in\{0,1\},z\in\mathcal{J}\}.$$
For simplicity, we relabel potential values of $G$ as $S(1)S(2)\ldots S(J)$. For example, with $J=4$ arms, $G=0111$ indicates the basic principal stratum with $S(1)=0$ and $S(2)=S(3)=S(4)=1$. We define $\mu_{\mathfrak{g}}(z)=E\{Y(z)| G=\mathfrak{g}\}$ as the mean of the potential outcome within stratum $\mathfrak{g}\in\mathcal{G}$. Importantly, $\mu_{\mathfrak{g}}(z)$ is well-defined if and only if the $z$-th coordinate of $\mathfrak{g}$ equals $1$ due to truncation by death. To enable simultaneous comparison among multiple treatments, our causal estimands are defined as the collection of pairwise SACEs:
\begin{equation}\label{def:pairPCE}
\Delta_{\mathfrak{g}}(z,z^\prime)=\mu_{\mathfrak{g}}(z)-\mu_{\mathfrak{g}}(z^\prime)=\ E\left\{Y(z)-Y(z^\prime)| G=\mathfrak{g}\right\},~~z\neq z'\in\mathcal{J},
\end{equation}
where stratum $\mathfrak{g}$ must satisfy $S(z)=S(z')=1$ to ensure that both $\mu_{\mathfrak{g}}(z)$ and $\mu_{\mathfrak{g}}(z^\prime)$ are well-defined.}

A few remarks are in order for the class of estimands in Equation \eqref{def:pairPCE}. First, the class of estimands is transitive such that $\Delta_\mathfrak{g}(z,z'')=\Delta_\mathfrak{g}(z,z')+\Delta_\mathfrak{g}(z',z'')$ if the form of $\mathfrak{g}$ satisfies $S(z)=S(z')=S(z'')=1$, and reflexive such that $\Delta_\mathfrak{g}(z,z')=-\Delta_\mathfrak{g}(z',z)$. Second, accounting for all possible combinations of treatment and strata, the cardinality of the class of estimands is $J(J-1)\times 2^{J-2}$ because there are $J(J-1)$ pairs of distinct $(z,z')$ in total and $2^{J-2}$ choices of stratum given the pair. Third, $\Delta_\mathfrak{g}\left(z,z^\prime\right)$ is identifiable if $\mu_\mathfrak{g}(z)$ is identifiable, and the cardinality of the class of estimands based on $\mu_\mathfrak{g}(z)$ is effectively reduced to $J\times 2^{J-1}$. In what follows, we focus on the identification and estimation of $\mu_\mathfrak{g}(z)$, $\forall z\in\mathcal{J}$, based on which all combinations of $\Delta_\mathfrak{g}(z,z')$ can be obtained. In a multi-arm randomized trial, we assume randomization such that $Z\perp\{S(1),\ldots,S(J),Y(1),\ldots,Y(J),\mathbf{X}\}$. {An extension to observational studies under ignorable treatment assignment is presented in Section \ref{sec:generalization-os}.} Next, we require the following two additional assumptions in order to point identify $\mu_\mathfrak{g}(z)$ under randomization.

\begin{assumption}[\emph{Monotonicity}]\label{assump:mono}
$S(z)\geq S(z^\prime)$ for $\forall~z\geq z^\prime\in\mathcal{J}$.
\end{assumption}
Assumption \ref{assump:mono} is commonly invoked for the point identification of SACEs \citep{DingandLu2016}, and is likely plausible when treatments are ordinal and higher dosages do not increase mortality. Under monotonicity, the number of principal strata is reduced from $2^J$ to $J+1$ due to the removal of the harmed stratum. Furthermore, under monotonicity, each element in $\mathcal{G}$ then takes the form of $\mathfrak{g}=0^{\otimes (J-g)}1^{\otimes g}$ with $g=0,\ldots,J$ {(i.e., $S(z)=0$ for $z\leq J-g$ and $S(z)=1$ for $z\geq J-g+1$). For notational simplicity, we will continue to use {the nonnegative integer $g$} to index each element in $\mathcal{G}$, versus the Fraktur notation `$\mathfrak{g}$' used in the original estimand definition \eqref{def:pairPCE} under more general settings. This also means that the estimand \eqref{def:pairPCE} is re-expressed as 
\begin{equation}\label{def:pairPCE_2}
\Delta_{g}(z,z^\prime)=\mu_{g}(z)-\mu_{g}(z^\prime)=\ E\left\{Y(z)-Y(z^\prime)| G=g\right\},~~z\neq z'\in\mathcal{J}.
\end{equation}}

Under this simplified notation structure, Table \ref{tab:G+S|Z} defines the latent principal strata $G$ and shows their relationship with survival status conditional on treatment arms, under monotonicity. The monotonicity assumption also implies that $\mu_g(z)$ is only defined within $g+z\geq J+1$ and that the contrast estimands in Equation \eqref{def:pairPCE} are only defined when both $g+z$ and $g+z'$ are not smaller than $J+1$. {To see this, we recall that $\mu_g(z)$ is well-defined only if the $z$-th coordinate of $\mathfrak{g}$ is one, which, by monotonicity, implies that all subsequent coordinates from $z+1$ to $J$ are also one. This then explains why the principal stratum $\mathfrak{g}$ must take the form $0^{\otimes (J-g)}1^{\otimes g}$ for some $g$ satisfying $g\geq J-z+1$.}

\begin{table}[htbp]
\centering
\caption{Correspondence between latent principal strata $G=g,g\in{\mathcal{Q}\equiv\mathcal{J}\cup\{0\}}$ and survivors conditional on treatment arms, $S=1|Z=z,z\in\mathcal{J}$, under monotonicity. The notation $\checkmark $ (yes) and $\times$ (no) denote whether the survivors in arm $z$ are a mixture of the principal strata $g$. The content of this table is adapted from Table 1 in \cite{luo2023causal}.} 
\label{tab:G+S|Z}
      \begin{adjustbox}{max width=\textwidth} 
\begin{tabular}{c|l|lllll}
\toprule
&Principal strata in shorthand notation&$Z=1$&$Z=2$&$\ldots$&$Z=J-1$&$Z=J$\\
\hline
\multirow{ 6}{*}{Definition of strata} & $g=0$&$S(1)=0$&$S(2)=0$&$\ldots$&$S(J-1)=0$&$S(J)=0$\\

&$g=1$&$S(1)=0$&$S(2)=0$&$\ldots$&$S(J-1)=0$&$S(J)=1$\\

&$g=2$&$S(1)=0$&$S(2)=0$&$\ldots$&$S(J-1)=1$&$S(J)=1$\\

&$\vdots$&$\vdots$&$\vdots$&$\ddots$&$\vdots$&$\vdots$\\

&$g=J-1$&$S(1)=0$&$S(2)=1$&$\ldots$&$S(J-1)=1$&$S(J)=1$\\

&$g=J$&$S(1)=1$&$S(2)=1$&$\ldots$&$S(J-1)=1$&$S(J)=1$\\
\hline

&Observed survivor subgroups &$g=0$&$g=1$&$\ldots$&$g=J-1$&$g=J$\\
\hline
\multirow{5}{*}{Mixture components} & $S=1|Z=1$&$\times$ & $\times$&$\ldots$&$\times$&$\checkmark$\\

& $S=1|Z=2$&$\times$&$\times$&$\ldots$&$\checkmark $&$\checkmark $\\

& $\vdots$&$\vdots$&$\vdots$&$\ddots$&$\vdots$&$\vdots$\\

& $S=1|Z=J-1$&$\times$&$\times$&$\ldots$&$\checkmark $&$\checkmark $\\

& $S=1|Z=J$&$\times$&$\checkmark $&$\ldots$&$\checkmark $&$\checkmark $\\
\bottomrule
\end{tabular}
\end{adjustbox}
\end{table}

\begin{assumption}[\emph{Principal Ignorability}]\label{assump:GPI}
For any $z\in\mathcal{J}$ {and any $\mathfrak{g},\mathfrak{g}'$ such that $S(z)=1$}, 
$E\{Y(z)|G=\mathfrak{g},\mathbf{X}\}=E\{Y(z)|G=\mathfrak{g}',\mathbf{X}\}$.
\end{assumption}

{Under monotonicity, the condition that $\mathfrak{g},\mathfrak{g}'$ satisfy $S(z)=1$ is equivalent to requiring $g,g'\in\{J-z+1,\ldots,J\}$.} Assumption \ref{assump:GPI} extends the principal ignorability assumption of \cite{DingandLu2016} and \cite{JiangJRSSB2022} to multiple treatments with $J\geq 2$. It posits that, conditional on measured covariates $\mathbf{X}$, the expectation of the potential outcome does not vary across the basic principal strata of survivors. In other words, $\mathbf{X}$ fully accounts for any confounding between the potential final non-mortality outcome and potential survival status. {To aid illustration, Example \ref{eg:4-arm} demonstrates the monotonicity and principal ignorability assumptions in the context of a four-arm clinical trial.} It is worth noting that, both monotonicity and principal ignorability assumptions involve cross-world conditions, and are therefore unverifiable from the observed data alone. In Section \ref{s:SA}, we present a sensitivity analysis framework to assess the impact of departure from these assumptions in multi-arm trials. 

\begin{example}\label{eg:4-arm}
{Consider a four-arm trial with $J=4$ ordinal treatment levels, where the principal strata can be denoted by a four-digit binary number $G=S(1)S(2)S(3)S(4)$. Monotonicity rules out individuals who would survive under lower treatment levels but die under higher treatment levels, thereby precluding existence of strata with $S(z)=1$ but $S(z+j)=0$, for some $1\leq z\leq J-1$ and $j \geq 1$. Therefore, under monotonicity, at most five strata exist: $G=0000$, $0001$, $0011$, $0111$, and $1111$.
These five strata characterize individuals (i) who would always not survive, regardless of the treatment level, (ii) who would survive only under the highest treatment level, (iii) who would survive under treatment level $3$ or above, (iv) who would survive under treatment level $2$ or above, (v) who would always survive, regardless of treatment levels. 
Further assuming principal ignorability, we require that the mean of counterfactual outcomes satisfy the following three homogeneity conditions (a)--(c):
\begin{enumerate}[itemsep=-3ex]
\item[(a)] $E[Y(2)|G=0111,\mathbf X]=E[Y(2)|G=1111,\mathbf X]$.
\item[(b)] $E[Y(3)|G=0011,\mathbf X]=E[Y(3)|G=0111,\mathbf X]=E[Y(3)|G=1111,\mathbf X]$.
\item[(c)] $E[Y(4)|G=0001,\mathbf X]=E[Y(4)|G=0011,\mathbf X]=E[Y(4)|G=0111,\mathbf X]=E[Y(4)|G=1111,\mathbf X]$.
\end{enumerate}
In words, the above conditions assume that conditional on covariates $\mathbf X$, the expected potential outcome under treatment level 2, 3, or 4 is exchangeable across all principal strata who would survive under treatment level 2, 3, or 4, respectively. 
}
\end{example}

\section{Identification and estimation}\label{s:iden&est}

\subsection{Principal score weighting estimator}\label{subs:psw}
We first consider the principal score weighting approach to estimate $\mu_g(z)$ \citep{DingandLu2016}. The principal score is defined as the probability of an individual belonging to the stratum $g$ conditional on baseline covariates $\mathbf{X}$: $e_g(\mathbf{X})=\Pr(G=g|\mathbf{X})$ for $g\in\mathcal{Q}=\{0,\ldots,J\}$. 
We also define $e_g=E\{e_g(\mathbf{X})\}$ as the marginal principal score for the stratum $g$. Note that $\mu_g(z)$ is well-defined only if $e_g>0$. Since $G$ is only partially observed, we leverage the information from the observed survival status and monotonicity to point-identify the principal score. Under Assumption \ref{assump:mono}, we show in the Supplementary Material that the principal score can be identified from the probability of survival conditional on the treatment and covariates, expressed through the following series of equations:
\begin{equation}\label{eq:PrinScoreIden}
e_{g}(\mathbf{X})=p_{J-g+1}(\mathbf{X})-p_{J-g}(\mathbf{X}),\quad g\in\mathcal{Q},
\end{equation}
where $p_z(\mathbf{X})=\Pr(S=1|Z=z,\mathbf{X})$ for $z\in\mathcal{J}$, and for completeness, we also define $p_0(\mathbf{X})=0$ and $p_{J+1}(\mathbf{X})=1$. Hereafter, we refer to $p_z(\mathbf{X})$ as the principal score because \eqref{eq:PrinScoreIden} defines a bijection between $\{e_0(\mathbf{X}),\ldots,e_J(\mathbf{X})\}$ and $\{p_1(\mathbf{X}),\ldots,p_{J+1}(\mathbf{X})\}$. We then define the following set of principal score weights 
\begin{equation}
\nonumber
    w_{zg}(\mathbf{X})=\left\{\frac{e_g}{\sum_{g'=J-z+1}^{J}e_{g'}}\right\}^{-1}\frac{e_g(\mathbf{X})}{\sum_{g'=J-z+1}^{J}e_{g'}(\mathbf{X})} ,~~~~z\in\mathcal{J},~~~~g\geq J-z+1.
\end{equation}
Based on \eqref{eq:PrinScoreIden}, one can write out $w_{zg}(\mathbf{X})$ as
\begin{equation}\nonumber
w_{zg}(\mathbf{X})=\left\{\frac{p_{J-g+1}-p_{J-g}}{p_z}\right\}^{-1}\frac{p_{J-g+1}(\mathbf{X})-p_{J-g}(\mathbf{X})}{p_z(\mathbf{X})},
\end{equation}
where $p_z=E\{p_z(\mathbf{X})\}$ {is the observed survival probability conditional on $Z=z$, marginalized over covariates $\mathbf X$.} Under Assumptions \ref{assump:mono} and \ref{assump:GPI}, $\mu_g(z)$ is then identified by
\begin{equation}\label{eq:iden_ps_weighting}
    \mu_g(z)=E\left\{w_{zg}(\mathbf{X})Y|Z=z,S=1\right\}
    ,
\end{equation}
which is an expectation of the observed outcome conditional on treatment $z$ and 
survivors, weighted by $w_{zg}(\mathbf{X})$. The weights $w_{zg}(\mathbf{X})$ are functions of principal scores and, more precisely, they are proportional to the ratio of the principal score for stratum $g$ and the total principal score for a set of strata whose members will all survive under arm $z$ or with $S(z)=1$. In fact, $w_{zg}(\mathbf{X})$ represents the importance sampling weights for the probability distribution of covariates conditional on the survivors, treatment, and principal stratum versus that conditional on the survivors and treatment only. The identification formula \eqref{eq:iden_ps_weighting} generalizes the results under binary treatment proposed by \cite{DingandLu2016} to $J\geq2$. 

The identification formula \eqref{eq:iden_ps_weighting} corresponds to a collection of balancing conditions for the arbitrary vector-valued function of covariates $h(\mathbf{X})$. That is, replacing the final outcome $Y$ in \eqref{eq:iden_ps_weighting} with an arbitrary $h(\mathbf{X})$ yields the balancing properties of the principal score weights. To see this, under Assumptions \ref{assump:mono} and \ref{assump:GPI}, for $\forall g$ and $\forall ~z\geq J+1-g$, we have 
$E\{h(\mathbf{X})|G=g\}=E\left\{w_{zg}(\mathbf{X})h(\mathbf{X})|Z=z,S=1\right\}$. 
Then just as one could check the adequacy of propensity score models in observational studies with multiple treatments \citep{li2019propensity}, the empirical counterparts corresponding to the covariate balancing conditions motivate natural criteria to check if the estimated principal scores sufficiently balance the covariates and are thus adequate. Operationally, one can follow Section 5.2 in \cite{cheng2023multiply} to construct a set of weighted standardized mean difference metrics and consider an iterative checking-fitting process to arrive at a final principal score model without peeking at the final outcome.


To implement the principal score weighting estimator, for any $z\in\mathcal{J}$, we can posit a parametric working model $p_z(\mathbf{X};\boldsymbol{\alpha}_z)$ with a vector of unknown parameters $\boldsymbol{\alpha}_z$ for $p_z(\mathbf{X})$, where $\widehat{\balpha}_z$ is obtained by solving a maximum likelihood score equation $\mathbb{P}_n\{\kappa_z(S,Z,\mathbf{X};{\balpha}_z)\}=\mathbf 0$. Here, $\kappa_z(S,Z,\mathbf{X};{\balpha}_z)$ is the score function of a binary regression model and $\mathbb{P}_n\{V\}=n^{-1}\sum_{i=1}^n V_i$ defines the empirical mean. 
We consider the following plug-in estimator $\widehat{p}_z(\mathbf{X})=p_z(\mathbf{X};\widehat{\balpha}_z)$. We note $p_z(\mathbf{X};\widetilde{\boldsymbol{\alpha}}_z)=p_z(\mathbf{X})$ when $p_z(\mathbf{X};\boldsymbol{\alpha}_z)$ is correctly specified, where $\widetilde{\boldsymbol{\alpha}}_z$ is the probability limit of $\widehat{\balpha}_z$. 
To reduce the dependence on the parametric working model, we then use a simple non-parametric estimator, $\widehat{p}_z=\mathbb{P}_n\{\mathbf{1}(Z=z)S\}/\pi_z,z\in\mathcal{J}$, where $\pi_z=\Pr(Z=z)$ is the treatment probability and is known by the study design. Then \eqref{eq:iden_ps_weighting} leads to the following weighting estimator, which is consistent when the principal score working model is correctly specified,
\begin{equation}\nonumber
    \widehat{\mu}^{\text{PSW}}_g(z)=
    \frac{\mathbb{P}_n\left\{\widehat{w}_{zg}(\mathbf{X})\mathbf{1}(Z=z)SY\right\}}{\mathbb{P}_n\left\{\mathbf{1}(Z=z)S\right\}}.
\end{equation}

\subsection{Outcome regression estimator}\label{subs:or}
Alternatively, we can estimate $\mu_g(z)$ by postulating non-mortality outcome models. Define the mean of the observed final outcome conditional on treatment, survivors, and covariates as $m_{z}(\mathbf{X})=E\{Y|Z=z,S=1,\mathbf{X}\}.$ Under Assumptions \ref{assump:mono} and \ref{assump:GPI}, we show in the Supplementary Material that the following identification formula for $\mu_g(z)$ holds for $g\in \mathcal J$,
\begin{equation}
\label{eq:iden_outcome_regression}
\mu_g(z)=E\left\{\frac{\mathbf{1}(Z=J-g+1)S/\pi_{J-g+1}-\mathbf{1}(Z=J-g)S/\pi_{J-g}}{p_{J-g+1}-p_{J-g}}m_{z}(\mathbf{X})\right\}.
\end{equation}
For completeness, we define $\mathbf{1}(Z=0)/\pi_{0} = 0$ when calculating $\mu_J(z)$.
{Similar to \eqref{eq:iden_ps_weighting},}
\eqref{eq:iden_outcome_regression} also motivates the balancing conditions by replacing $m_{z}(\mathbf{X})$ with arbitrary vector-valued random functions of covariates $h(\mathbf{X})$. That is, under Assumptions \ref{assump:mono} and \ref{assump:GPI},
$$E\left\{\frac{\mathbf{1}(Z=J-g+1)S/\pi_{J-g+1}-\mathbf{1}(Z=J-g)S/\pi_{J-g}}{p_{J-g+1}-p_{J-g}}h(\mathbf{X})\right\}=E\left\{h(\mathbf{X})|G=g\right\}.$$
To implement this estimator, we posit a parametric working model $m_{z}(\mathbf{X};\boldsymbol{\gamma}_{z})$ for $m_{z}(\mathbf{X})$, where $\boldsymbol{\gamma}_{z}$ is a vector of unknown parameters. Analogously,  $\widehat{\boldsymbol{\gamma}}_{z}$ can be obtained by solving a generalized estimating equation $\mathbb{P}_n\{\tau_z(\mathbf{V};{\boldsymbol\gamma}_z)\}=\mathbf 0$, where $\mathbf{V}=(Y,S,Z,\mathbf{X}^\top)^\top$ is the observed data vector and $\tau_z(\mathbf{V};{\boldsymbol\gamma}_z)$ are the unbiased estimating function determined by the outcome model specification (for example, the score function). We define the probability limit for $\widehat{\boldsymbol{\gamma}}_{z}$ as $\widetilde{\boldsymbol{\gamma}}_{z}$, and under the true working model and suitable regularity conditions, $m_{z}(\mathbf{X};\widetilde{\boldsymbol{\gamma}}_{z})=m_{z}(\mathbf{X})$. 
We then propose the following estimators based on the empirical counterparts of \eqref{eq:iden_outcome_regression}
\begin{equation}
\nonumber
\widehat{\mu}^{\text{OR}}_g(z)=\mathbb{P}_n\left\{\frac{\mathbf{1}(Z=J-g+1)S/\pi_{J-g+1}-\mathbf{1}(Z=J-g)S/\pi_{J-g}}{\widehat{p}_{J-g+1}-\widehat{p}_{J-g}}\widehat{m}_{z}(\mathbf{X})\right\},
\end{equation}
for $g\in \mathcal J$, where  $\widehat{m}_{z}(\mathbf{X})=m_{z}(\mathbf{X};\widehat{\boldsymbol\gamma}_{z})$.
In its current form, $\widehat{\mu}^{\text{OR}}_g(z)$ is a g-computation formula estimator that standardizes the outcome model estimate to the target principal strata subpopulation, and $\widehat{\mu}^{\text{OR}}_g(z)$ is consistent if $m_{z}(\mathbf{X};\boldsymbol{\gamma}_{z})$ is correctly specified. 

\subsection{Doubly robust and locally efficient estimator}\label{subs:dre}
To further improve upon the weighting and regression estimators, we first derive the efficient influence function for $\mu_g(z)$ under the nonparametric model $\mathcal M_{np}$ of the observed data $\mathbf{V}$ in a sense that we place no restrictions on $\mathcal M_{np}$. 
Derivation of the efficient influence function follows the standard procedure established under the general semiparametric efficiency theory \citep{bickel1993efficient}, and generalizes the derivation from \cite{JiangJRSSB2022} from a binary treatment to multiple treatments. To proceed, for $z\in\mathcal{J}$, we first define the following  quantity for any function $F(Y,S,\mathbf{X})$:
$$\psi_{F(Y,S,\mathbf{X}),z}=\frac{\mathbf{1}(Z=z)}{\pi_z}\Big\{F(Y,S,\mathbf{X})-E\{F(Y,S,\mathbf{X})|Z=z,\mathbf{X}\}\Big\}+E\{F(Y,S,\mathbf{X})|Z=z,\mathbf{X}\}.$$
To facilitate exposition, we also define $\psi_{F(Y,S,\mathbf{X}),0}=0$ and $\psi_{F(Y,S,\mathbf{X}),J+1}=1$. In addition, we define two quantities that appear in the efficient influence function as 
\begin{align*}
\psi_{S,z}=&\frac{\mathbf{1}(Z=z)}{\pi_z}\{S-p_z(\mathbf{X})\}+p_z(\mathbf{X}),\\
\psi_{YS,z}=&\frac{\mathbf{1}(Z=z)}{\pi_z}\{YS-m_{z}(\mathbf{X})p_z(\mathbf{X})\}+m_{z}(\mathbf{X})p_z(\mathbf{X}).
\end{align*}
{These two functions can be seen as the uncentered efficient influence functions for estimating $E\{S(z)\}$ and $E\{Y(z)S(z)\}$, respectively.} Next, Theorem \ref{thm:eif} gives the form of the efficient influence function for $\mu_g(z)$.
\begin{theorem}[\emph{Efficient Influence Function}]\label{thm:eif}
       For any $z\in\mathcal{J}$ and $g\geq J-z+1$, the efficient influence function for $
       \mu_g(z)$ under the nonparametric model $\mathcal M_{np}$ is $\Psi_{zg}(\mathbf{V}) = {\xi_{zg}(\mathbf{V})}/{(p_{J-g+1}-p_{J-g}})$, where 
       $$\xi_{zg}(\mathbf{V})=\displaystyle \frac{\{p_{J-g+1}(\mathbf{X})-p_{J-g}(\mathbf{X})\}\{\psi_{YS,z}-m_{z}(\mathbf{X})\psi_{S,z}\}}{p_z(\mathbf{X})}+\{m_{z}(\mathbf{X})-\mu_g(z)\}(\psi_{S,J-g+1}-\psi_{S,J-g}).$$
       Therefore, the semiparametric efficiency  bound for estimating $
       \mu_g(z)$ is $E\{[\Psi_{zg}(\mathbf{V})]^2\}$.
\end{theorem}
Theorem \ref{thm:eif} suggests a new estimator, $\widehat\mu_g^{\text{DR}}(z)$, by solving the efficient influence function based estimating equation in terms of $\mu_g(z)$, where the unknown nuisance functions, $\{p_z(\mathbf{X}),m_z(\mathbf{X})\}$ are estimated by parametric working models as in Sections \ref{subs:psw} and \ref{subs:or}. Because the denominator of the efficient influence function is a constant with respect to the estimand, $\widehat\mu_g^{\text{DR}}(z)$ is the solution of $\mathbb{P}_n\left\{\xi_{zg}(\mathbf{V};\mu_g(z),\widehat{\bm\alpha}_{J-g+1},\widehat{\bm\alpha}_{J-g},\widehat{\bm\alpha}_{z},\widehat{\bm\gamma}_z)\right\}=0$ in $\mu_g(z)$, where $\xi_{zg}(\mathbf{V};\mu_g(z),{\bm\alpha}_{J-g+1},\allowbreak {\bm\alpha}_{J-g},{\bm\alpha}_{z},{\bm\gamma}_z)$ is $\xi_{zg}(\mathbf{V})$ evaluated based on the parametric working models. 
After some algebraic simplifications, we obtain 
\begin{equation}\nonumber
    \widehat{\mu}_g^{\text{DR}}(z)=\frac{ \mathbb{P}_n\left\{\displaystyle \frac{\widehat{p}_{J-g+1}(\mathbf{X})-\widehat{p}_{J-g}(\mathbf{X})}{\widehat{p}_z(\mathbf{X})}\left\{\widehat\psi_{YS,z}-\widehat m_{z}(\mathbf{X})\widehat \psi_{S,z}\right\}+\widehat{m}_{z}(\mathbf{X})(\widehat{\psi}_{S,J-g+1}-\widehat{\psi}_{S,J-g})\right\}}{\mathbb{P}_n\{\widehat{\psi}_{S,J-g+1}-\widehat{\psi}_{S,J-g}\}},
\end{equation}
where $\{\widehat \psi_{S,z},\widehat \psi_{YS,z}\}$ are $\{\psi_{S,z},\psi_{YS,z}\}$ evaluated based on  $\widehat p_z(\mathbf{X})$ and $\widehat m_z(\mathbf{X})$. 
Finally, as a further improvement for estimating $p_z$, an augmented estimator, $\mathbb{P}_n\{\widehat{\psi}_{S,z}\}$, is used in $\widehat{\mu}_g^{\text{DR}}(z)$, as it is always consistent for $p_z$ even under arbitrary misspecifications of $p_z(\mathbf{X};\boldsymbol{\alpha}_z)$, due to randomization. We summarize the large-sample properties of $\widehat{\mu}_g^{\text{DR}}(z)$ in Theorem \ref{thm:triply robustness} below. 
\begin{theorem}[\emph{Double Robustness and Local Efficiency}]\label{thm:triply robustness}
    Suppose that Assumptions \ref{assump:mono} and \ref{assump:GPI} hold and $\{p_z(\mathbf{X};\widetilde{\boldsymbol{\alpha}}_z),p_z(\mathbf{X};\widehat{\boldsymbol{\alpha}}_z)\}$ are uniformly bounded away from 0 and 1. Then, $\widehat{\mu}^{\text{DR}}_g(z)$ is consistent and asymptotically normal if either $p_z(\mathbf{X};\boldsymbol{\alpha}_z)$ or $m_{z}(\mathbf{X};\boldsymbol{\gamma}_{z})$ is correctly specified. If both models are correctly specified, the asymptotic variance of $\widehat{\mu}_g^{\text{DR}}(z)$ achieves the efficiency lower bound and $\widehat{\mu}_g^{\text{DR}}(z)$ is locally efficient. 
\end{theorem}
By Theorem \ref{thm:triply robustness}, $\widehat{\mu}^{\text{DR}}_g(z)$ is doubly robust in a sense that the bias is asymptotically negligible if either $p_z(\mathbf{X};\boldsymbol{\alpha}_z)$ or $m_z(\mathbf{X};\boldsymbol{\gamma}_z)$ is correct, but not necessarily both. When both are correctly specified, $\widehat{\mu}^{\text{DR}}_g(z)$ is locally efficient in the sense with asymptotic variance $E\{[\Psi_{zg}(\mathbf{V})]^2\}$ and is an optimal estimator among the class of regular and asymptotically linear estimators for the same target estimand $\mu_g(z)$. {In Section \ref{sec:generalization-os}, we extend the doubly robust estimator from randomized trials to observational settings, where the assignment mechanisim is unknown and must be estimated. In this case, three working models need to be specified: one for the propensity score, one for the principal score, and one for the outcome mean. This gives a triply robust estimator, which is consistent if any two out of the three working models are correctly specified. 
When the propensity score is known, the triply robust estimator automatically reduces to the doubly robust estimator. See Section \ref{sec:generalization-os} for further discussions.}

\subsection{Variance estimation}\label{subs:var_est}

The SACEs are estimated by $\widehat\Delta_g^{\text{PSW}}(z,z^\prime)=\widehat{\mu}^{\text{PSW}}_g(z)-\widehat{\mu}^{\text{PSW}}_g(z^\prime)$, $\widehat\Delta_g^{\text{OR}}(z,z^\prime)=\widehat{\mu}^{\text{OR}}_g(z)-\widehat{\mu}^{\text{OR}}_g(z^\prime)$, and $\widehat\Delta_g^{\text{DR}}(z,z^\prime)=\widehat{\mu}^{\text{DR}}_g(z)-\widehat{\mu}^{\text{DR}}_g(z^\prime)$, if the principal score weighting, outcome regression, and doubly robust approach are used. We propose to use the sandwich variance approach to estimate their asymptotic variances, and construct a Wald confidence interval for statistical inference. Below, we describe the variance estimator for $\widehat\Delta_g^{\text{DR}}(z,z')$, and the remaining variance estimators follow a similar construction and are provided in the Supplementary Material. 
Define $\btheta^{\text{DR}}=(\mu_g(z),\mu_g(z'),\bm\alpha_{J-g+1}^\top,\bm\alpha_{J-g}^\top,\bm\alpha_{z}^\top,\bm\alpha_{z'}^\top,{\bm\gamma}_z^\top,{\bm\gamma}_{z'}^\top)^\top$ that includes all parameters used to construct $\widehat{\Delta}_g^{\text{DR}}(z,z')$. Thus, $\widehat{\btheta}^{\text{DR}}=(\widehat\mu_g^{\text{DR}}(z),\widehat\mu_g^{\text{DR}}(z'),\widehat{\bm\alpha}_{J-g+1}^\top,\\\widehat{\bm\alpha}_{J-g}^\top,\widehat{\bm\alpha}_{z}^\top,\widehat{\bm\alpha}_{z'}^\top,\widehat{\bm\gamma}_z^\top,\widehat{\bm\gamma}_{z'}^\top)^\top$ can be treated as the solution of the joint estimating equation $\mathbb{P}_n\{\Phi(\mathbf{V};\btheta^{\text{DR}})\}=\mathbf{0}$ with
\begin{align*}
\Phi(\mathbf{V};\btheta^{\text{DR}}) = \begin{pmatrix}
\xi_{zg}(\mathbf V;\mu_g(z),{\bm\alpha}_{J-g+1}, {\bm\alpha}_{J-g},{\bm\alpha}_{z},{\bm\gamma}_z) \\
\xi_{z'g}(\mathbf V;\mu_g(z'),{\bm\alpha}_{J-g+1}, {\bm\alpha}_{J-g},{\bm\alpha}_{z'},{\bm\gamma}_{z'}) \\
\xi_{\text{nuisance}}(\mathbf V;\bm\alpha_{J-g+1},\bm\alpha_{J-g},\bm\alpha_{z},\bm\alpha_{z'},{\bm\gamma}_z,{\bm\gamma}_{z'})
\end{pmatrix},
\end{align*}
where $\xi_{\text{nuisance}} \equiv (\kappa_{J-g+1}^\top(S,Z,\mathbf{X};{\balpha}_{J-g+1}),\kappa_{J-g}^\top(S,Z,\mathbf{X};{\balpha}_{J-g}),\kappa_z^\top(S,Z,\mathbf{X};{\balpha}_z),\kappa_{z^\prime}^\top(S,Z,\mathbf{X};{\balpha}_{z^\prime}),\\\tau_z^{\top}(\mathbf{V};{\bgamma}_{z}),\tau_{z^\prime}^\top(\mathbf{V};{\bgamma}_{z^\prime}))^\top$ is the collection of score vectors of the nuisance parameters, and the first element in $\xi_{\text{nuisance}}$ is excluded if $J-g+1=z\text{ or }z^\prime$ and the second element in $\xi_{\text{nuisance}}$ is discarded if $g=J$. The doubly robust SACE estimator is therefore $\widehat{\Delta}_g^{\text{DR}}(z,z') = {\bm{\lambda}}^\top \widehat{\btheta}^{\text{DR}}$ where $\bm{\lambda} = (1,-1,{\bm{0}}^\top)^\top$ is a vector with the first element 1, second element $-1$, and all other elements 0. Following  regularity conditions in Theorem 5.41 in \cite{van2000asymptotic}, $\sqrt{n}(\widehat{\btheta}^{\text{DR}}-\widetilde{\btheta}^{\text{DR}})$ converges to a mean-zero normal distribution with the variance consistently estimated by 
$$\widehat{\mathbb{V}}(\widehat{\btheta}^{\text{DR}})\equiv\mathbb{P}_n\left\{\frac{\partial \Phi(\mathbf{V};\widehat{\btheta}^{\text{DR}})}{\partial {\btheta^{\text{DR}}}^\top}\right\}^{-1}\mathbb{P}_n\left\{\Phi(\mathbf{V};\widehat{\btheta}^{\text{DR}})\Phi^\top(\mathbf{V};\widehat{\btheta}^{\text{DR}})\right\} 
    {\mathbb{P}_n\left\{\frac{\partial \Phi(\mathbf{V};\widehat{\btheta}^{\text{DR}})}{\partial {\btheta^{\text{DR}}}^\top}\right\}^{-\top}},$$ 
where $\widetilde{\btheta}^{\text{DR}}$ is the unique solution to $E\{\Phi(\mathbf{V};\btheta^{\text{DR}})\}=\bm 0$. By the delta method, the sandwich variance estimator of $\widehat{\Delta}_g^{\text{DR}}(z,z')$ is $n^{-1}\bm\lambda^\top \widehat{\mathbb{V}}(\widehat{\btheta}^{\text{DR}})\bm\lambda$. The finite-sample performance of the proposed variance estimators is investigated in Section \ref{subs:sim_study}.

\section{Sensitivity analysis methods under violations of causal assumptions}\label{s:SA}

Since the validity of the estimators in Section \ref{s:iden&est} depends on Assumptions \ref{assump:mono} and \ref{assump:GPI}, we further develop sensitivity analysis methods under violations of these two structural assumptions. To focus ideas, when we investigate sensitivity under departure from one assumption, we assume the other assumption holds. 

\subsection{Sensitivity analysis for principal ignorability}\label{s:SA;subsec:PI}

Let $\widetilde g\in \mathcal J$ be a reference stratum. We suppose that,  
$E\{Y(z)|G=\widetilde{g},\mathbf{X}\}\neq0$ almost surely  $\forall~z\geq J-\widetilde{g}+1$. We then define the following set of sensitivity functions with respect to the reference stratum $\widetilde{g}=J$,
\begin{equation}\label{eq:sensitivity}
\delta_{zg}(\mathbf{X})=\displaystyle\frac{E\{Y(z)|G=g,\mathbf{X}\}}{E\{Y(z)|G=J,\mathbf{X}\}},~g\geq J-z+1,~z\in\mathcal{J},
\end{equation}
where $\delta_{zJ}(\mathbf{X})=1$ by construction, and the cardinality of the set of non-trivial sensitivity functions is $J\times (J-1)/2$. Of note, the reference stratum can be user-defined; for example, one may pick any $\widetilde{g}\geq J-z+1$ as a reference group, and then define
\begin{equation}\label{eq:sensitivity_userdefined}
\delta'_{zg}(\mathbf{X})=\displaystyle\frac{E\{Y(z)|G=g,\mathbf{X}\}}{E\{Y(z)|G=\widetilde{g},\mathbf{X}\}},~g\geq J-z+1,~g\neq \widetilde{g},~z\in\mathcal{J},
\end{equation}
as a general set of sensitivity functions. Then $\delta_{zg}(\mathbf{X})$ can be recovered from the quantities in \eqref{eq:sensitivity_userdefined} with $\delta_{zg}(\mathbf{X})=\delta'_{zg}(\mathbf{X})/\delta'_{zJ}(\mathbf{X})$. Therefore, we take 
$\widetilde{g}=J$ as the reference stratum without loss of generality, but for the simplicity of presentation. 

Recall that Assumption \ref{assump:GPI} is equivalent to $\delta_{zg}(\mathbf{X})=1$ for $\forall z,g$. However, when Assumption \ref{assump:GPI} is violated, at least one sensitivity function $\delta_{zg}(\mathbf{X})$ deviates from unity. Suppose that monotonicity assumption holds and the sensitivity functions $\delta_{zg}(\mathbf{X})$ are known. Then, for $z\in\mathcal{J}$ and $g\geq J-z+1$, $\mu_g(z)$ can be identified in \eqref{eq:iden_ps_weighting} by replacing the (standard) principal score weight $w_{zg}(\mathbf{X})$ with the following bias-corrected principal score weight
\begin{equation}\label{eq:weights_without_PI}
    w_{zg}^{\text{BC-PI}}(\mathbf{X})= w_{zg}(\mathbf{X})\Omega_{zg}(\mathbf{X}).
\end{equation}
Here, $\Omega_{zg}(\mathbf{X})$, referred to as the \textit{sensitivity weight}, is defined as
\begin{equation}\nonumber
   \Omega_{zg}(\mathbf{X})=
 \frac{\delta_{zg}(\mathbf{X})p_z(\mathbf{X})}
 {\sum_{g^\prime\geq J+1-z }\delta_{zg^\prime}(\mathbf{X})\left\{p_{J-g^\prime+1}(\mathbf{X})-p_{J-g^\prime}(\mathbf{X})\right\}}, \quad \text{$z\in\mathcal{J}$ and $g\geq J-z+1$,}
\end{equation}
which depends on the sensitivity functions, $\delta_{zg}(\mathbf{X})$, and the principal scores, $p_z(\mathbf{X})$. The sensitivity weight arises naturally through an algebraic transformation of the original identification formula when principal ignorability does not hold. Evidently, when principal ignorability holds, $\delta_{zg}(\mathbf{X})=1$ for all $(z,g)$, implying $\Omega_{zg}(\mathbf{X})=1$, and therefore the bias-corrected principal score weight $w_{zg}^{\text{BC-PI}}$ degenerates to the standard principal score weight. Otherwise, $\Omega_{zg}(\mathbf{X}) \neq 1$, and the adjustment is applied to remove the bias in the identification formula \eqref{eq:iden_ps_weighting}.

Similarly, the identification formulas based on outcome regression are also multiplied by $\Omega_{zg}(\mathbf{X})$ within the expectation, which gives, for $g=1,\ldots,J-1$,
\begin{equation}\label{eq:iden_outcome_regression_withoutPI}
\mu_g(z)=E\left\{\frac{\mathbf{1}(Z=J-g+1)S/\pi_{J-g+1}-\mathbf{1}(Z=J-g)S/\pi_{J-g}}{p_{J-g+1}-p_{J-g}}\Omega_{zg}(\mathbf{X})m_{z}(\mathbf{X})\right\}.
\end{equation}
In practice, specifying particular functional forms in $\mathbf{X}$ is subject to accurate domain knowledge, and a convenient choice is to specify each sensitivity function as a constant. As noted by \cite{JiangJRSSB2022} with a binary treatment, constant tilting functions correspond to a log-linear model for the potential outcome $Y(z)$ conditional on the latent stratum variable $G$ and covariates $\mathbf{X}$. The constructions of principal score weighting estimators and outcome regression estimators are simply by replacing unknown parameters with plug-in estimators in the empirical versions of \eqref{eq:weights_without_PI} and \eqref{eq:iden_outcome_regression_withoutPI}. Furthermore, the efficient influence function under assumed departure from principal ignorability is given by
\begin{align}
    \Psi_{zg}^{\text{PI}}(\mathbf{V})=&\frac{w_{zg}(\mathbf{X})}{p_z}\left\{\psi_{YS,z}-\displaystyle\frac{\Omega_{zg}(\mathbf{X})}{\delta_{zg}(\mathbf{X})}m_z(\mathbf{X})\sum_{g^\prime\geq J+1-z}\delta_{zg^\prime}(\mathbf{X})(\psi_{S,J-g^\prime+1}-\psi_{S,J-g^\prime})\right\}+\nonumber\\
       &\displaystyle\frac{\left\{\Omega_{zg}(\mathbf{X})m_{z}(\mathbf{X})-\mu_g(z)\right\}(\psi_{S,J-g+1}-\psi_{S,J-g})}{p_{J-g+1}-p_{J-g}}\label{eq:EIF_withoutPI}.
\end{align}
This motivates a bias-corrected estimator under violation of the principal ignorability as $$\widehat{\mu}_g^{\text{BC-PI}}(z) = \mathbb{P}_n\{\widehat{\Xi}^{\text{PI}}(\mathbf{V})\}/\mathbb{P}_n\{\widehat{\psi}_{S,J-g+1}-\widehat{\psi}_{S,J-g}\},$$ where 
\begin{align*}
\widehat{\Xi}^{\text{PI}}(\mathbf{V})=& \displaystyle \frac{\widehat{\Omega}_{zg}(\mathbf{X})(\widehat{p}_{J-g+1}(\mathbf{X})-\widehat{p}_{J-g}(\mathbf{X}))}{\widehat{p}_z(\mathbf{X})}\left\{\widehat\psi_{YS,z}-\displaystyle\frac{\widehat{\Omega}_{zg}(\mathbf{X})}{\delta_{zg}(\mathbf{X})}\widehat{m}_z(\mathbf{X})\sum_{g'\geq J+1-z}\delta_{zg'}(\mathbf{X})(\widehat\psi_{S,J-g'+1}-\widehat\psi_{S,J-g'})\right\}\\
& + \widehat{\Omega}_{zg}(\mathbf{X})\widehat{m}_{z}(\mathbf{X})(\widehat{\psi}_{S,J-g+1}-\widehat{\psi}_{S,J-g}).
\end{align*}
Yet, different from $\widehat{\mu}_g^{\text{DR}}(z)$, $\widehat{\mu}_g^{\text{BC-PI}}(z)$ is no longer doubly robust because the correction factor $\Omega_{zg}(\mathbf{X})$ in $\widehat{\Xi}^{\text{PI}}(\mathbf{V})$ does not allow a factorization of the difference between the true principal score $p_{z}(\mathbf{X})$ and the estimated one $\widehat p_{z}(\mathbf{X})$. Thus, with assumed knowledge of the sensitivity functions, $\widehat{\mu}_g^{\text{BC-PI}}(z)$ is consistent and asymptotically normal only if the principal score model is correctly specified, regardless of whether the outcome model is correctly specified. 

\subsection{Sensitivity analysis for monotonicity}\label{s:SA;subsec:Monotonicity}

Recall that the collection of all possible principal strata without monotonicity is defined as 
${\mathcal{G}}=\{(S(1),\ldots,S(J)):S(z)\in\{0,1\},z\in\mathcal{J}\}$. 
When the monotonicity assumption is violated, two types of strata arise: (i) strata satisfying monotonicity, of the form $\mathfrak{g}=0^{\otimes(J-g)}1^{\otimes g}$ and uniquely indexed by some $g\in\{0,\ldots,J\}$, and (ii) harmed strata that violate monotonicity in certain directions. To align with previous notation, we continue to index the monotonicity-satisfying strata by the nonnegative integer $g$ with $0\leq g\leq J$, and denote their collection as $\mathcal{Q}=\{0,\ldots,J\}$. For example, if $g\in\mathcal{Q}$, it corresponds to the stratum of the form $\mathfrak{g}=0^{\otimes(J-g)}1^{\otimes g}$. That is, with a slight abuse of notation, $\mathcal{Q}=\{0,\ldots,J\}$ is also understood as the set $\{\mathfrak{g}\in\mathcal{G}:\mathfrak{g}=0^{\otimes(J-g)}1^{\otimes g},0\leq g\leq J\}$. Finally, we denote the disjoint set of harmed strata as $\mathcal{G}\setminus\mathcal{Q}$.
We further define for $z\in\mathcal{J}$, $\mathcal{G}_{z}=\{\mathfrak{g}\in\mathcal{G}:S(z)=1\}$, which contains all the elements in $\mathcal{G}$ whose $z$-th coordinate is $1$. 
For a given user-defined reference group $r\in\mathcal{Q}$, we define the set of sensitivity functions
\begin{equation}\nonumber
   \rho_{\mathfrak{g}}(\mathbf{X})=\frac{\Pr(G=\mathfrak{g}|\mathbf{X})}{\Pr(G=r|\mathbf{X})}, \text{ for } \mathfrak{g}\in \mathcal{G}\backslash\mathcal{Q},
\end{equation}
provided that $\Pr(G=r|\mathbf{X})>0$ almost surely. 
Here, $\rho_{\mathfrak{g}}(\mathbf{X})$ measures the deviation from monotonicity for the harmed stratum $\mathfrak{g}$. The monotonicity assumption is satisfied if $\rho_{\mathfrak{g}}(\mathbf{X})=0$ for $\forall \mathfrak{g}\in \mathcal{G}\backslash\mathcal{Q}$. Otherwise, monotonicity is violated if $\rho_{\mathfrak{g}}(\mathbf{X})>0$ for some of $\mathfrak{g}\in \mathcal{G}\backslash\mathcal{Q}$. Our framework is a generalization of the sensitivity analysis methodology in \cite{JiangJRSSB2022} from a binary treatment to multiple treatments, and an expansion of \cite{luo2023causal} by allowing for more general non-monotonicity beyond violations only between adjacent strata. 

Under violation of monotonicity, and provided that 
$e_{\mathfrak{g}}\geq0,\forall \mathfrak{g}\in\mathcal{G}$, we show in the Supplementary Material that the principal score $e_{\mathfrak{g}}(\mathbf{X})$ can be identified by the following series of equations,
{
\begin{align}
    &e_{\mathfrak{g}}(\mathbf{X})=\begin{cases}
        p_{J-g+1}(\mathbf{X})-p_{J-g}(\mathbf{X})-(q_{J-g+1}(\mathbf{X})-q_{J-g}(\mathbf{X}))\displaystyle\frac{p_{J-r+1}(\mathbf{X})-p_{J-r}(\mathbf{X})}{1+q_{J-r+1}(\mathbf{X})-q_{J-r}(\mathbf{X})},&g\in\mathcal{Q}\\
        \rho_{\mathfrak{g}}(\mathbf{X})\displaystyle \frac{p_{J-r+1}(\mathbf{X})-p_{J-r}(\mathbf{X})}{1+q_{J-r+1}(\mathbf{X})-q_{J-r}(\mathbf{X})},&\mathfrak{g}\in \mathcal{G}\backslash\mathcal{Q}
    \end{cases}\label{eq:psiden_without_mono},
\end{align}
}
where $q_0(\mathbf{X})=0$, $q_z(\mathbf{X})=\sum_{\mathfrak{g}'\in \mathcal{G}_{z}\backslash\mathcal{Q}}\rho_{\mathfrak{g}'}(\mathbf{X})$ (summation taken over all the violating strata whose $z$-th coordinate is $1$) for $z\in\mathcal{J}$, and $q_{J+1}(\mathbf{X})=\sum_{\mathfrak{g}'\in\mathcal{G}\backslash\mathcal{Q}}\rho_{\mathfrak{g}'}(\mathbf{X})$ (summation taken over all the violating strata). 
Given the identifiability of the principal score $e_{\mathfrak{g}}(\mathbf{X})$, we can identify $\mu_\mathfrak{g}(z)$ only if Assumption \ref{assump:GPI} is strengthened, as follows. 
\begin{assumption}[\emph{Extended Principal Ignorability}]\label{assump:stronger PI for sens_mono}For $z\in\mathcal{J}$, 
$E\{Y(z)|G=\mathfrak{g}^\prime,\mathbf{X}\}=E\{Y(z)|G=\mathfrak{g},\mathbf{X}\}$ for $\forall \mathfrak{g},\mathfrak{g}^\prime\in \mathcal{G}_{z}$.     
\end{assumption}
Assumption \ref{assump:GPI} and Assumption \ref{assump:stronger PI for sens_mono} share the same spirit in the sense that conditional on baseline covariates, the expected potential outcome remains the same across the collection of principal strata of survivors. {Unsurprisingly, Assumption \ref{assump:stronger PI for sens_mono} is stronger than the standard principal ignorability (Assumption \ref{assump:GPI}) because the former requires the homogeneity condition of expected potential outcomes across a much larger set of strata. Specifically, Assumption \ref{assump:stronger PI for sens_mono} requires the homogeneity condition to hold for $\mathcal{G}_{z}$, a set of $2^{J-1}$ strata. By contrast, the monotonicity assumption makes Assumption \ref{assump:stronger PI for sens_mono} less strict (reducing to Assumption \ref{assump:GPI}), as it only requires the condition to hold for $z$ strata. In cases where prior knowledge suggests the extended principal ignorability assumption is partially violated, a sensitivity analysis analogous to our approach for principal ignorability can be constructed. Below, we provide a concrete example to interpret the extended principal ignorability assumption with a four-arm trial in which monotonicity fails.}

\begin{example}
{Following Example \ref{eg:4-arm}, we consider a four-arm trial. If monotonicity fails to hold,  we have a total of $16$ strata, with each stratum represented by a four-digit binary number $G=S(1)S(2)S(3)S(4)$. Then, the extended principal ignorability requires that, for $z\in\{1,2,3,4\}$, $E[Y(z)|G=\mathfrak{g},\mathbf X]$ is identical across all principal strata $\mathfrak{g}\in \mathcal{G}_{z}$. 
In other words, the extended principal ignorability requires the following four homogeneity conditions (a')--(d'):
\begin{enumerate}[itemsep=-3ex]
\item[(a')] $E[Y(1)|G=\mathfrak{g},\mathbf X]$ is identical across all $\mathfrak{g} \in \mathcal{G}_{1}=\{1000,1100,1010,1001,1110,1101,1011,1111\}$.
\item[(b')] $E[Y(2)|G=\mathfrak{g},\mathbf X]$ is identical across all $\mathfrak{g} \in \mathcal{G}_{2}=\{0100,1100,0110,0101,1110,1101,0111,1111\}$.
\item[(c')] $E[Y(3)|G=\mathfrak{g},\mathbf X]$ is identical across  all $\mathfrak{g} \in \mathcal{G}_{3}=\{0010,1010,0110,0011,1110,1011,0111,1111\}$.
\item[(d')] $E[Y(4)|G=\mathfrak{g},\mathbf X]$ is identical across all $\mathfrak{g} \in \mathcal{G}_{4}=\{0001,1001,0101,0011,1101,1011,0111,1111\}$.
\end{enumerate}
In contrast, with a four-arm trial, the standard principal ignorability assumption only requires three homogeneity conditions across a smaller number of principal strata, as demonstrated in conditions (a)--(c) in Example \ref{eg:4-arm}. 
}
\end{example}

{We begin by proposing bias-corrected identification formulas, which form the basis for the weighting and outcome regression estimators.} Based on the sensitivity functions $\rho_{\mathfrak{g}}(\mathbf{X})$, the principal score weight now becomes 
$$w_{z\mathfrak{g}}(\mathbf{X})=\left\{\frac{p_z(\mathbf{X})}{p_z}\right\}^{-1}\frac{e_{\mathfrak{g}}(\mathbf{X})}{e_{\mathfrak{g}}},$$ 
with $e_{\mathfrak{g}}(\mathbf{X})$ is given by \eqref{eq:psiden_without_mono} and $e_{\mathfrak{g}}=E[e_{\mathfrak{g}}(\mathbf{X})]$. Replacing $p_z(\mathbf{X})$ with $\mathbf{1}(Z=z)S/\pi_z$ in \eqref{eq:psiden_without_mono} and plugging into $\mu_{\mathfrak{g}}(z)=E\{e_{\mathfrak{g}}(\mathbf{X})/e_{\mathfrak{g}}\times m_z(\mathbf{X})\}$ yields the identification formulas based on outcome regression.
Constructions of the estimators similar to $\widehat{\mu}^{\text{PSW}}_{\mathfrak{g}}(z)$ and $\widehat{\mu}^{\text{OR}}_{\mathfrak{g}}(z)$ are straightforward by using the empirical counterparts based on the bias-corrected identification formulas introduced above. Under Assumption \ref{assump:stronger PI for sens_mono} and assumed sensitivity functions $\rho_{\mathfrak{g}}(\mathbf{X})$, the efficient influence function for $\mu_{\mathfrak{g}}(z)$ is given by
$$\Psi_{z\mathfrak{g}}^{\text{MO}}(\mathbf{V})=\frac{w_{z\mathfrak{g}}(\mathbf{X})\left\{\psi_{YS,z}-m_z(\mathbf{X})\psi_{S,z}\right\}}{p_z}+\frac{\left\{m_{z}(\mathbf{X})-\mu_{\mathfrak{g}}(z)\right\}{\psi}^\ast_{\mathfrak{g}}}{e_{\mathfrak{g}}},$$
{where ${\psi}^\ast_{\mathfrak{g}}$ is given by
\begin{align*}
    &{\psi}^\ast_{\mathfrak{g}}=\begin{cases}
        \psi_{S,J-g+1}-\psi_{S,J-g}-(q_{J-g+1}(\mathbf{X})-q_{J-g}(\mathbf{X}))\displaystyle\frac{\psi_{S,J-r+1}-\psi_{S,J-r}}{1+q_{J-r+1}(\mathbf{X})-q_{J-r}(\mathbf{X})},&g\in\mathcal{Q}\\
        \rho_{\mathfrak{g}}(\mathbf{X})\displaystyle\frac{\psi_{S,J-r+1}-\psi_{S,J-r}}{1+q_{J-r+1}(\mathbf{X})-q_{J-r}(\mathbf{X})},&\mathfrak{g}\in \mathcal{G}\backslash\mathcal{Q}
    \end{cases}.
\end{align*}}
Similarly, the efficient influence function induces a bias-correct estimator of $\mu_{\mathfrak{g}}(z)$, 
\begin{equation}\nonumber
    \widehat{\mu}^{\text{BC-MO}}_{\mathfrak{g}}(z)={ \mathbb{P}_n\left\{\displaystyle \frac{\widehat{e}_{\mathfrak{g}}(\mathbf{X})\mathbf{1}(Z=z)S}{\widehat{p}_z(\mathbf{X})\pi_z}\left(Y-\widehat{m}_{z}(\mathbf{X})\right)+\widehat{m}_{z}(\mathbf{X})\widehat{{\psi}}^\ast_{\mathfrak{g}}\right\}}\Big/{\mathbb{P}_n\{\widehat{{\psi}}^\ast_{\mathfrak{g}}\}},
\end{equation}
where $\widehat{{\psi}}^\ast_{\mathfrak{g}}$ is the plug-in estimator for ${\psi}^\ast_{\mathfrak{g}}$. 
In the Supplementary Material, we show that, due to the construction of the sensitivity function, the bias-corrected estimator $\widehat{\mu}^{\text{BC-MO}}_{\mathfrak{g}}(z)$ remains doubly robust; that is, it is consistent and asymptotically normal if either the principal score model or the outcome regression model is correctly specified, but not necessarily both. 

{
\section{Extension to non-Randomized observational settings}\label{sec:generalization-os}
Although we primarily focus on randomized clinical trials, the proposed methods can be extended to observational studies. In an observational study with multiple treatments, in place of randomization, we assume the following condition.
\begin{assumption}[\emph{Treatment Ignorability}]\label{Assump:TrtIg} $Z\perp\{S(1),\ldots,S(J),Y(1),\ldots,Y(J)\}|\mathbf{X}.$
\end{assumption}
Let $\pi_z(\bX) = \Pr(Z = z | \bX),z\in\mathcal{J}$ denote the generalized treatment propensity score \citep{li2019propensity}. Under Assumptions \ref{assump:mono}, \ref{assump:GPI}, and \ref{Assump:TrtIg}, $\mu_g(z)$ can be identified via three alternative formulas: 
\begin{align}
    &\mu_g(z)=E\left\{\frac{p_{J-g+1}(\mathbf{X})-p_{J-g}(\mathbf{X})}{p_{J-g+1}-p_{J-g}}\frac{S}{p_z(\mathbf{X})}\frac{\mathbf{1}(Z=z)}{\pi_z(\bX)}Y\right\}\label{eq:iden_ps_weighting_observational}
    ,\\
    &\mu_g(z)=E\left\{\frac{\mathbf{1}(Z=J-g+1)S/\pi_{J-g+1}(\bX)-\mathbf{1}(Z=J-g)S/\pi_{J-g}(\bX)}{p_{J-g+1}-p_{J-g}}m_{z}(\mathbf{X})\right\}\label{eq:iden_outcome_regression_observational},\\
    &\mu_g(z)=E\left\{\frac{p_{J-g+1}(\mathbf{X})-p_{J-g}(\mathbf{X})}{p_{J-g+1}-p_{J-g}}m_{z}(\mathbf{X})\right\}\label{eq:iden_ps+om},
\end{align}
for any $g\in\mathcal{Q}$ and any $z\geq J+1-g$.
Of note, identification formula \eqref{eq:iden_ps_weighting_observational} is derived using both principal score and propensity score weighting; formula \eqref{eq:iden_outcome_regression_observational} leverages the propensity score and outcome regression; and formula \eqref{eq:iden_ps+om} combines principal score and outcome regression. 
In Section \ref{sec:CovBalance} of the Supplementary Material, we derive a similar set of balancing conditions and three estimators based on the moment conditions in Equations \eqref{eq:iden_ps_weighting_observational}--\eqref{eq:iden_ps+om}. 
All estimators motivated by the moment conditions are consistent, provided that the corresponding working models are correctly specified.} {The semiparametrically efficient estimator, denoted as $\widehat{\mu}^{\text{TR}}_g(z)$, can be constructed similarly, except that the propensity score is unknown and must be estimated. To proceed, we posit a parametric working model $\pi_z(\bX;\bbeta_z)$ for the propensity score, where $\bbeta_z$ is a vector of unknown parameters and its estimator $\widehat{\bbeta}_z$ is obtained by solving the maximum likelihood score equations, $\mathbb{P}_n\{\iota(Z, \mathbf{X}; \boldsymbol{\beta})\} = \mathbf{0}$, where $\bbeta = (\bbeta_1^\top, \ldots, \bbeta_J^\top)$ denotes the collection of all parameters in the treatment propensity score model. For example, $\iota(Z, \bX; \bbeta)$ denotes the score function associated with the ordinal regression model. We define the probability limit of $\widehat{\bbeta}_z$ as $\widetilde{\bbeta}_z$. Under a correctly specified working model and suitable regularity conditions, the probability limit satisfies $\pi_z(\bX; \widetilde{\bbeta}_z) = \pi_z(\bX)$. We summarize the triple robustness property of $\widehat{\mu}^{\text{TR}}_g(z)$ in the proposition below, which parallels the binary setup in \cite{JiangJRSSB2022}.
\begin{proposition}[\emph{Triple Robustness}]\label{prop:triply robustness}
    Suppose that Assumptions \ref{assump:mono}, \ref{assump:GPI}, and \ref{Assump:TrtIg} hold and $\{{\pi_z(\bX;\widetilde{\bbeta}_z),\pi_z(\bX;\widehat{\bbeta}_z),}p_z(\mathbf{X};\widetilde{\boldsymbol{\alpha}}_z),p_z(\mathbf{X};\widehat{\boldsymbol{\alpha}}_z)\}$ are uniformly bounded away from 0 and 1. Then, $\widehat{\mu}^{\text{TR}}_g(z)$ is consistent and asymptotically normal {if any two of the three working models in $\{\pi_z(\bX;\bbeta_z),p_z(\bX;\balpha_z),m_z(\bX;\bgamma_z)\}$ are correctly specified.} If all three working models are correctly specified, $\widehat{\mu}_g^{\text{TR}}(z)$ is locally efficient in the sense that its asymptotic variance achieves the efficiency lower bound, i.e., the variance of the efficient influence function. 
\end{proposition}
By Proposition \ref{prop:triply robustness}, $\widehat{\mu}^{\text{TR}}_g(z)$ is triply robust in the sense that the bias is asymptotically negligible if any two of the working models in $\{\pi_z(\bX;\bbeta_z),p_z(\bX;\balpha_z),m_z(\bX;\bgamma_z)\}$ are correctly specified, but not necessarily all. In the special scenario of randomized trials where the propensity score $\pi_z(\bX)=\pi_z$ is known, the working model $\pi_z(\bX;\bbeta_z)$ is always correctly specified. In that case, the robustness of $\widehat{\mu}^{\text{TR}}_g(z)$ would align with the robustness property of the proposed doubly robust estimator.  
To compute the robust sandwich variance estimator of $\widehat{\mu}^{\text{TR}}_g(z)$, the estimating equations should be expanded to include the estimating equations, $\iota^\top(Z,\bX;\bbeta)$, on the propensity score model. The sensitivity analysis can be modified by replacing the known treatment probability with the estimated propensity score, while the remaining procedures remain the same as in the randomized trial setup. 
For the bias-corrected estimators in the sensitivity analysis for principal ignorability and monotonicity assumptions, their robustness properties differ slightly. When monotonicity is violated, $\widehat{\mu}_{\mathfrak{g}}^{\text{BC-MO}}(z)$ is still triply robust. When principal ignorability is violated, $\widehat{\mu}_g^{\text{BC-PI}}(z)$ is consistent and asymptotically normal only if the principal score model is correctly specified. In other words, $\widehat{\mu}_g^{\text{BC-PI}}(z)$ is conditionally doubly robust: its consistency requires that the principal score model is correctly specified and at least one of  the propensity score model or the outcome mean model is correctly specified. 
}

\section{Simulation studies}\label{subs:sim_study}
\subsection{Connection and comparison with Luo et al.}\label{sec:connect}
{We conduct simulations to investigate the finite-sample performance of the proposed approach in the context of randomized trials. Our comparisons focus on an existing method developed by \cite{luo2023causal}, with the overall goal of understanding when each estimator may be preferable. Through this process, we highlight their relative advantages, practical uses, and potential limitations, thereby helping investigators make more informed choices among the available estimators in the broader toolbox for principal stratification with multiple treatments. To begin with, we provide a brief review of the \cite{luo2023causal} approach.} {\cite{luo2023causal} propose two identification strategies. The first relies on a scalar instrument $A$, taken as a component of the covariates $\bX=(A,\mathbf{C}^\top)^\top$, which influences the outcome $Y$ only through the latent principal stratum $G$, i.e., $Y \perp A \mid Z,G,\mathbf{C}$; under this assumption, they establish nonparametric identification. The second strategy adopts a parametric approach by specifying a linear structural model for the conditional mean of the observed outcome given treatment, covariates, and principal stratum, 
$$m^\ast_{zg}(\mathbf{X})\equiv E\{Y|Z=z,G=g,\mathbf{X}\}.$$  However, their estimation strategy primarily focuses on the latter, which specifies a linear model for $m^\ast_{zg}(\mathbf{X})$, since the former may be impractical when the covariate dimension is high. In particular, their estimation requires specifying two models: (i) a principal score model and (ii) a model for $m^\ast_{zg}(\mathbf{X})$.} 
By the law of total expectation, $m_z(\mathbf{X})$ is a weighted sum of $m^\ast_{zg}(\mathbf{X})$, and under principal ignorability, $m^\ast_{zg}(\mathbf{X})=m_z(\mathbf{X})$  for $\forall g\geq J-z+1$. The point estimator of $\mu_g(z)$ proposed by \cite{luo2023causal} is  
\begin{align}\label{eq:estimator_luo}
\widehat{\mu}_g(z)=\frac{\mathbb{P}_n\{e_g(\mathbf{X};\widehat{\balpha^\ast}_g)m_{zg}^\ast(\mathbf{X};\widehat{\bgamma^\ast}_{zg})\}}{\mathbb{P}_n\{e_g(\mathbf{X};\widehat{\balpha^\ast}_g)\}},
\end{align}
where $m_{zg}^\ast(\mathbf{X};\bgamma^\ast_{zg})$ is the parametric working model for $m^\ast_{zg}(\mathbf{X})$ with unknown parameters $\bgamma^\ast_{zg}$ and $e_g(\mathbf{X};\balpha^\ast_g)$ is the parametric working model for the principal score with unknown parameters $\balpha^\ast_g$. Their approach diverges from ours in two key dimensions. First, \cite{luo2023causal} posit a parametric working model for the principal score directly and estimate the associated unknown parameters using the Expectation-Maximization algorithm, which is a direct generalization of methods used in \cite{DingandLu2016}. In contrast, our methods posit parametric working models for $p_z(\mathbf{X})$ and estimate the principal score using  $e_g(\mathbf{X})=p_{J-g+1}(\mathbf{X})-p_{J-g}(\mathbf{X})$ under monotonicity. Second, their outcome regression $m^\ast_{zg}(\mathbf{X};\bgamma^\ast_{zg})$ is conditional on the latent strata variable rather than the observed survival status.
Consequently, more unknown parameters in their outcome working model need to be estimated. Moreover, $m^\ast_{zg}(\mathbf{X})$ and $m_z(\mathbf{X})$ are connected through
\begin{align}\label{eq:connection_two_outcome_regressions}m_z(\mathbf{X})=\sum_{g=J-z+1}^J\left\{\frac{e_g(\mathbf{X})}{\sum_{g'=J-z+1}^Je_{g'}(\mathbf{X})}\right\}m^\ast_{zg}(\mathbf{X}).
\end{align}
Based on Equation \eqref{eq:connection_two_outcome_regressions}, \cite{luo2023causal} employed the generalized method of moments (GMM) to estimate $\bgamma^\ast_{zg}$. Under monotonicity but without principal ignorability, the estimator \eqref{eq:estimator_luo} is valid if two working models, $e_g(\mathbf{X};\balpha^\ast_g)$ and $m^\ast_{zg}(\mathbf{X};\bgamma^\ast_{zg})$, are both correctly specified. Further assuming principal ignorability, Equation \eqref{eq:connection_two_outcome_regressions} implies $m^\ast_{zg}(\mathbf{X})=m_z(\mathbf{X})$ and $\bgamma^\ast_{zg}=\bgamma_z$ for $\forall g\geq J-z+1$. 

Finally, to ensure a fair comparison, we shall recycle 
$p_{z}(\mathbf{X};\widehat{\bm\alpha}_z)$ from our proposed doubly robust estimator to update the estimator in Equation \eqref{eq:estimator_luo} by \cite{luo2023causal}. Based on the identity $e_g(\mathbf{X}) = p_{J-g+1}(\mathbf{X}) - p_{J-g}(\mathbf{X})$, we substitute $e_g(\mathbf{X};\widehat{\balpha^\ast}_g)$ with the expression $p_{J-g+1}(\mathbf{X};\widehat{\balpha}_{J-g+1}) - p_{J-g}(\mathbf{X};\widehat{\balpha}_{J-g})$. 
This eventually leads to the following estimator:
\begin{align*}
    \widehat{\mu}^{\text{Luo}}_g(z)=\frac{\mathbb{P}_n\{(p_{J-g+1}(\mathbf{X};\widehat{\balpha}_{J-g+1})-p_{J-g}(\mathbf{X};\widehat{\balpha}_{J-g}))m_{zg}^\ast(\mathbf{X};\widehat{\bgamma^\ast}_{zg})\}}{\mathbb{P}_n\{p_{J-g+1}(\mathbf{X};\widehat{\balpha}_{J-g+1})-p_{J-g}(\mathbf{X};\widehat{\balpha}_{J-g})\}},
\end{align*}
where $m_{zg}^\ast(\mathbf{X};\widehat{\bgamma^\ast}_{zg})$ is specified as in \cite{luo2023causal} and estimated through GMM. However, in scenarios when principal ignorability holds, we additionally restrict the specification of $m_{zg}^\ast(\mathbf{X};\widehat{\bgamma^\ast}_{zg})$ to  $m_{z}(\mathbf{X};\widehat{\bgamma}_{z})$ because principal ignorability implies $m_{zg}^*(\mathbf{X})=m_{z}(\mathbf{X})$ for all $g$. 

\subsection{Simulation designs}
We conduct three simulation studies to evaluate the method under different scenarios. The first study assumes that our causal identification assumptions are satisfied (Section \ref{sss:sim-design-all-correct}). The second and third studies explore scenarios in which the principal ignorability assumption (Section \ref{sss:sim-design-pi-violated}) and the monotonicity assumption (Section \ref{sss:sim-design-monotonicity-violated}) are violated, respectively.


\subsubsection{Simulation design under monotonicity and principal ignorability}\label{sss:sim-design-all-correct}
In Section \ref{sss:sim-design-all-correct}, we conduct a simulation study to assess the empirical performance of the proposed estimators with the following three objectives: (i) evaluating the validity and relative efficiency among $\widehat \Delta_{g}^{\text{PSW}}(z,z')$, $\widehat \Delta_{g}^{\text{OR}}(z,z')$, and $\widehat \Delta_{g}^{\text{DR}}(z,z')$, under correct and incorrect specifications of the  principal score and outcome regression models; (ii) investigating the performance of the proposed sandwich variance estimator in finite samples; (iii) comparing our proposed estimators to an existing method by \cite{luo2023causal} to study the relative merits and limitations of different approaches.



We consider a three-arm randomized trial ($J=3$) with a small or large sample size ($n=500$ or $2000$), with balanced assignment such that $\Pr(Z=1)=\Pr(Z=2)=\Pr(Z=3)=1/3$. Four baseline covariates $\mathbf{X}=(X_1,X_2,X_3,X_4)^\top$ are generated from $X_j=|\widetilde{X}_j|$ with $\widetilde{X}_j\sim\mathcal{N}(0,1)$ for $j\in\{1,2,3\}$ and $X_4\sim\text{Bernoulli}(0.5)$. We generate the principal strata membership $G\in\{0,1,2,3\}$ based on a categorical distribution with $e_{0}(\mathbf{X})=1-\text{expit}(\boldsymbol{\alpha}_{3}^\top\mathbf{X})$, $e_{g}(\mathbf{X})=\text{expit}(\boldsymbol{\alpha}_{4-g}^\top\mathbf{X})-\text{expit}(\boldsymbol{\alpha}_{3-g}^\top\mathbf{X}),~g\in\{1,2\}$, and $e_{3}(\mathbf{X})=\text{expit}(\boldsymbol{\alpha}_{1}^\top\mathbf{X})$, where $\boldsymbol{\alpha}_z=(-0.8+0.3 z,-0.8+0.4 z,-0.8+0.5 z,-0.8+0.4 z)$, for $z\in\{1,2,3\}$ and $\text{expit}(x)=(1+e^{-x})^{-1}$. Then the observed survival status is given by $S=\mathbf{1}(G+Z\geq J+1)$. Given $G$ and $\mathbf{X}$, the potential outcome $Y(z)$ is generated by 
\begin{align*}
    &Y(1)|\left\{\mathbf{X},G=3\right\}\sim\mathcal{N}(X_1+3X_2+3X_3+3X_4+2,1),\\
    &Y(2)|\left\{\mathbf{X},G\in\{2,3\}\right\}\sim\mathcal{N}(X_1+2X_2+2X_3+2X_4+2,1),\\
    &Y(3)|\left\{\mathbf{X},G\in\{1,2,3\}\right\}\sim\mathcal{N}\left(\sum_{i=1}^4 X_i+3,1\right),
\end{align*}
and $Y(z)$ within $G=g< J+1-z$ is undefined due to truncation by death. We consider all possible causal contrast parameters $\{\Delta_2(2,3),\Delta_3(1,2),\Delta_3(1,3),\Delta_3(2,3)\}$ that are well-defined. The observed outcome is $Y=\sum_{z=1}^3 Y(z)\mathbf{1}(Z= z)$. Of note, Assumptions \ref{assump:mono} and \ref{assump:GPI} hold under the above data generation process, and by construction, the principal score $p_z(\mathbf{X})=\text{expit}(\boldsymbol{\alpha}_{z}^\top\mathbf{X})$ and the outcome model $m_z(\mathbf{X})$ is a linear function of $\mathbf{X}$.

For estimation, we specify a logistic regression for $p_z(\mathbf{X};\bm\alpha_z)$ with $\balpha_z^\top\mathbf{X}$ as linear predictors. For the outcome model $m_z(\mathbf{X};\bm\gamma_z)$, we fit a linear regression adjusting for $\{Z,\mathbf{X}\}$ and their interaction.
i.e., specifying 
\begin{align*}
    E(Y|Z,S=1,\mathbf{X})=&\gamma_0+\gamma_{1}\mathbf{1}(Z=1)+\gamma_{2}\mathbf{1}(Z=2)+\sum_{j=1}^4\gamma_{j+2} X_j+\sum_{j=1}^4\gamma_{j+6}\mathbf{1}(Z=1)X_j+\\
    &\sum_{j=1}^4\gamma_{j+10}\mathbf{1}(Z=2)X_j.
\end{align*}
We conduct 1,000 simulations and calculate the bias, Monte Carlo standard deviation, average standard error estimates based on the proposed variance estimators ($500$ bootstrap samples are used to obtain standard error estimates for \cite{luo2023causal}), and empirical coverage of the 95\% Wald confidence interval (using normal approximation). The true value of $\mu_g(z)$ is approximated by the empirical mean of the potential outcome $Y(z)$ within subgroup $g$ based on a sufficiently large super-population of size $n=250,000$.  
We consider all combinations of correctly or incorrectly specified principal score and outcome models, where the misspecified model is obtained by ignoring $X_2,X_3,X_4$, and fitting regression models only on a transformed covariate, $\text{cos}(X_1)$. 

\subsubsection{Simulation design under violation of principal ignorability}\label{sss:sim-design-pi-violated}
We conduct an additional simulation study to examine the scenario when principal ignorability is violated. Theoretically, this violation is expected to introduce bias in our estimators. The data generation process follows the approach described in Section \ref{sss:sim-design-all-correct}, with some modifications. That is, the true principal score is now defined as $\Pr(G=g|\bX)=\Pr(G=g) = 0.1 + 0.1 \times g,~g \in \{1,2,3\}$. Additionally, the potential non-mortality outcome follows
\begin{align*}
    &Y(2)|\{\mathbf{X},G=2\}\sim\mathcal{N}(2+X_1+2X_2+2X_3+2X_4,1),\\
    &Y(2)|\{\mathbf{X},G=3\}\sim\mathcal{N}(1+X_1+2X_2+2X_3+2X_4,1),\\
    &Y(3)|\left\{\mathbf{X},G\in\{1,3\}\right\}\sim\mathcal{N}\left(3+\sum_{i=1}^4 X_i,1\right),~~~~
    Y(3)|\{\mathbf{X},G=2\}\sim\mathcal{N}\left(4+\sum_{i=1}^4 X_i,1\right).
\end{align*}

Principal ignorability is violated under this new data-generating process. However, logistic regression remains the correct model for $p_z(\mathbf{X})$, and linear regressions are still the correct models for $m_z(\mathbf{X})$ and $m^\ast_{zg}(\mathbf{X})$. Beyond our proposed estimator and \cite{luo2023causal} estimator, we also implement our proposed sensitivity analysis method in Section \ref{s:SA;subsec:PI} for bias correction. By construction, the true sensitivity functions are given by
\begin{align*}
    &\delta_{22}(\mathbf{X})=1+\frac{1}{1+X_1+2X_2+2X_3+2X_4},~~~~
    \delta_{31}(\mathbf{X})=1,~~~~
    \delta_{32}(\mathbf{X})=1+\frac{1}{3+\sum_{i=1}^4 X_i}.
\end{align*}

{In practice, it may be challenging to fully specify the covariate-dependent forms of the sensitivity functions, and hence a common simplification is to consider a constant value approximation. This motivates us to study the impact of misspecifying the sensitivity function by its average across the covariate distribution. Additionally, we further consider an alternative data-generating process in which the true sensitivity functions are indeed constant such that $\delta_{zg}(\mathbf X)\equiv \delta_{zg}$ and there is no sensitivity function misspecification. In that scenario, the potential non-mortality outcomes are generated via the following process:}
{\begin{align*}
    &Y(2)|\{\mathbf{X},G=2\}\sim\mathcal{N}\left({\delta_{22}(1+X_1+2X_2+2X_3+2X_4)},1\right),\\
    &Y(2)|\{\mathbf{X},G=3\}\sim\mathcal{N}\left(1+X_1+2X_2+2X_3+2X_4,1\right),\\
    &Y(3)|\{\mathbf{X},G=g\}\sim\mathcal{N}\left(\left(\mathbf{1}(g=1)\delta_{31}+\mathbf{1}(g=2)\delta_{32}+\mathbf{1}(g=3)\right)\left(3+\sum_{i=1}^4 X_i\right),1\right).
\end{align*}
For simplicity, we also assume that the true sensitivity functions do not depend on $z$ such that $\delta_1=\delta_{31}$ and $\delta_2 \equiv \delta_{22}= \delta_{32}$. We consider the two scenarios with $\{\delta_1, \delta_2\} \in \{(0.5, 0.5),(2, 2)\}$. Such specifications of the sensitivity functions also align with our real data analysis in the Section \ref{s:DE;subs:SA}. }

{\subsubsection{Simulation design under non-monotonicity}\label{sss:sim-design-monotonicity-violated}
We conduct an final set of simulations to evaluate the performance of the proposed sensitivity method under non-monotonicity. Motivated by the data analysis in Section \ref{s:DE;subs:SA}, we take $r=0$ as the reference stratum and assume that the four principal strata violating monotonicity are constant and occur in equal proportion relative to the reference. Specifically, we set $\rho_{010}(\bX)=\rho_{100}(\bX)=\rho_{101}(\bX)=\rho_{110}(\bX)=\rho$ for some nonnegative constant $\rho$. We generate the principal strata variable by sampling from the following categorical distribution specified by $\rho$:
\begin{align*}
    &e_0(\bX)=(1+3\rho)^{-1}(1-p_3(\bX)),\\
    &e_1(\bX)=p_3(\bX)-p_2(\bX)+\rho(1+3\rho)^{-1}(1-p_3(\bX)),\\
    &e_2(\bX)=p_2(\bX)-p_1(\bX)+\rho(1+3\rho)^{-1}(1-p_3(\bX)),\\
    &e_3(\bX)=p_1(\bX)-3\rho(1+3\rho)^{-1}(1-p_3(\bX)),\\
    &e_{\mathfrak{g}}(\bX)=\rho (1+3\rho)^{-1}(1-p_3(\bX)),~~~\mathfrak{g}\in\mathcal{G}\backslash\mathcal{Q},
\end{align*}
where $p_z(\bX)\equiv0.2+0.2 z$ for $z\in\{1,2,3\}$. The potential outcome $Y(z)$ is generated by 
\begin{align*}
    &Y(1)|\{\mathbf{X},G\in\mathcal{G}_{1}\}\sim\mathcal{N}(2+X_1+3X_2+3X_3+3X_4,1),\\
    &Y(2)|\{\mathbf{X},G\in\mathcal{G}_{2}\}\sim\mathcal{N}(2+X_1+2X_2+2X_3+2X_4,1),\\
    &Y(3)|\{\mathbf{X},G\in\mathcal{G}_{3}\}\sim\mathcal{N}\left(3+\sum_{i=1}^4 X_i,1\right).
\end{align*}
All other aspects of the data-generating process remain identical to those described in Section \ref{sss:sim-design-all-correct}. Under this setup, $\rho$ takes values within the interval $[0,\infty)$, which is consistent with our subsequent data application. We set $\rho \in \{0.2, 5\}$ to represent mild and substantial violations of the monotonicity assumption, respectively. 
}

\subsection{Simulation results}

\begin{table}[htbp]
\caption{Bias, Monte Carlo standard deviations (`MCSD'), average empirical standard errors (`AESE') based on robust sandwich variance estimators, and empirical coverage (`CP') using AESE for all possible contrasts $\Delta_g(z,z^\prime)$, based on the principal score weighting estimator (`PSW'), outcome regression estimator (`OR'), doubly robust estimators (`DR'), and estimator in \cite{luo2023causal} (`Luo') when the sample size is $500$. For the column of ps (or om),  we set $\checkmark$ and $\times$ to indicate the correct and incorrect specification of the principal score model (or outcome regression), respectively. The symbol ``$\backslash$" indicates that the principal score weighting estimator and the outcome regression estimator are independent of the outcome mean model and the principal score model, respectively. {The data-generating process assumes that both principal ignorability and monotonicity hold.}
}
\label{tab:simulation500}
\begin{adjustbox}{width=1\textwidth}
\begin{tabular}{lllll cccc cccc cccc cccc}
    \hline
            &   &   &   & &\multicolumn{4}{c}{BIAS} & \multicolumn{4}{c}{CP} & \multicolumn{4}{c}{MCSD} & \multicolumn{4}{c}{AESE} \\ 
           \cmidrule(lr){6-9}\cmidrule(lr){10-13}\cmidrule(lr){14-17}\cmidrule(lr){18-21}
       $g$&$z$&$z^\prime$   & ps & om           & PSW  & OR   & DR   & Luo  & PSW  & OR   & DR   & Luo  & PSW  & OR   & DR   & Luo  & PSW  & OR   & DR    & Luo\\ \hline
        2 & 2 & 3 & $\checkmark$ & $\checkmark$ & 0.05       & 0.01       & 0.01    & 0.04       & 97.3       & 96.8       & 96.8 & 97.2 & 0.91 & 0.37 & 0.28 & 0.29 & 1.04 & 0.37 & 0.29 & 0.50 \\ 
        ~ & ~ & ~ & $\checkmark$ & $\times$     &$\backslash$& $-$0.36    & 0.00    & $-$0.48    &$\backslash$& 76.9       & 96.0 & 78.4 &$\backslash$& 0.28 & 0.35 & 0.41 &$\backslash$& 0.30 & 0.37 & 0.50 \\ 
        ~ & ~ & ~ & $\times$     & $\checkmark$ & 0.42       &$\backslash$& 0.02    & $-$0.32    & 97.0       &$\backslash$& 95.1 & 51.1 & 0.82 &$\backslash$& 0.29 & 0.16 & 0.77 &$\backslash$& 0.29 & 0.19 \\ 
        ~ & ~ & ~ & $\times$     & $\times$     &$\backslash$&$\backslash$& $-$0.37 & $-$0.37    &$\backslash$&$\backslash$& 74.6 & 78.5 &$\backslash$&$\backslash$& 0.29 & 0.28 &$\backslash$&$\backslash$& 0.29 & 0.35 \\ \hline
        3 & 1 & 2 & $\checkmark$ & $\checkmark$ & 0.03       & $-$0.01    & 0.00    & 0.02       & 98.1       & 92.9       & 93.6 & 94.9 & 0.79 & 0.25 & 0.24 & 0.25 & 1.25  & 0.24 & 0.24 & 0.25 \\ 
        ~ & ~ & ~ & $\checkmark$ & $\times$                &$\backslash$& $-$0.55    & 0.00    & $-$0.41    &$\backslash$& 79.6       & 94.8 & 88.9 &$\backslash$& 0.54 & 0.37 & 0.59 &$\backslash$& 0.52 & 0.38 & 0.58 \\ 
        ~ & ~ & ~ & $\times$ & $\checkmark$     & 0.64       &$\backslash$& 0.00    & 0.24       & 93.2       &$\backslash$& 93.6 & 82.5 & 0.91 &$\backslash$& 0.25 & 0.22 & 0.95  &$\backslash$& 0.24 & 0.24 \\ 
        ~ & ~ & ~ & $\times$ & $\times$                &$\backslash$&$\backslash$& $-$0.57 & $-$0.55    &$\backslash$&$\backslash$& 78.9 & 80.0 &$\backslash$&$\backslash$& 0.52 & 0.54 &$\backslash$&$\backslash$& 0.52 & 0.53 \\ 
        ~ & 1 & 3 & $\checkmark$ & $\checkmark$ & 0.03       & $-$0.01    & $-$0.01 & 0.04       & 98.4       & 94.0       & 93.3 & 95.6 & 0.71 & 0.34 & 0.33 & 0.33 & 1.22  & 0.34 & 0.32 & 0.34 \\ 
        ~ & ~ & ~ & $\checkmark$ & $\times$                &$\backslash$& $-$0.35    & 0.01    & $-$0.38    &$\backslash$& 85.4       & 94.2 & 86.5 &$\backslash$& 0.48 & 0.40 & 0.51 &$\backslash$& 0.47 & 0.39 & 0.50 \\ 
        ~ & ~ & ~ & $\times$ & $\checkmark$     & 0.41       &$\backslash$& 0.00    & 0.50       & 94.9       &$\backslash$& 94.5 & 43.6 & 0.79 &$\backslash$& 0.33 & 0.23 & 0.82  &$\backslash$& 0.32 & 0.24 \\ 
        ~ & ~ & ~ & $\times$ & $\times$                &$\backslash$&$\backslash$& $-$0.36 & $-$0.37    &$\backslash$&$\backslash$& 86.6 & 85.0 &$\backslash$&$\backslash$& 0.46 & 0.49 &$\backslash$&$\backslash$& 0.47 & 0.48 \\ 
        ~ & 2 & 3 & $\checkmark$ & $\checkmark$ & 0.01       & 0.00       & 0.00    & $-$0.01    & 98.1       & 94.4       & 94.7 & 95.4 & 0.76 & 0.22 & 0.21 & 0.21 & 1.02  & 0.21 & 0.21 & 0.22 \\ 
        ~ & ~ & ~ & $\checkmark$ & $\times$                &$\backslash$& 0.19       & $-$0.01 & 0.07       &$\backslash$& 90.4       & 96.0 & 94.2 &$\backslash$& 0.28 & 0.28 & 0.42 &$\backslash$& 0.28 & 0.29 & 0.41 \\ 
        ~ & ~ & ~ & $\times$ & $\checkmark$     & 0.22       &$\backslash$& 0.01    & 0.25       & 95.4       &$\backslash$& 94.4 & 66.4 & 0.77 &$\backslash$& 0.21 & 0.16 & 0.79  &$\backslash$& 0.21 & 0.16 \\ 
        ~ & ~ & ~ & $\times$ & $\times$                &$\backslash$&$\backslash$& 0.21    & 0.20       &$\backslash$&$\backslash$& 89.3 & 91.3 &$\backslash$&$\backslash$& 0.28 & 0.28 &$\backslash$&$\backslash$& 0.28 & 0.29 \\ \hline

    \end{tabular}
\end{adjustbox}
\end{table}

{Under the simulation design in Section \ref{sss:sim-design-all-correct},} the results under sample size $n=500$ are given in Table \ref{tab:simulation500}. 
First, the empirical bias of all estimators is minimal when both working models are correctly specified. The bias of $\widehat\Delta_g^{\text{DR}}(z,z')$ remains negligible when either the principal score model or outcome model is incorrectly specified, which empirically verifies the double robustness property in Theorem \ref{thm:triply robustness}. Second, the proposed sandwich variance estimator for $\widehat\Delta_g^{\text{PSW}}(z,z')$ tends to overestimate the true variance, but the variance estimators for $\widehat\Delta_g^{\text{OR}}(z,z')$ and $\widehat\Delta_g^{\text{DR}}(z,z')$ are centered around the empirical variance, showing adequate performance in finite samples. Third, we observe that the coverage for $\widehat\Delta_g^{\text{PSW}}(z,z')$ does not deviate too much from the nominal level under model misspecification. We further explore this phenomenon in Supplementary Material Figure 1, by visualizing the empirical distribution of  $\widehat\Delta_g^{\text{PSW}}(z,z')$ over 1,000 simulations. In principle, under model misspecification, the standardized principal score weighting estimator—defined as the estimate minus the truth divided by its standard error—approximately follows a normal distribution with mean equal to the bias-to-standard-error ratio and variance one. However, the figure indicates that the empirical distribution of this estimator is more concentrated than a mean-shifted standard normal, suggesting that the normal approximation may be conservative for the weighting estimator in small samples. Fourth, $\widehat\Delta_g^{\text{OR}}(z,z')$ and $\widehat\Delta_g^{\text{DR}}(z,z')$ are almost equally efficient for estimating most of the causal contrasts, and they are both substantially more efficient than $\widehat\Delta_g^{\text{PSW}}(z,z')$ irrespective of model misspecification. Finally, we empirically confirm that consistency of the estimator in \cite{luo2023causal} requires the correct specifications of both models and it is nearly as efficient as $\widehat\Delta_g^{\text{OR}}(z,z')$; however, in contrast to the doubly robust estimator, we observe bias and substantial undercoverage of the approach in \cite{luo2023causal} when either the principal score model or the outcome mean model is misspecified. Table \ref{tab:simulation2000} 
presents the simulation results under a larger sample size of $n=2,000$, where the patterns are qualitatively similar. 

Under the simulation design in Section \ref{sss:sim-design-pi-violated} where principal ignorability is violated, Table \ref{tab:simulationcase2} demonstrates that the \cite{luo2023causal} estimator is unbiased under correct model specification, whereas the proposed doubly robust estimator is subject to bias. Table \ref{tab:simulationcase2} further shows that our proposed sensitivity method (with correctly specified sensitivity functions) can effectively correct the bias due to violation of principal ignorability, restoring the validity of causal inference with minimal bias and nominal coverage. Interestingly, the proposed bias-corrected estimator based on the efficient influence function appears to substantially improve the efficiency over \cite{luo2023causal} for estimating all causal estimands regardless of sample size configurations. {As an additional exploration under the same data-generating process, Supplementary Material Table 1 presents the results when the sensitivity functions are misspecified as a constant (equal to the mean values of the true sensitivity functions). It is observed that this type of misspecification has little effect on bias, although the empirical coverage probabilities for both the bias-corrected outcome regression estimator and doubly robust estimator sometimes fall slightly below their nominal levels. 
As a final check, Supplementary Material Tables 2 and 3 present the simulation results under the ideal scenario when the true sensitivity functions are constant. Under this setup, our bias-corrected estimators indeed carry minimal bias and achieve nominal coverage throughout. Collectively, these findings provide some support for implementing the bias-corrected estimator in the data application of Section \ref{s:DE;subs:SA}.}

{Supplementary Material Tables 4 and 5 present the simulation results under violations of the monotonicity assumption, based on the simulation design in Section \ref{sss:sim-design-monotonicity-violated}. First, the bias‐correction method from Section \ref{s:SA;subsec:Monotonicity} generally delivers unbiased estimates with nominal coverage. However, at $n=500$ with a high degree of monotonicity violation, the bias of weighting estimator increases slightly, but such bias disappears when $n=2000$. This suggests that the bias-corrected weighting estimator may require a larger sample size to provide stable estimates. Second, under a higher degree of monotonicity violation, the bias-corrected doubly robust estimator becomes notably more efficient than the bias-corrected weighting and regression estimators. This contrasts with our earlier result that the doubly robust estimator was nearly as efficient as the outcome regression estimator when both models are correctly specified. 
}

\begin{table}[htbp]
\caption{Bias, Monte Carlo standard deviations (`MCSD'), average empirical standard errors (`AESE') based on robust sandwich variance estimators, and empirical coverage (`CP') using AESE for all possible contrasts $\Delta_g(z,z^\prime)$, based on the principal score weighting estimator (`PSW'), outcome regression estimator (`OR'), doubly robust estimators (`DR'), and estimator in \cite{luo2023causal} (`Luo') when the sample size is $2000$. For the column of ps (or om),  we set $\checkmark$ and $\times$ to indicate correct and incorrect specification of the principal score model (or outcome regression), respectively.  $\backslash$ indicates that the principal score weighting estimator and the outcome regression estimator are independent of the outcome mean model and the principal score model, respectively. {The data-generating process assumes that both principal ignorability and monotonicity hold.}
}
\label{tab:simulation2000}
\begin{adjustbox}{width=1\textwidth}
\begin{tabular}{lllllrrrrllllllllllll}
    \hline
            &   &   &   & &\multicolumn{4}{c}{BIAS} & \multicolumn{4}{c}{CP} & \multicolumn{4}{c}{MCSD} & \multicolumn{4}{c}{AESE} \\ 
           \cmidrule(lr){6-9}\cmidrule(lr){10-13}\cmidrule(lr){14-17}\cmidrule(lr){18-21}
        $g$ & $z$ & $z^\prime$ & ps & om & PSW & OR & DR & Luo & PSW & OR & DR & Luo & PSW & OR & DR & Luo & PSW & OR & DR & Luo\\ \hline
         2 & 2 & 3 & $\checkmark$ & $\checkmark$ & 0.00       & 0.01       & 0.01 & 0.01   & 91.9        & 96.5       & 95.7 & 97.1 & 0.43       & 0.16       & 0.13 & 0.13 & 0.38       & 0.16 & 0.13 & 0.14 \\ 
        ~ & ~ & ~ & $\checkmark$ & $\times$                 &$\backslash$& -0.36      & 0.01 & -0.50  &$\backslash$ & 28.6       & 96.4 & 41.3 &$\backslash$& 0.14       & 0.17 & 0.23 &$\backslash$& 0.14 & 0.17 & 0.23 \\ 
        ~ & ~ & ~ & $\times$ & $\checkmark$      & -0.36      &$\backslash$& 0.00 & -0.32  & 86.2        &$\backslash$& 96.4 & 1.20 & 0.35       &$\backslash$& 0.13 & 0.08 & 0.35       &$\backslash$& 0.13 & 0.08 \\ 
        ~ & ~ & ~ & $\times$ & $\times$                 &$\backslash$&$\backslash$& -0.36 & -0.36 &$\backslash$ &$\backslash$& 27.6 & 27.0 &$\backslash$&$\backslash$& 0.14 & 0.14 &$\backslash$&$\backslash$& 0.14 & 0.14 \\ \hline
        3 & 1 & 2 & $\checkmark$ & $\checkmark$  & 0.00       & 0.00       & 0.00 & 0.02   & 99.1        & 95.3       & 94.8 & 94.7 & 0.38       & 0.12       & 0.12 & 0.13 & 0.57       & 0.12 & 0.12 & 0.13 \\ 
        ~ & ~ & ~ & $\checkmark$ & $\times$                 &$\backslash$& -0.57      & 0.01 & -0.37  &$\backslash$ & 41.2       & 95.3 & 80.8 &$\backslash$& 0.26       & 0.18 & 0.31 &$\backslash$& 0.26 & 0.18 & 0.31 \\ 
        ~ & ~ & ~ & $\times$ & $\checkmark$      & -0.57      &$\backslash$& -0.01 & 0.25  & 78.9        &$\backslash$& 95.1 & 37.9 & 0.46       &$\backslash$& 0.12 & 0.11 & 0.47       &$\backslash$& 0.12 & 0.11 \\ 
        ~ & ~ & ~ & $\times$ & $\times$                 &$\backslash$&$\backslash$& -0.57 & -0.57 &$\backslash$ &$\backslash$& 40.2 & 40.6 &$\backslash$&$\backslash$& 0.26 & 0.26 &$\backslash$&$\backslash$& 0.26 & 0.26 \\ 
        ~ & 1 & 3 & $\checkmark$ & $\checkmark$  & -0.02      & 0.01       & 0.01 & 0.03   & 99.2        & 95.5       & 94.7 & 94.5 & 0.35       & 0.17       & 0.16 & 0.17 & 0.55       & 0.17 & 0.16 & 0.17 \\ 
        ~ & ~ & ~ & $\checkmark$ & $\times$                 &$\backslash$& -0.37      & 0.00 & -0.34  &$\backslash$ & 63.9       & 95.3 & 71.4 &$\backslash$& 0.24       & 0.19 & 0.25 &$\backslash$& 0.24 & 0.19 & 0.25 \\ 
        ~ & ~ & ~ & $\times$ & $\checkmark$      & -0.36      &$\backslash$& 0.01 & 0.49   & 88.1        &$\backslash$& 94.5 & 0.90 & 0.40       &$\backslash$& 0.16 & 0.11 & 0.40       &$\backslash$& 0.16 & 0.11 \\ 
        ~ & ~ & ~ & $\times$ & $\times$                 &$\backslash$&$\backslash$& -0.35 & -0.37 &$\backslash$ &$\backslash$& 68.3 & 66.1 &$\backslash$&$\backslash$& 0.24 & 0.24 &$\backslash$&$\backslash$& 0.24 & 0.24 \\ 
        ~ & 2 & 3 & $\checkmark$  & $\checkmark$ & 0.01       & 0.00       & 0.00 & 0.00   & 98.2        & 96.2       & 95.3 & 95.1 & 0.36       & 0.10       & 0.10 & 0.10 & 0.47       & 0.10 & 0.10 & 0.10 \\ 
        ~ & ~ & ~ & $\checkmark$ & $\times$                 &$\backslash$& 0.20       & 0.00 & 0.00   &$\backslash$ & 69.6       & 95.2 & 94.4 &$\backslash$& 0.14       & 0.14 & 0.23 &$\backslash$& 0.14 & 0.14 & 0.22 \\ 
        ~ & ~ & ~ & $\times$ & $\checkmark$      & 0.20       &$\backslash$& 0.00 & 0.25   & 93.4        &$\backslash$& 93.3 & 10.1 & 0.38       &$\backslash$& 0.10 & 0.08 & 0.39       &$\backslash$& 0.10 & 0.08 \\ 
        ~ & ~ & ~ & $\times$ & $\times$                 &$\backslash$&$\backslash$& 0.21 & 0.19   &$\backslash$ &$\backslash$& 69.0 & 72.3 &$\backslash$&$\backslash$& 0.14 & 0.14 &$\backslash$&$\backslash$& 0.14 & 0.14 \\ \hline
    \end{tabular}
\end{adjustbox}
\end{table}

\begin{table}[htbp]
\caption{Bias, Monte Carlo standard deviations (`MCSD'), average empirical standard errors (`AESE') based on robust sandwich variance estimators, and empirical coverage (`CP') using AESE for all possible contrasts $\Delta_g(z,z^\prime)$, based on the principal score weighting estimator (`PSW'), principal score weighting estimator with bias-correction (`PSW-BC'), outcome regression estimator (`OR'), outcome regression estimator with bias-correction (`OR-BC'), doubly robust estimator (`DR'), doubly robust estimator with bias-correction (`DR-BC'), and estimator in \cite{luo2023causal} (`Luo'). {The data-generating process assumes that principal ignorability does not hold but monotonicity holds. The associated working models for each estimator are assumed to be correctly specified, or compatible with the true data-generating process.}
}
\label{tab:simulationcase2}
\begin{adjustbox}{width=1\textwidth}
\begin{tabular}{llll rrrrrrr cccccccc}
    \toprule
            &&   &   &\multicolumn{7}{c}{BIAS} & \multicolumn{7}{c}{CP} \\ 
           \cmidrule(lr){5-11}\cmidrule(lr){12-18}
       $n$&$g$&$z$&$z^\prime$             & PSW & PSW-BC  & OR &OR-BC   & DR  &DR-BC  & Luo  & PSW & PSW-BC  & OR &OR-BC   & DR  &DR-BC & Luo  \\ \hline
        500 & 2 & 2 & 3 & 0.03 & -0.06 & 0.08 & 0.02 & 0.09 & 0.01 & 0.01 & 95.2 & 94.8 & 95.0 & 97.6 & 94.1 & 94.2 & 96.6 \\ 
        ~ & 3 & 1 & 2 & -0.45 & -0.10 & -0.43 & -0.03 & -0.43 & -0.03 & -0.08 & 91.0 & 94.6 & 45.8 & 95.6 & 40.9 & 94.5 & 95.9 \\ 
        ~ & ~ & ~ & 3 & -0.36 & -0.06 & -0.32 & -0.02 & -0.32 & -0.01 & -0.02 & 93.6 & 94.8 & 78.5 & 95.5 & 75.3 & 95.1 & 96.8 \\ 
        ~ & ~ & 2 & 3 & 0.06 & 0.00 & 0.11 & 0.01 & 0.11 & 0.01 & 0.07 & 95.4 & 95.2 & 90.5 & 95.1 & 89.4 & 96.6 & 96.9 \\ \hline
        2000 & 2 & 2 & 3 & 0.09 & -0.02 & 0.10 & 0.00 & 0.10 & 0.00 & 0.06 & 94.7 & 95.2 & 89.0 & 94.9 & 86.7 & 94.9 & 98.1 \\ 
        ~ & 3 & 1 & 2 & -0.43 & -0.02 & -0.43 & -0.01 & -0.43 & -0.01 & 0.02 & 80.5 & 95.6 & 0.7 & 95.3 & 0.4 & 95.5 & 97.7 \\ 
        ~ & ~ & ~ & 3 & -0.36 & -0.01 & -0.34 & -0.01 & -0.33 & 0.00 & 0.04 & 84.2 & 95.7 & 33.6 & 94.8 & 27.1 & 95.6 & 98.9 \\ 
        ~ & ~ & 2 & 3 & 0.08 & 0.01 & 0.09 & 0.00 & 0.10 & 0.00 & 0.03 & 94.8 & 95.6 & 81.6 & 95.4 & 79.7 & 94.9 & 98.1 \\ \hline
                   &&   &   &\multicolumn{7}{c}{MCSD} & \multicolumn{7}{c}{AESE} \\ 
           \cmidrule(lr){5-11}\cmidrule(lr){12-18}
       $n$&$g$&$z$&$z^\prime$             & PSW & PSW-BC  & OR &OR-BC   & DR  &DR-BC  & Luo  & PSW & PSW-BC  & OR &OR-BC   & DR  &DR-BC & Luo  \\ \hline
  500 & 2 & 2 & 3 & 0.86 & 0.93 & 0.35 & 0.44 & 0.24 & 0.25 & 0.90 & 0.96 & 1.01 & 0.35 & 0.42 & 0.24 & 0.25 & 1.20 \\ 
        ~ & 3 & 1 & 2 & 0.78 & 0.79 & 0.21 & 0.21 & 0.20 & 0.20 & 0.59 & 0.83 & 0.79 & 0.20 & 0.21 & 0.19 & 0.20 & 0.53 \\ 
        ~ & ~ & ~ & 3 & 0.82 & 0.75 & 0.28 & 0.29 & 0.25 & 0.27 & 0.55 & 0.82 & 0.77 & 0.29 & 0.29 & 0.26 & 0.26 & 0.68 \\ 
        ~ & ~ & 2 & 3 & 0.78 & 0.73 & 0.19 & 0.18 & 0.18 & 0.19 & 0.81 & 0.78 & 0.73 & 0.19 & 0.18 & 0.18 & 0.21 & 0.83 \\ \hline
        2000 & 2 & 2 & 3 & 0.37 & 0.41 & 0.15 & 0.16 & 0.12 & 0.12 & 0.48 & 0.38 & 0.42 & 0.16 & 0.16 & 0.11 & 0.12 & 0.63 \\ 
        ~ & 3 & 1 & 2 & 0.39 & 0.37 & 0.10 & 0.10 & 0.09 & 0.10 & 0.27 & 0.40 & 0.37 & 0.10 & 0.10 & 0.10 & 0.10 & 0.35 \\
        ~ & ~ & ~ & 3 & 0.39 & 0.37 & 0.14 & 0.14 & 0.13 & 0.13 & 0.26 & 0.39 & 0.37 & 0.14 & 0.14 & 0.13 & 0.13 & 0.37 \\ 
        ~ & ~ & 2 & 3 & 0.39 & 0.36 & 0.09 & 0.09 & 0.09 & 0.08 & 0.38 & 0.37 & 0.35 & 0.09 & 0.09 & 0.09 & 0.08 & 0.49 \\ \bottomrule
    \end{tabular}
\end{adjustbox}
\end{table}

\section{Illustrative Data application}\label{s:DE}

We apply the proposed methods to an animal antimony trioxide inhalation study conducted by the National Toxicology Program (NTP). The two-year antimony trioxide inhalation study randomized 800 Wistar Han rats and B6C3F1/N mice into four-level (0, 3, 10 or 30 $mg/m^3$) exposure to whole-body inhalation of antimony trioxide \citep{National_Toxicology_Program2017-qs}. Since it was a toxicity study, we follow the convention to encode higher exposure levels into lower treatment values, i.e., $Z\in\{1,2,3,4\}$ represents the dosages $\{30,10,3,0\}$ respectively. We consider the logarithmic transformed animal body weight after two years as the final outcome, which is truncated by death occurred before the end of the study. We consider four covariates in our analysis including the animal body weight in the first week, sex of rats or mice, species (i.e., rats or mice), and the interaction between sex and species. Similar to \cite{luo2023causal}, domain knowledge from toxicity studies and summary statistics of survival rates suggest no conflict with the monotonicity assumption, i.e., that lower toxicity dosage implies no worse survival. {Therefore, for the main analysis, we first estimate all possible SACE estimands (whenever they are well-defined) under the monotonicity and principal ignorability assumptions. However, to provide a more focused discussion in the sensitivity analyses, we will concentrate only on the estimands for the most stringent always-survivor stratum (i.e., $\Delta_{4}(z,z^\prime)$). This stratum is typically of primary interest as it allows for transitive pairwise comparisons among all treatments and, in this application, this most stringent always-survivor stratum is also expected to be the largest.}

\subsection{Main analysis under monotonicity and principal ignorability}\label{s:DE;subs:MPI}

\begin{table}[htbp]
\caption{Point estimates and associated quantile-based 95\% confidence intervals using $50,000$ times bootstrap for all marginal principal scores for the NTP data set based on augmented (`AUG') estimators or nonparametric (`NP') estimators. 
}
\label{tab:eg}
\centering
\begin{tabular}{lllllll}
    \toprule
        ~ & $e_0$ & $e_1$ & $e_2$ & $e_3$ & $e_4$  \\ \hline
        AUG & 0.29~(0.22,~0.35) & 0.07~(0.00,~0.16) & 0.10~(0.01,~0.20) & 0.20~(0.10,~0.29) & 0.34~(0.28,~0.41)  \\ 
        NP & 0.29~(0.18,~0.39) & 0.07~(0.00,~0.23) & 0.10~(0.00,~0.26) & 0.20~(0.06,~0.32) & 0.35~(0.27,~0.42)  \\  \bottomrule
    \end{tabular}
\end{table}

We first estimate the marginal principal score $e_g$ based on two approaches: (i) a simple nonparametric estimator $\widehat{e}_g^{\text{NP}} = \widehat{p}_{J-g+1}-\widehat{p}_{J-g}$ with $\widehat{p}_z=\mathbb{P}_n\{\mathbf{1}(Z=z)S\}/\pi_z$ and (ii) an augmented estimator $\widehat{e}_g^{\text{AUG}} = \widehat{p}_{J-g+1}-\widehat{p}_{J-g}$ with $\widehat{p}_z=\mathbb{P}_n\left\{{\mathbf{1}(Z=z)\{S-\widehat{p}_z(\mathbf{X})\}}/{\pi_z}+\widehat{p}_z(\mathbf{X})\right\}$. 
The estimated marginal principal scores and associated quantile-based 95\% confidence intervals using $50,000$ times non-parametric bootstrap are provided in Table \ref{tab:eg}. 
Of note, the intervals based on augmented estimators are generally narrower than those based on simple proportions. Further, assuming principal ignorability, we obtain the point estimates and corresponding 95\% Wald confidence intervals based on the proposed sandwich variance estimators using weighting, outcome regression, and doubly robust methods from a logistic principal score model and a linear conditional outcome mean model. The results are summarized in Table \ref{tab:ntp}. First, the principal score weighting estimator has a much wider confidence interval than the other two estimators in general, which aligns with results in our simulation study. Thus, most intervals based on weighting alone fail to exclude the null, while the other two methods produce narrower intervals that exclude zero. This demonstrates the potential efficiency gain with an additional outcome model. Second, the point and interval estimates are similar when using either the outcome regression or the doubly robust approach, suggesting that the conditional outcome mean model is likely adequately specified. Overall, the findings suggest that higher antimony trioxide dosage negatively affects body weight in the tested rats and mice, under the assumptions of principal ignorability and monotonicity.

\begin{table}[htbp]
\caption{Point estimates and associated Wald 95\% confidence intervals based on principal score weighting (`PSW'), outcome regression (`OW'), and doubly robust estimators (`DR') for estimating all possible SACEs on different principal strata, $g\in\{2,3,4\}$, for NTP data set. Notice that each causal contrast $\Delta_g(z,z^\prime)$ is based on comparing higher dosage (lower value of $z$) with lower dosage (higher value of $z$). 
}
\label{tab:ntp}
\centering
\begin{adjustbox}{width=1\textwidth}
\begin{tabular}{ccccrcrcrcc}
    \toprule
        $g$ & $z$ & $z^\prime$ & Estimand &\multicolumn{2}{c}{PSW}  & \multicolumn{2}{c}{OR}  & \multicolumn{2}{c}{DR}  \\ \hline
        2 & 3 & 4 & $\Delta_2(3,4)$ & 0.042   &($-$0.293,~0.377)  & $-$0.100&($-$0.174,~$-$0.026)  & $-$0.096&($-$0.151,~$-$0.041)  \\ 
        3 & 2 & 3 & $\Delta_3(2,3)$ & $-$0.039&($-$0.255,~0.177)  & $-$0.058&($-$0.119,~0.003)  & $-$0.056&($-$0.109,~$-$0.003) \\ 
        ~ & ~ & 4 & $\Delta_3(2,4)$ & $-$0.142&($-$0.393,~0.109)  & $-$0.129&($-$0.190,~$-$0.068)  & $-$0.130&($-$0.183,~$-$0.077)  \\ 
        ~ & 3 & 4 & $\Delta_3(3,4)$ & $-$0.103&($-$0.334,~0.128)  & $-$0.071&($-$0.122,~$-$0.020)  & $-$0.074&($-$0.117,~$-$0.031)  \\ 
        4 & 1 & 2 & $\Delta_4(1,2)$ & $-$0.110&($-$0.304,~0.084)  & $-$0.127&($-$0.178,~$-$0.076)  & $-$0.125&($-$0.176,~$-$0.074)  \\ 
        ~ & ~ & 3 & $\Delta_4(1,3)$ & $-$0.179&($-$0.383,~0.025)  & $-$0.187&($-$0.236,~$-$0.138)  & $-$0.185&($-$0.234,~$-$0.136)  \\ 
        ~ & ~ & 4 & $\Delta_4(1,4)$ & $-$0.242&($-$0.463,~$-$0.021)  & $-$0.268&($-$0.315,~$-$0.221)  & $-$0.265&($-$0.312,~$-$0.218)  \\ 
        ~ & 2 & 3 & $\Delta_4(2,3)$ & $-$0.069&($-$0.265,~0.127)  & $-$0.059&($-$0.102,~$-$0.016)  & $-$0.060&($-$0.103,~$-$0.017)  \\ 
        ~ & ~ & 4 & $\Delta_4(2,4)$ & $-$0.132&($-$0.334,~0.070)  & $-$0.140&($-$0.181,~$-$0.099)  & $-$0.140&($-$0.181,~$-$0.099)  \\ 
        ~ & 3 & 4 & $\Delta_4(3,4)$ & $-$0.063&($-$0.273,~0.147) & $-$0.081&($-$0.118,~$-$0.044)  & $-$0.080&($-$0.117,~$-$0.043)  \\ \bottomrule
    \end{tabular}
\end{adjustbox}
\end{table}

\subsection{Sensitivity analysis under violation of principal ignorability}\label{s:DE;subs:SA}

\begin{figure}
    \centering
    \includegraphics[scale=0.6]{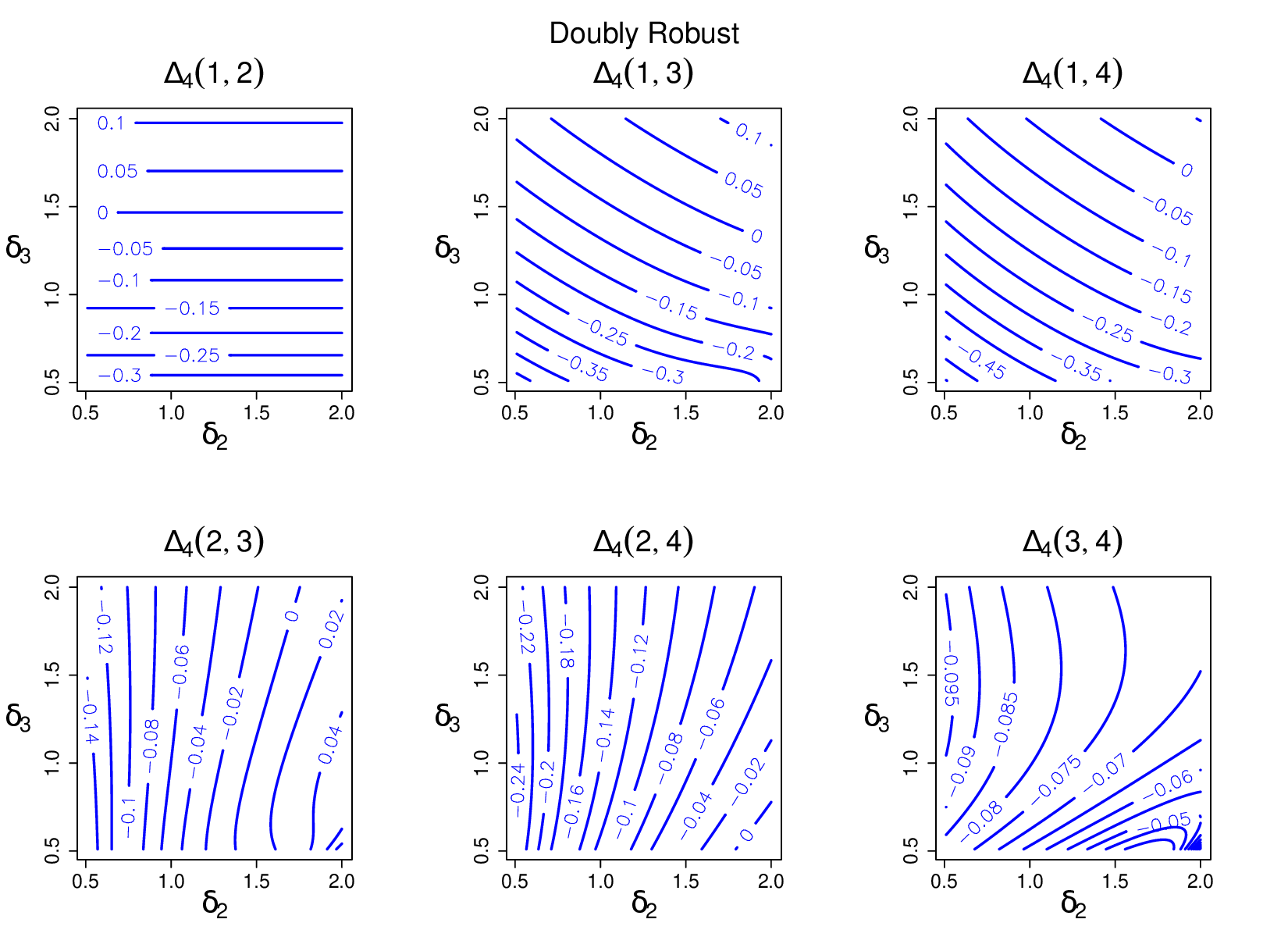}
    \caption{The contour plots for the point estimates of SACEs within the stratum $g=4$ in NTP study using the bias-corrected doubly robust estimator given equal conditional mean potential outcomes between the stratum $g=1$ and the stratum $g=4$, i.e., $\delta_1=1$, and the ratios of conditional mean potential outcome for the stratum $g=2$ or $g=3$ with respect to the stratum $g=4$ varying from half to twice, i.e., $\delta_2,\delta_3\in[0.50,2.00]$. As explained in Section \ref{s:SA;subsec:PI}, the bias-corrected doubly robust estimator is in fact singly robust as it requires correct specification of the principal score model. However, we retain the ``doubly robust'' in the estimator name to differentiate it from the simple weighting and regression estimators.}
    \label{fig:sa_pi_delta1}
\end{figure}

We investigate the sensitivity of the results when principal ignorability is violated. For simplicity, we assume that $\delta_{zg}(\mathbf{X})=\delta_{zg}$ does not depend on $\mathbf{X}$. Recall that the estimation of $\mu_g(z)$ requires specification of the sensitivity parameters in each row of the following right matrix
  \begin{align}\label{eq:sa_pi_matrix}
    \begin{pmatrix}
\ast&\ast&\mu_3(2)&\mu_4(2)\\\ast&\mu_2(3)&\mu_3(3)&\mu_4(3)\\\mu_1(4)&\mu_2(4)&\mu_3(4)&\mu_4(4)
    \end{pmatrix}   \Leftarrow \begin{pmatrix}
         \ast&\ast&\delta_{23}\\
         \ast&\delta_{32}&\delta_{33}\\
         \delta_{41}&\delta_{42}&\delta_{43}
     \end{pmatrix};
 \end{align}
and the estimation of $\mu_4(1)$ is unaffected due to the choice of the reference stratum. To focus ideas, we focus on assessing $\Delta_4(z,z')$ using the bias-corrected doubly robust estimator, and further assume that the sensitivity parameters are independent of the treatment assignment, i.e., $\delta_{41}=\delta_1$, $\delta_{32}=\delta_{42}=\delta_2$, and $\delta_{23}=\delta_{33}=\delta_{43}=\delta_3$; that is, elements in each column of the matrix in \eqref{eq:sa_pi_matrix} equal. Under this simplification, the total sensitivity parameters become $\{\delta_1,\delta_2,\delta_3\}$. We consider $3$ Scenarios; for Scenario $k\in\{1,2,3\}$, we fix $\delta_k=1$ and vary the other two sensitivity parameters between $0.5$ and $2$. For example, in Scenario 1, we set $\delta_1=1$ and vary $\delta_2$ and $\delta_3$ between 0.5 and 2. This corresponds to a setting where the expected potential (logarithmic) body weights of the mice or rats in stratum $g=1$ are the same as what would have been observed in stratum $g=4$, whereas the expected potential (logarithmic) body weights in strata $g=2$ and $g=3$ vary within a biologically plausible range between half and twice the (logarithmic) body weights that would have been observed in stratum $g=4$, adjusting for all measured covariates.

Figure \ref{fig:sa_pi_delta1} presents the sensitivity results under Scenario 1 with $\delta_1=1$ and $\{\delta_2,\delta_3\}\in[0.5,2]^{\otimes 2}$. Within the given ranges of $\delta_2$ and $\delta_3$, the signs of the point estimates of $\Delta_4(1,2)$, $\Delta_4(1,3)$ and $\Delta_4(1,4)$ are reversed only on a minor proportion of the sensitivity parameter space, suggesting that our SACE estimates are relatively robust to the violation of principal ignorability; this is especially the case for $\Delta_4(2,3),\Delta_4(2,4),\Delta_4(1,4)$. Similar patterns are observed in the Supplementary Material Figures 2-3 under Scenarios 2 and 3, respectively.

\subsection{Sensitivity analysis for monotonicity}\label{s:DE;subs:SA;subsub:mono}

\begin{figure}
    \centering
    \includegraphics[scale=0.48]{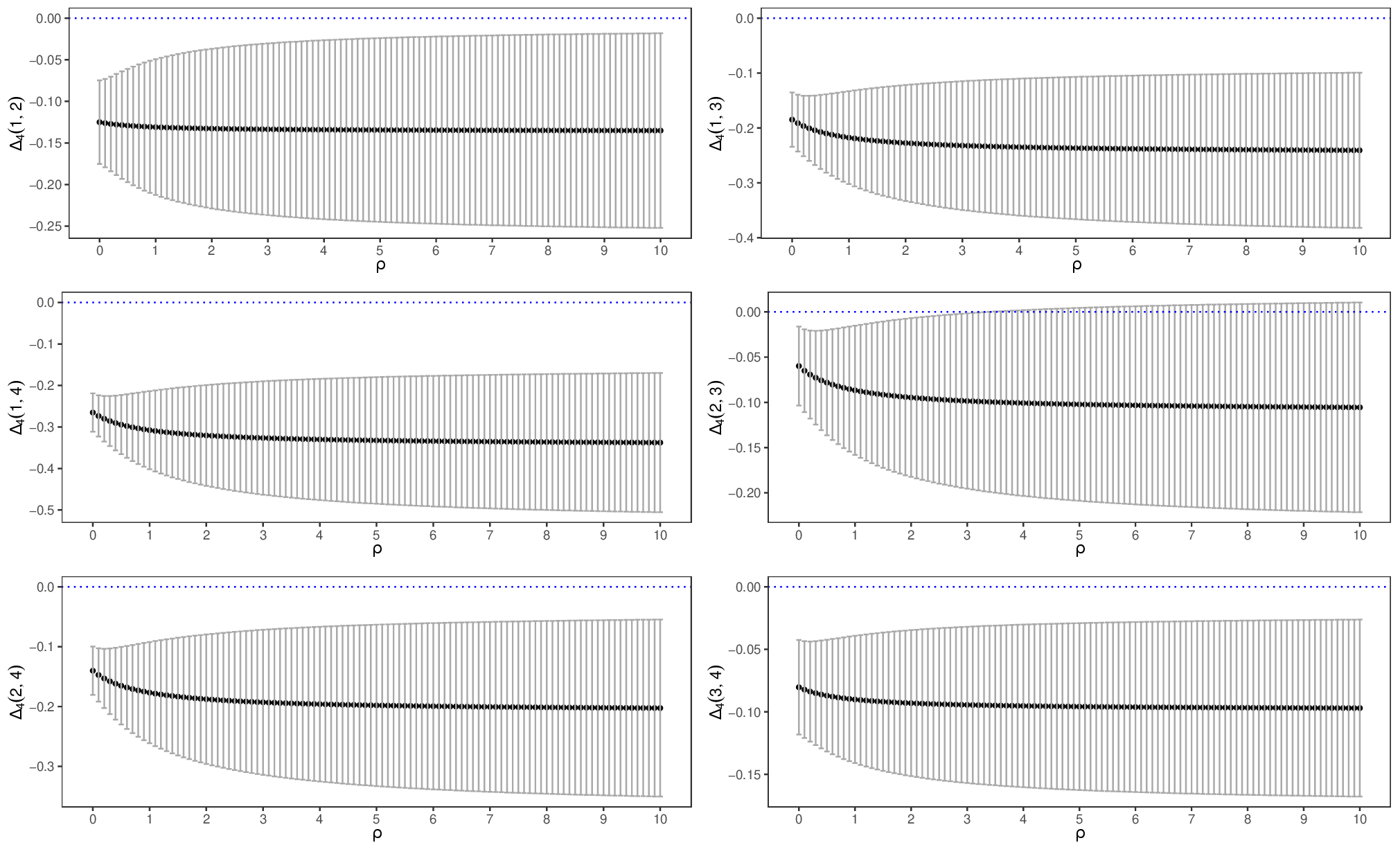}
    \caption{The point estimates and the associated 95\% Wald confidence intervals for the bias-corrected doubly robust estimator of $\Delta_4(z,z^\prime)$ when the monotonicity is violated with sensitivity parameters $\rho\in[0,10]$. Here, the parameter $\rho$ measures the magnitude of deviation from the monotonicity assumption. The blue dotted line indicates the null.}
    \label{fig:sa_mono}
\end{figure}

We next assess the sensitivity of our conclusions under assumed departure from the monotonicity assumption. Without monotonicity, there exists at most $11$ additional principal strata and we define them with respect to the reference group $r=0$ because $\widehat{e}_0$ is estimated to be the second largest principal stratum. To make the procedure practically operationalizable, we make a simplification by assuming that all $11$ sensitivity parameters are constant and equal, and denote them as $\rho_{0010}(\bX)=\rho_{0100}(\bX)=\rho_{1000}(\bX)=\rho_{0101}(\bX)=\rho_{1001}(\bX)=\rho_{1010}(\bX)=\rho_{0110}(\bX)=\rho_{1100}(\bX)=\rho_{1011}(\bX)=\rho_{1101}(\bX)=\rho_{1110}(\bX)=\rho$, 
where $\rho\geq0$ 
satisfies the constraints $e_{g}\geq0$ for $\forall g\in\mathcal{Q}$ based on $\widehat{e}^\text{AUG}_{g}$ in Table \ref{tab:eg}. For example, $\rho=0$ implies that no harmed strata exist, while $\rho>0$ implies the existence of all additional harmed principal strata by redistributing the members originally in strata $g=0$ and $g=4$. In addition, Equations \eqref{eq:psiden_without_mono} imply that the marginal principal scores for the unharmed strata, i.e., $g\in\{0,1,2,3,4\}$, converge to $\{0,0.11,0.14,0.24,0.05\}$ when $\rho\to\infty$.

Figure \ref{fig:sa_mono} and Supplementary Material Figures 4--5 show the point estimates with 95\% Wald confidence intervals based on the proposed sandwich variance estimators for all the contrasts within stratum $g=4$ using the bias-corrected doubly robust estimator, bias-corrected principal score weighting estimator, and bias-corrected outcome regression estimator, respectively, under violation of monotonicity within the range $\rho\in[0,10]$. First, the signs and the statistical significance remain unchanged when varying the sensitivity parameter, except for $\Delta_4(2,3)$ under $\rho>3$; this generally supports the robustness of the final estimates to the non-monotonicity with respect to harmed strata. Second, the interval estimates widen as the sensitivity parameter $\rho$ increases; this is because the uncertainty increases with larger values of $\rho$. For instance, the interval estimate for the expected decrement in (logarithmic) body weights of the mice or rats widens from $(-0.181,-0.099)$ to $(-0.350,-0.055)$ as the proportion of harmed strata increases, if the toxicity level increases from $0$ to $10$ $mg/m^3$. Third, the bias-corrected doubly robust estimator and the bias-corrected outcome regression estimator remain more efficient than the bias-corrected weighting estimator when monotonicity is violated, in alignment with findings under monotonicity. 

To offer additional comparisons with the exploration in \cite{luo2023causal}, we also consider a more restricted scenario of partial deviation from monotonicity, i.e., only three additional harmed strata, $\{1011,0101,0010\}$, may exist. Similarly, we define them with respect to the reference group $g=0$. For ease of representation, we further assume three sensitivity parameters equal, and denote them as $\rho_{1011}=\rho_{0101}=\rho_{0010}=\rho$, where $\rho$ can only take values in $[0,0.526]$ due to the constraints $e_g\geq0$ for $\forall g\in\mathcal{Q}$ based on $\widehat{e}^\text{AUG}_g$ in Table \ref{tab:eg}. Results are reported in Supplementary Material Figures 6-8, and similarly show that our estimates remain robust to the partial violation between adjacent strata.

\section{Discussion}\label{s:Dis}

In this article, we addressed the identification and estimation of SACEs in multi-arm randomized trials under truncation by death. We proposed the principal score weighting estimator and the outcome regression estimator based on simple moment conditions, and the doubly robust estimator based on the efficient influence function. The doubly robust estimator is consistent if either the principal score model or the  outcome mean model is correctly specified, and is locally efficient under correct specifications of both models. We also proposed the sandwich variance estimators for each estimator when the nuisance models are estimated by parametric regression. As the proposed estimators depend on the principal ignorability and monotonicity assumptions, we further articulated a sensitivity function approach to address violation of each assumption, and operationalized our methods in a four-arm toxicity study. {For completeness, an extension of our approach to observational studies under ignorable assignment was also presented in Section \ref{sec:generalization-os}.}


{In the context of multi-arm studies, our method should be viewed as a strong alternative to the approach proposed by \cite{luo2023causal}, with each having distinct strengths and limitations. First, when principal ignorability holds, our doubly robust estimator offers greater protection against working model misspecification. In comparable simulation scenarios when principal ignorability holds, the estimator by \cite{luo2023causal} is inconsistent if either the principal score or the outcome model is misspecified. Our simulations further show that when both models are correctly specified, our estimator is at least as efficient as theirs in most cases. Second, it is important to acknowledge that when principal ignorability does not hold but the assumptions required by Luo et al. are satisfied, their estimator remains valid whereas our estimator may be biased. Therefore, our simulations demonstrate that both methods may incur bias depending on which assumptions are violated. Interestingly, under violation of principal ignorability, our bias-corrected estimator (with correctly specified sensitivity functions) yields bias comparable to that of \cite{luo2023causal} while substantially improving efficiency. Third, we provide a computationally efficient sandwich variance estimator that is more scalable to larger datasets compared to their bootstrap-based variance calculation, and may be faster to implement in practice. Finally, it is worth pointing out that both approaches rely on monotonicity to reduce the number of strata. Our sensitivity analysis, however, extends beyond \cite{luo2023causal} by accommodating more general departures from this assumption. Taken together, we recommend that in practice analysts consider both methods as complementary, using each as a possible sensitivity analysis for the other to assess the robustness of conclusions to alternative causal identification assumptions.}

A possible limitation of this work is that we have primarily focused on parametric modeling of the nuisance parameters, following common practice in analyzing clinical trials in practice. More flexible modeling strategies, such as data-adaptive machine learning methods, may have advantages in estimating the principal score and conditional outcome functions, especially if baseline covariates are high-dimensional or include several continuous components; thus, these flexible regression models can effectively reduce model misspecification bias, when the required causal identification assumptions hold. Because flexible modeling strategies often converge to the true model at a rate slower than root-$n$, they are best combined with our doubly robust or multiply robust estimators to arrive at a debiased machine learning estimator; see, for example, the developments in \cite{Chernozhukov2018} for general theory, and \cite{JiangJRSSB2022} and \cite{cheng2025identification} for machine-learning based principal stratification with a binary treatment. It would be useful to explore this type of causal machine learning development in the multiple treatments setting with a binary intermediate outcome in future work.

\section*{Data Availability Statement}
The data set analyzed in Section 6 of this article is publicly available at \url{https://cebs.niehs.nih.gov/cebs/publication/TR-590}. 

\bigskip
\begin{center}
{\large\bf SUPPLEMENTARY MATERIAL}
\end{center}

\section{Summary}
\label{sec:intro_supple}
For greater generality, all proofs in this supplementary material are based on the non‐randomized observational study setup, in which the randomized case can be viewed as a special case such that $\pi_z(\bX)=\pi_z$ is a known constant. This supplementary material is organized as follows. Section \ref{sec:CovBalance} formally states the balancing properties of principal scores. Section \ref{sec:proof_main} provides the proof of the main results under monotonicity and principal ignorability. Sections \ref{sec:proof_without_PI} and \ref{sec:proof_without_mono} prove the results when principal ignorability and monotonicity are violated, respectively. 
Section \ref{ss:additional_sim} provides supplementary details for the simulation study.
We attach Supplementary Material tables and figures in Section \ref{sec:tables}.

\section{Additional statistical results}\label{sec:CovBalance}
\subsection{Balancing properties of principal scores}
The below proposition characterizes a class of balancing properties motivated by the identification formulas in the main manuscript.
\begin{prop}\label{prop:balance property}
Under treatment ignorability but without Assumptions 1--2, for $\forall z\in\mathcal{J}$ and arbitrary vector-valued random functions of covariates, $h(\mathbf{X})$, we have that
\begin{align*}
    &E\left\{\frac{p_{J-g+1}(\mathbf{X})-p_{J-g}(\mathbf{X})}{p_{J-g+1}-p_{J-g}}\frac{S}{p_z(\mathbf{X})}\frac{\mathbf{1}(Z=z)}{\pi_z(\bX)}h(\bX)\right\}\\
    =&E\left\{\frac{\mathbf{1}(Z=J-g+1)S/\pi_{J-g+1}(\bX)-\mathbf{1}(Z=J-g)S/\pi_{J-g}(\bX)}{p_{J-g+1}-p_{J-g}}h(\mathbf{X})\right\}\\
    =&E\left\{\frac{p_{J-g+1}(\mathbf{X})-p_{J-g}(\mathbf{X})}{p_{J-g+1}-p_{J-g}}h(\mathbf{X})\right\}.
\end{align*}
Furthermore, if Assumption 1 holds, they also equal to
$$E\left\{h(\mathbf{X})|G=g\right\},$$
provided $E\left\{h(\mathbf{X})|G=g\right\}<\infty$.
\end{prop}
The proof is given in Section \ref{sec:proof_main}. Proposition \ref{prop:balance property} is a direct generalization of balancing properties in \cite{JiangJRSSB2022} (see Supplementary Material S1) to multiple treatments and it is parallel to the classic covariates balancing property of propensity score in \cite{rosenbaum1983central}. 
Proposition \ref{prop:balance property} says that the weighted functions of covariates are balanced in expectation across each treatment arm even without monotonicity and PI, and this weighted expectation can be further characterized by its conditional mean within the stratum $g$ if monotonicity holds. 

\subsection{Estimators based on moment conditions}
The moment conditions in Equations \eqref{eq:iden_ps_weighting}–\eqref{eq:iden_ps+om} of the main manuscript motivate three natural estimators, which can be expressed as follows: for any $g\in\mathcal{J}$ and any $z\geq J+1-g$,
\begin{align*}
    &\widehat{\mu}^{\text{tp+ps}}_g(z)=\mP_n\left\{\frac{\widehat{p}_{J-g+1}(\mathbf{X})-\widehat{p}_{J-g}(\mathbf{X})}{\widehat{p}_{J-g+1}-\widehat{p}_{J-g}}\frac{S}{\widehat{p}_z(\mathbf{X})}\frac{\mathbf{1}(Z=z)}{\widehat{\pi}_z(\bX)}Y\right\}
    ,\\
    &\widehat{\mu}_g^{\text{tp+or}}(z)=\mP_n\left\{\frac{\mathbf{1}(Z=J-g+1)S/\widehat{\pi}_{J-g+1}(\bX)-\mathbf{1}(Z=J-g)S/\widehat{\pi}_{J-g}(\bX)}{\widehat{p}^\ast_{J-g+1}-\widehat{p}^\ast_{J-g}}\widehat{m}_{z}(\mathbf{X})\right\},\\
    &\widehat{\mu}_g^{\text{ps+or}}(z)=\mP_n\left\{\frac{\widehat{p}_{J-g+1}(\mathbf{X})-\widehat{p}_{J-g}(\mathbf{X})}{\widehat{p}_{J-g+1}-\widehat{p}_{J-g}}\widehat{m}_{z}(\mathbf{X})\right\},
\end{align*}
where $\widehat{p}_z=\mP_n\{\widehat{p}_z(\bX)\}$ and $\widehat{p}^\ast_z=\mP_n\{S\mathbf{1}(Z=z)/\widehat{\pi}_Z(\bX)\}$ for all $z\in\mathcal{J}$. Here, we use the superscript `tp' to denote the treatment probability model $\pi_z(\bX; \bbeta_z)$, `ps' to denote the principal score model $p_z(\bX; \balpha_z)$, and `or' to denote the outcome regression model $m_z(\bX; \bgamma_z)$. Then, $\widehat{\mu}_g^{\text{tp+ps}}(z)$ is the weighting estimator based on the propensity score and principal score models; $\widehat{\mu}_g^{\text{tp+or}}(z)$ combines the propensity score and outcome regression models; and $\widehat{\mu}_g^{\text{ps+or}}(z)$ combines the principal score and outcome regression models.
It is clear that $\widehat{\mu}_g^{\text{tp+or}}(z)$ and $\widehat{\mu}_g^{\text{ps+or}}(z)$ are g-computation formula estimators, which standardize the outcome regression estimates to the target principal stratum subpopulation.

\subsection{Triply robust estimators based on EIF}
The triply robust estimators take the same form as the doubly robust estimators under a randomized trial, except that $\psi_{F(Y,S,\bX),z}$ is replaced by the following:
\begin{align*}
\psi_{F(Y,S,\mathbf{X}),z}=\frac{\mathbf{1}(Z=z)}{\pi_z(\bX)}\Big\{F(Y,S,\mathbf{X})-E\{F(Y,S,\mathbf{X})|Z=z,\mathbf{X}\}\Big\}+E\{F(Y,S,\mathbf{X})|Z=z,\mathbf{X}\}.
\end{align*}

\section{Proof of the main results under monotonicity and principal ignorability}\label{sec:proof_main}
\subsection{Proof of the identification formulas for principal scores}
According to Table \ref{tab:G+S|Z}, the observed stratum $S=1|Z=z$ is a mixture of latent strata $G=J-z+1,\ldots,J$, which shows that the event $S=1|Z=z$ is a union of events $\underset{g=J-z+1,\ldots,J}{\cup} G=g|Z=z$. As a result, 
\begin{equation}\label{appenS1:eq:diff}
p_z(\mathbf{X})=\Pr(S=1|Z=z,\mathbf{X})=\sum_{g=J-z+1}^J\Pr(G=g|Z=z,\mathbf{X})=\sum_{g=J-z+1}^J\Pr(G=g|\mathbf{X}),
\end{equation}
where the last equality is due to treatment ignorability.
Noting that the system of Equations in \eqref{appenS1:eq:diff} is linear, solving \eqref{appenS1:eq:diff} by Gaussian eliminations yields the characterizations of principal scores with respect to estimable quantity $p_z(\mathbf{X})$ in Equation \eqref{eq:PrinScoreIden} in the main manuscript. 

\subsection{Proof of the identification formulas for $\mu_g(z)$ based on moment conditions}\label{ss:proof_iden}
Our proof relies on the following $4$ lemmas. 
\begin{lemma}[Importance Sampling]\label{iden:lemma1}
    Assume $X\sim f_X$ and $Y\sim f_Y$ are random variables (possibly random vectors) with $P_X\ll P_Y$. Then for arbitrary scalar function $h$ such that $E\{h(X)\}<\infty$, 
$$E\{h(X)\}=E\left\{\frac{f_X(Y)}{f_Y(Y)}h(Y)\right\}.$$
\end{lemma}
\begin{proof}
We assume the underlying probability measures for $X,Y$ are both dominated by the Lebesgue measure $P$. Then 
$$E\{h(X)\}=\int h(x)f_X(x)dP=\int h(y)\frac{f_X(y)}{f_Y(y)}f_Y(y)dP=E\left\{\frac{f_X(Y)}{f_Y(Y)}h(Y)\right\},$$
where ${f_X(y)}/{f_Y(y)}$ is well-defined on the support of $X$ because $P_X\ll P_Y$.
\end{proof}
\begin{lemma}\label{iden:lemma2}
    For $g\in\mathcal{Q}\equiv\{0,\dots,J\}$ and arbitrary vector-valued function $h$,
$$E\{h(\mathbf{X})|G=g\}=E\left\{\frac{e_g(\mathbf{X})}{e_g}h(\mathbf{X})\right\}=E\left\{\frac{p_{J-g+1}(\mathbf{X})-p_{J-g}(\mathbf{X})}{p_{J-g+1}-p_{J-g}}h(\mathbf{X})\right\}.$$
\end{lemma}

\begin{proof}
By Bayes' theorem, we have that
$$f_{\mathbf{X}|G=g}=\frac{\Pr(G=g|
\mathbf{X})f_{\mathbf{X}}}{\Pr(G=g)}=\frac{e_g(\mathbf{X})}{e_g}f_{\mathbf{X}},$$
where $f_{\mathbf{X}|G=g}$ is the conditional density of covariates given stratum $G=g$ and $f_{\mathbf{X}}$ is the marginal density of covariates. Applying Lemma \ref{iden:lemma1} and Equation \eqref{eq:PrinScoreIden} in the main manuscript completes the proof. 
\end{proof}

\begin{lemma}\label{lemma:withoutmono_iden1}
    For $\forall z\in\mathcal{J}$ and arbitrary vector-valued function $h$,
    $$E\{p_z(\mathbf{X})\times h(\mathbf{X})\}=E\left\{\frac{S\mathbf{1}(Z=z)}{\pi_z(\bX)}\times h(\mathbf{X})\right\}.$$
\end{lemma}
\begin{proof}
  By the law of total expectation (LOTE) and treatment ignorability,
  $$E\left\{\frac{S\mathbf{1}(Z=z)}{\pi_z(\bX)}\times h(\mathbf{X})\right\}=E\left\{\Pr(S=1,Z=z|\mathbf{X})h(\mathbf{X})/\pi_z(\bX)\right\}=E\{p_z(\mathbf{X})\times h(\mathbf{X})\}.$$
\end{proof}
\begin{lemma}\label{iden:lemma3}
    For arbitrary vector-valued function $h$,
$$E\{h(\mathbf{X})|G=g\}=E\left\{\left(\frac{S\mathbf{1}(Z=J-g+1)}{\pi_{J-g+1}(\bX)}-\frac{S\mathbf{1}(Z=J-g)}{\pi_{J-g}(\bX)}\right )\frac{h(\mathbf{X})}{p_{J-g+1}-p_{J-g}}\right\}.$$
\end{lemma}
\begin{proof}
It follows from Lemma \ref{iden:lemma2} and Lemma \ref{lemma:withoutmono_iden1}. 
\end{proof}
For $z\in\mathcal{J}$, 
we define $U_z=\{J-z+1,\ldots,J\}$. 
Then we have
\begin{align}
    \mu_g(z)&=E\{Y(z)|G=g\}=E\{E\{Y(z)|G=g,\mathbf{X}\}|G=g\}\quad \text{(by LOTE)}\nonumber\\
    &=E\{E\{Y(z)|G\in U_{z},\mathbf{X}\}|G=g\}\quad \text{(by principal ignorability)}\nonumber\\
    &=E\{E\{Y|Z=z,G\in U_{z},\mathbf{X}\}|G=g\}\quad \text{(by treatment ignorability and SUTVA)}\nonumber\\
    &
    =E\{m_{z}(\mathbf{X})|G=g\}\quad \text{(Table \ref{tab:G+S|Z})}\label{eq:table2implication}\\
    &=E\left\{\frac{p_{J-g+1}(\mathbf{X})-p_{J-g}(\mathbf{X})}{p_{J-g+1}-p_{J-g}}m_{z}(\mathbf{X})\right\}\quad \text{(by Lemma \ref{iden:lemma2})},\label{eq:alt_iden}
\end{align}
which corresponds to the identification formula \eqref{eq:iden_ps+om} in the main manuscript. Then, we apply Lemma \ref{iden:lemma3} to Equation \eqref{eq:table2implication}, leading to identification formula \eqref{eq:iden_outcome_regression} in the main manuscript.
Next, we show the identification formula \eqref{eq:iden_ps_weighting}, using both propensity score and principal score weighting. By LOTE, we induce that 
\begin{align}\label{appenS2:eq:IdenPCEs9(c)pre}
    E\{S\mathbf{1}(Z=z)Y|\mathbf{X}\}&=E\{E\{S\mathbf{1}(Z=z)Y|S\mathbf{1}(Z=z),\mathbf{X}\}|\mathbf{X}\}\nonumber\\
    &=E\{\Pr(S=1,Z=z|\mathbf{X})E\{Y|Z=z,S=1,\mathbf{X}\}|\mathbf{X}\}\nonumber\\
    &=p_z(\mathbf{X})\pi_{z}(\bX)m_{z}(\mathbf{X}).
\end{align}
By LOTE and treatment ignorability, one obtains
\begin{align}
    &E\left\{\frac{p_{J-g+1}(\mathbf{X})-p_{J-g}(\mathbf{X})}{p_{J-g+1}-p_{J-g}}\frac{S\mathbf{1}(Z=z)}{p_z(\mathbf{X})\pi_{z}(\bX)}Y\right\}\nonumber\\
    =&E\left\{E\left\{\frac{p_{J-g+1}(\mathbf{X})-p_{J-g}(\mathbf{X})}{p_{J-g+1}-p_{J-g}}\frac{S\mathbf{1}(Z=z)}{p_z(\mathbf{X})\pi_{z}(\bX)}Y\rvert\mathbf{X}\right\}\right\}\nonumber\\
    =&E\left\{\frac{p_{J-g+1}(\mathbf{X})-p_{J-g}(\mathbf{X})}{p_{J-g+1}-p_{J-g}}\frac{E\{S\mathbf{1}(Z=z)Y\rvert\mathbf{X}\}}{p_z(\mathbf{X})\pi_{z}(\bX)}\right\}\nonumber\\
    =&E\left\{\frac{p_{J-g+1}(\mathbf{X})-p_{J-g}(\mathbf{X})}{p_{J-g+1}-p_{J-g}}m_{z}(\mathbf{X})\right\}\quad \text{(by Equation \eqref{appenS2:eq:IdenPCEs9(c)pre})}\label{eq:iden_mugz_ps+om not dr}.
\end{align}

\subsection{Proof of Proposition \ref{prop:balance property}}
It follows from the proof of identification formulas given in Section \ref{ss:proof_iden} in the main manuscript by replacing $Y$ with $h(\mathbf{X})$ provided $E\left\{h(\mathbf{X})|G=g\right\}<\infty$. 

\subsection{Proof of Theorem \ref{thm:eif}}\label{sec:main;ss:eif}
Our proof is based on Chapter 3 and Chapter 4 in \cite{tsiatis2006semiparametric}. According to the identification formulas, we can derive the efficient influence function (EIF) based on the joint density of observed data vector $\mathbf{V}$. We derive EIF in the non-parametric sense, i.e., we impose no restrictions on the joint density of observed vector $\mathbf V$. Denote $f(\mathbf{V})$ as the joint density function of $\mathbf{V}$. Consider the following factorization
\begin{equation*}f(\mathbf{V})=f(\mathbf{X})f(Z|\mathbf{X})f(S|Z,\mathbf{X})f(Y|S,Z,\mathbf{X}).
\end{equation*}
By Theorem 4.4 and Theorem 4.5 in \cite{tsiatis2006semiparametric}, the tangent space $\mathcal{F}$ is the entire Hilbert space $\mathcal{H}$, i.e., the collection of all $1$ dimensional random functions of $\mathbf{V}$ with mean zero and finite variance, and furthermore, 
\begin{equation*}
 \mathcal{F}=\mathcal{F}_1\oplus\mathcal{F}_2\oplus\mathcal{F}_3\oplus\mathcal{F}_4,
\end{equation*}
where $\{\mathcal{F}_1,\mathcal{F}_2,\mathcal{F}_3,\mathcal{F}_4\}$ are mutually orthogonal with
\begin{align*}
    &\mathcal{F}_1=\{h(\mathbf{X})\in\mathcal{H}:E\{h(\mathbf{X})\})=0\},\\
    &\mathcal{F}_2=\{h(Z,\mathbf{X})\in\mathcal{H}:E\{h(Z,\mathbf{X})\}|\mathbf{X})=0\},\\
    &\mathcal{F}_3=\{h(S,Z,\mathbf{X})\in\mathcal{H}:E\{h(S,Z,\mathbf{X})|Z,\mathbf{X}\})=0\},\\
    &\mathcal{F}_4=\{h(\mathbf{V})\in\mathcal{H}:E\{h(\mathbf{V})|S,Z,\mathbf{X}\})=0\}.
\end{align*}
Consider a parametric sub-model with Euclidean parameters $\boldsymbol{\theta}$ and the density $f_{\boldsymbol{\theta}}(\mathbf{V})$. Assume $f_{\boldsymbol{\theta}}(\mathbf{V})$ attains the truth at $\boldsymbol{\theta}=\boldsymbol{\theta}_0$ and we write $f_{\boldsymbol{\theta}_0}=f$ and $E_{\boldsymbol{\theta}_0}=E$ for ease of notation. 
Consider the following orthogonal decomposition of the score vector
\begin{equation*}S(\mathbf{V})=S(\mathbf{X})+S(Z|\mathbf{X})+S(S|Z,\mathbf{X})+S(Y|S,Z,\mathbf{X}),
\end{equation*}
where 
\begin{align*}
    &S(\mathbf{V})=\partial \text{log}f_{\boldsymbol{\theta}}(\mathbf{V})/\partial \boldsymbol{\theta}|_{\boldsymbol{\theta}=\boldsymbol{\theta}_0},~S(Y|S,Z,\mathbf{X})=\partial \text{log}f_{\boldsymbol{\theta}}(Y|S,Z,\mathbf{X})/\partial \boldsymbol{\theta}|_{\boldsymbol{\theta}=\boldsymbol{\theta}_0},\\
    &S(Z|\mathbf{X})=\partial \text{log}f_{\boldsymbol{\theta}}(Z|\mathbf{X})/\partial \boldsymbol{\theta}|_{\boldsymbol{\theta}=\boldsymbol{\theta}_0},~S(\mathbf{X})=\partial \text{log}f_{\boldsymbol{\theta}}(\mathbf{X})/\partial \boldsymbol{\theta}|_{\boldsymbol{\theta}=\boldsymbol{\theta}_0}.
\end{align*}
We define $\beta(\btheta)\equiv\mu_g^{(\btheta)}(z)$ as the value of $\mu_g(z)$ in the sub-model and the truth $\mu_g(z)=\beta(\btheta_0)=\beta$. By Theorem 3.2 in \cite{tsiatis2006semiparametric}, the influence function $\Psi_{zg}(\mathbf{V})\in\mathcal{H}$ for the sub-model can be characterized by 
\begin{equation}\label{eq:influfuncchara}
  E\{\Psi_{zg}(\mathbf{V})S(\mathbf{V})\}=\frac{\partial \beta(\boldsymbol{\theta})}{\partial \boldsymbol{\theta}}|_{\boldsymbol{\theta}=\boldsymbol{\theta}_0}.
\end{equation}
Hereafter, we shall use $\dot{\beta}(\btheta)|_{\boldsymbol{\theta}=\boldsymbol{\theta}_0}$ to denote $\frac{\partial \beta(\boldsymbol{\theta})}{\partial \boldsymbol{\theta}}|_{\boldsymbol{\theta}=\boldsymbol{\theta}_0}$ and apply it to all pathwise partial derivatives with respect to $\boldsymbol{\theta}$. 
\cite{kennedy2023semiparametric} showed that there is at most one solution to the differential equation \eqref{eq:influfuncchara} under $\mathcal{M}_{np}$. By Theorem 4.3 in \cite{tsiatis2006semiparametric}, the EIF is indeed $\Psi_{zg}(\mathbf{V})$ because the tangent space is the entire Hilbert space. As a result, EIF is given by the solution to Equation \eqref{eq:influfuncchara}. 
By Equation \eqref{eq:iden_mugz_ps+om not dr}, $\beta={N}\times {D^{-1}}$ with 
$$N=E\{(p_{J-g+1}(\mathbf{X})-p_{J-g}(\mathbf{X}))m_{z}(\mathbf{X})\},\quad D=p_{J-g+1}-p_{J-g}.$$
Let $\Psi_N(\mathbf{V})$ and $\Psi_D(\mathbf{V})$ be the influence function of $N$ and $D$, respectively. By \cite{kennedy2023semiparametric} or Lemma S2 in the Supplementary Material of \cite{JiangJRSSB2022}, if both $\Psi_N(\mathbf{V})$ and $\Psi_D(\mathbf{V})$ are known, the influence function of $\mu_g(z)$ can be explicitly given by 
$$\Psi_{zg}(\mathbf{V})=\frac{1}{D}\Psi_N(\mathbf{V})-\frac{N}{D^2}\Psi_D(\mathbf{V}),$$
where $E\{\Psi_N(\mathbf{V})S(\mathbf{V})\}=\dot{N}_{\btheta}|_{\btheta=\btheta_0}$ and $E\{\Psi_D(\mathbf{V})S(\mathbf{V})\}=\dot{D}_{\btheta}|_{\btheta=\btheta_0}$. This is called the \verb"quotient rule" for influence function operator (similar to the quotient rule for calculus). Therefore, $\Psi_{zg}(\mathbf{V})$ is obtained once we know $\Psi_N(\mathbf{V})$ and $\Psi_D(\mathbf{V})$. Below, we present three lemmas to facilitate our proof.
\begin{lemma}\label{eif:lemma1}
    Suppose $F(Y,S,\mathbf{X})$ is any integrable random function of $(Y,S,\mathbf{X})$. Define $\mu_{z,F(Y,S,\mathbf{X}),\boldsymbol{\theta}}(\mathbf{X})=E_{\boldsymbol{\theta}}[F(Y,S,\mathbf{X})|Z=z,\mathbf{X}\}$. Then, we have that
    $$\dot{\mu}_{z,F(Y,S,\mathbf{X}),\boldsymbol{\theta}}(\mathbf{X})|_{\boldsymbol{\theta}=\boldsymbol{\theta}_0}=E\{(\psi_{F(Y,S,\mathbf{X}),z}-\mu_{z,F(Y,S,\mathbf{X})}(\mathbf{X}))S(Y,S|Z,\mathbf{X})|\mathbf{X}\},$$
where $\mu_{z,F(Y,S,\mathbf{X})}(\mathbf{X})=\mu_{z,F(Y,S,\mathbf{X}),\btheta_0}(\mathbf{X})$.
\end{lemma}

\begin{proof}
We define $S(Y,S|Z=z,\mathbf{X})=\partial \text{log}f_{\btheta}(Y,S|Z=z,\mathbf{X})/\partial \btheta|_{\btheta=\btheta_0}$ as the score vector with respect to conditional density $f(Y,S|Z=z,\mathbf{X})$ evaluated at the truth, and hereafter, we will use similar notations with respect to other conditional densities. Then, 
\begin{align*}
    \dot{\mu}_{z,F(Y,S,\mathbf{X}),\boldsymbol{\theta}}(\mathbf{X})|_{\btheta=\boldsymbol{\theta}_0}&=E\{F(Y,S,\mathbf{X})S(Y,S|Z=z,\mathbf{X})|Z=z,\mathbf{X}\}\quad\\
    &=E\{(F(Y,S,\mathbf{X})-\mu_{z,F(Y,S,\mathbf{X})}(\mathbf{X}))S(Y,S|Z=z,\mathbf{X})|Z=z,\mathbf{X}\}\\
    &=E\left \{\frac{\mathbf{1}(Z=z)\{F(Y,S,\mathbf{X})-\mu_{z,F(Y,S,\mathbf{X})}(\mathbf{X}))\}}{\Pr(Z=z|\mathbf{X})}S(Y,S|Z,\mathbf{X})|\mathbf{X}\right\}\\
    &=E\{\{\psi_{F(Y,S,\mathbf{X}),z}-\mu_{z,F(Y,S,\mathbf{X})}(\mathbf{X})\}S(Y,S|Z,\mathbf{X})|\mathbf{X}\},
\end{align*}
where the second equality holds because the score function has mean zero, the third equality follows from the LOTE, and the last equality follows from the definition of $\psi_{F(Y,S,\mathbf{X}),z}$.

\end{proof}

\begin{lemma}\label{eif:lemma2}
    Suppose $F(Y,S,\mathbf{X})$ is any integrable random function in $(Y,S,\mathbf{X})$. Define $\mu_{z,F(Y,S,\mathbf{X}),\boldsymbol{\theta}}=E_{\boldsymbol{\theta}}\{\mu_{z,F(Y,S,\mathbf{X}),\boldsymbol{\theta}}(\mathbf{X})\}$. Then
    $$\dot{\mu}_{z,F(Y,S,\mathbf{X}),\boldsymbol{\theta}}|_{\boldsymbol{\theta}=\boldsymbol{\theta}_0}=E\{(\psi_{F(Y,S,\mathbf{X}),z}-\mu_{z,F(Y,S,\mathbf{X})})S(\mathbf{V})\},$$
    where $\mu_{z,F(Y,S,\mathbf{X})}=\mu_{z,F(Y,S,\mathbf{X}),\btheta_0}$ and $\psi_{F(Y,S,\mathbf{X}),z}-\mu_{z,F(Y,S,\mathbf{X})}\in\mathcal{H}$.
\end{lemma}
\begin{proof}
Note
\begin{align}
    E\{(\psi_{F(Y,S,\mathbf{X}),z}-\mu_{z,F(Y,S,\mathbf{X})}(\mathbf{X}))S(Z,\mathbf{X})\}&=E\{(E\{\psi_{F(Y,S,\mathbf{X}),z}|Z,\mathbf{X}\}-\mu_{z,F(Y,S,\mathbf{X})}(\mathbf{X}))S(Z,\mathbf{X})\}\nonumber\\
    &=0,\label{eq:eqfor lemma6}
\end{align}
where the first equality follows by LOTE and the second equality follows by the definition of $\psi_{F(Y,S,\mathbf{X}),z}$. Then, we have that
\begin{align*}
    \dot{\mu}_{z,F(Y,S,\mathbf{X}),\boldsymbol{\theta}}|_{\boldsymbol{\theta}=\boldsymbol{\theta}_0}&=E\{\mu_{z,F(Y,S,\mathbf{X})}(\mathbf{X})S(\mathbf{V})\}+E\{\dot{\mu}_{z,F(Y,S,\mathbf{X}),\boldsymbol{\theta}}(\mathbf{X})|_{\boldsymbol{\theta}=\boldsymbol{\theta}_0}\}\\
    &=E\{\mu_{z,F(Y,S,\mathbf{X})}(\mathbf{X})S(\mathbf{V})\}+E\{(\psi_{F(Y,S,\mathbf{X}),z}-\mu_{z,F(Y,S,\mathbf{X})}(\mathbf{X}))S(Y,S|Z,\mathbf{X})\}\\
    &=E\{\psi_{F(Y,S,\mathbf{X}),z}S(\mathbf{V})\}\quad \text{(Equation \eqref{eq:eqfor lemma6})}\\
    &=E\{(\psi_{F(Y,S,\mathbf{X}),z}-\mu_{z,F(Y,S,\mathbf{X})})S(\mathbf{V})\},
\end{align*}
where the first equality follows by the chain rule, the second equality follows by Lemma \ref{eif:lemma1}, the third equality follows by Equation \eqref{eq:eqfor lemma6}, the last equality holds because $E\{S(\mathbf{V})\}=0$. 
Moreover, $E\{\psi_{F(Y,S,\mathbf{X}),z}\}=\mu_{z,F(Y,S,\mathbf{X})}$ implies that $\psi_{F(Y,S,\mathbf{X}),z}-\mu_{z,F(Y,S,\mathbf{X})}\in\mathcal{H}$. This completes the proof.
\end{proof}

\begin{lemma}\label{eif:lemma3}
 $$\dot{m}_{z,\boldsymbol{\theta}}(\mathbf{X})|_{\boldsymbol{\theta}=\boldsymbol{\theta}_0}=E\left \{\frac{\psi_{YS,z}-m_z(\mathbf{X})\psi_{S,z}}{p_z(\mathbf{X})}S(Y|S,Z,\mathbf{X})|\mathbf{X}\right \}$$
\end{lemma}
\begin{proof}
Note $m_z(\mathbf{X})$ can be written as a ratio:
$$m_z(\mathbf{X})=\frac{E\{YS|Z=z,\mathbf{X}\}}{p_z(\mathbf{X})}\equiv\frac{N^\prime}{D^\prime}.$$
By Lemma \ref{eif:lemma1},
\begin{align*}
    &\dot{N}_{\btheta}^\prime|_{\boldsymbol{\theta}=\boldsymbol{\theta}_0}=E\{(\psi_{YS,z}-D^\prime m_z(\mathbf{X}))S(Y,S|Z,\mathbf{X})|\mathbf{X}\},\\
    &\dot{p}_{z,\btheta}(\mathbf{X})|_{\boldsymbol{\theta}=\boldsymbol{\theta}_0}=E\{(\psi_{S,z}-p_z(\mathbf{X}))S(Y,S|Z,\mathbf{X})|\mathbf{X}\}.
\end{align*}
    Combining  this with the quotient rule of influence function implies that
    $$\dot{m}_{z,\boldsymbol{\theta}}(\mathbf{X})|_{\boldsymbol{\theta}=\boldsymbol{\theta}_0}=E\left \{\frac{\psi_{YS,z}-m_z(\mathbf{X})\psi_{S,z}}{p_z(\mathbf{X})}S(Y,S|Z,\mathbf{X})|\mathbf{X}\right \}.$$
We then conclude the proof by observing
    $$E\{(\psi_{YS,z}-m_z(\mathbf{X}))\psi_{S,z}S(S|Z,\mathbf{X})|\mathbf{X}\}=E\{E\{\psi_{YS,z}-m_z(\mathbf{X})\psi_{S,z}|S,Z,\mathbf{X}\}S(S|Z,\mathbf{X})|\mathbf{X}\}=0.$$

\end{proof}

We now begin the proof of EIF. Specifically, Lemma \ref{eif:lemma2} implies that 
$$\dot{p}_{J-g,\boldsymbol{\theta}}|_{\boldsymbol{\theta}=\boldsymbol{\theta}_0}=E\{(\psi_{S,J-g}-p_{J-g})S(\mathbf{V})\},\quad \dot{p}_{J-g+1,\boldsymbol{\theta}}|_{\boldsymbol{\theta}=\boldsymbol{\theta}_0}=E\{(\psi_{S,J-g+1}-p_{J-g+1})S(\mathbf{V})\},$$
which concludes 
$$\Psi_D(\mathbf{V})=(\psi_{S,J-g+1}-\psi_{S,J-g})-D.$$
By the chain rule, we further obtain 
\begin{align}
    \dot{N}_{\btheta}|_{\btheta=\btheta_0}=&E\{(p_{J-g+1}(\mathbf{X})-p_{J-g}(\mathbf{X}))m_z(\mathbf{X})S(\mathbf{X})\} \label{eq:Nhat_1}\\
    &+ E\{(\dot{p}_{J-g+1,\btheta}(\mathbf{X})|_{\btheta=\btheta_0}-\dot{p}_{J-g,\btheta}(\mathbf{X})|_{\btheta=\btheta_0})m_{z}(\mathbf{X})\} \label{eq:Nhat_2} \\
    &+ E\{(p_{J-g+1}(\mathbf{X})-p_{J-g}(\mathbf{X}))\dot{m}_{z,\btheta}(\mathbf{X})|_{\btheta=\btheta_0}\} \label{eq:Nhat_3}.
\end{align}
Because $E\{NS(\mathbf{X})\}=0$, we conclude that
$$\eqref{eq:Nhat_1}=E\{[(p_{J-g+1}(\mathbf{X})-p_{J-g}(\mathbf{X}))m_{z}(\mathbf{X})-N]S(\mathbf{X})\}.$$
In addition, by Lemma \ref{eif:lemma1} and observing that $E\{(\psi_{S,z}-p_{z}(\mathbf{X}))S(Y|Z,S,\mathbf{X})|\mathbf{X}\}=0$, we can show that
\begin{equation}\label{eq:implication of lemma 4}
\dot{p}_{z,\btheta}(\mathbf{X})|_{\btheta=\btheta_0}=E\{(\psi_{S,z}-p_z(\mathbf{X}))S(S|Z,\mathbf{X})|\mathbf{X}\}.
\end{equation}
This further indicates that
$$\eqref{eq:Nhat_2}=E\{[\psi_{S,J-g+1}-\psi_{S,J-g}-(p_{J-g+1}(\mathbf{X})-p_{J-g}(\mathbf{X}))]m_{z}(\mathbf{X})S(S|Z,\mathbf{X})\}.$$
Moreover, Lemma \ref{eif:lemma3} suggests that 
$$\eqref{eq:Nhat_3}=E\left \{(p_{J-g+1}(\mathbf{X})-p_{J-g}(\mathbf{X}))\frac{\psi_{YS,z}-m_z(\mathbf{X})\psi_{S,z}}{p_z(\mathbf{X})}S(Y|S,Z,\mathbf{X})\right \}.$$
It is straightforward to verify that
\begin{align*}
    &(p_{J-g+1}(\mathbf{X})-p_{J-g}(\mathbf{X}))m_z(\mathbf{X})-N\in \mathcal{F}_1,\\
    &[\psi_{S,J-g+1}-\psi_{S,J-g}-(p_{J-g+1}(\mathbf{X})-p_{J-g}(\mathbf{X}))]m_z(\mathbf{X})\in \mathcal{F}_3,\\
    &(p_{J-g+1}(\mathbf{X})-p_{J-g}(\mathbf{X}))\frac{\psi_{YS,z}-m_z(\mathbf{X})\psi_{S,z}}{p_z(\mathbf{X})}\in \mathcal{F}_4.
\end{align*}
Because $\{\mathcal{F}_1,\ldots,\mathcal{F}_4\}$ are mutually orthogonal, we conclude that
\begin{align*}
    \dot{N}_{\boldsymbol{\theta}}|_{\boldsymbol{\theta}=\boldsymbol{\theta}_0}=&E\Bigg\{\Big\{(p_{J-g+1}(\mathbf{X})-p_{J-g}(\mathbf{X}))m_z(\mathbf{X})-N+[\psi_{S,J-g+1}-\psi_{S,J-g}-(p_{J-g+1}(\mathbf{X})-p_{J-g}(\mathbf{X}))]m_z(\mathbf{X})\\
    &+(p_{J-g+1}(\mathbf{X})-p_{J-g}(\mathbf{X}))\frac{\psi_{YS,z}-m_z(\mathbf{X})\psi_{S,z}}{p_z(\mathbf{X})} \Big\}S(\mathbf{V})
 \Bigg\},
\end{align*}
which implies that the EIF of $N$ is
$$\Psi_N(\mathbf{V})=\frac{p_{J-g+1}(\mathbf{X})-p_{J-g}(\mathbf{X})}{p_z(\mathbf{X})}\psi_{YS,z}-N+m_{z}(\mathbf{X})\left\{\psi_{S,J-g+1}-\psi_{S,J-g}-\frac{p_{J-g+1}(\mathbf{X})-p_{J-g}(\mathbf{X})}{p_z(\mathbf{X})}\psi_{S,z}\right\}.$$
This, together with the quotient rule of influence function, concludes the expression of the EIF shown in Theorem 1.

\subsection{Proof of Theorem \ref{thm:triply robustness}}\label{subsec:proof_dr}
We first show the triple robustness property of $\widehat\mu_g^{\text{DR}}(z)$. Consider the ratio representation 
\begin{equation*}
    \mu_g(z)=\frac{E\{Y(z)\mathbf{1}(G=g)\}}{E\{S(J-g+1)-S(J-g)\}}.
\end{equation*}
Following the standard arguments on doubly robust estimation of average treatment effect (see, for example, \cite{bang2005doubly}), one can show that the denominator of $\widehat\mu_g^{\text{DR}}(z)$, $\mathbb{P}_n\{\widehat{\psi}_{S,J-g+1}-\widehat{\psi}_{S,J-g}\}$, is consistent for $p_{J-g+1}-p_{J-g}=E\{S(J-g+1)-S(J-g)\}$ whenever either the propensity score model or the principal score model is correctly specified.
Next, we show consistency of the numerator of $\widehat\mu_g^{\text{DR}}(z)$, $\mathbb{P}_n\{\widehat{\xi}_{zg}(\mathbf V)\}$, with $\widehat{\xi}_{zg}(\mathbf V)$ defined as
\begin{equation*}
\widehat{\xi}_{zg}(\mathbf V)=(p_{J-g+1}(\mathbf{X};\widehat{\boldsymbol{\alpha}}_{J-g+1})-p_{J-g}(\mathbf{X};\widehat{\boldsymbol{\alpha}}_{J-g}))\frac{S\mathbf{1}(Z=z)}{p_z(\mathbf{X};\widehat{\boldsymbol{\alpha}}_z)\pi_z(\bX;\widehat{\bbeta}_z)}(Y-m_z(\mathbf{X};\widehat{\boldsymbol{\gamma}}_z)) +m_z(\mathbf{X};\widehat{\boldsymbol{\gamma}}_z)(\widehat\psi_{S,J-g+1}-\widehat \psi_{S,J-g}),
\end{equation*}
Therefore, $\mathbb{P}_n\{\widehat{\xi}_{zg}(\mathbf V)\}$ converges in probability to 
\begin{align*}
   &E\Bigg\{(p_{J-g+1}(\mathbf{X};\widetilde{\boldsymbol{\alpha}}_{J-g+1})-p_{J-g}(\mathbf{X};\widetilde{\boldsymbol{\alpha}}_{J-g}))\frac{S\mathbf{1}(Z=z)}{p_z(\mathbf{X};\widetilde{\boldsymbol{\alpha}}_z)\pi_z(\bX;\widetilde{\bbeta}_z)}(Y-m_z(\mathbf{X};\widetilde{\boldsymbol{\gamma}}_z)) +m_z(\mathbf{X};\widetilde{\boldsymbol{\gamma}}_z)(\psi_{S,J-g+1}-\psi_{S,J-g})\Bigg\}.
\end{align*}
By LOTE, we have that
\begin{align}
    &E\left\{(p_{J-g+1}(\mathbf{X};\widetilde{\boldsymbol{\alpha}}_{J-g+1})-p_{J-g}(\mathbf{X};\widetilde{\boldsymbol{\alpha}}_{J-g}))\frac{S\mathbf{1}(Z=z)}{p_z(\mathbf{X};\widetilde{\boldsymbol{\alpha}}_z)\pi_z(\bX;\widetilde{\bbeta}_z)}(Y-m_z(\mathbf{X};\widetilde{\boldsymbol{\gamma}}_z)) \right\}=\nonumber\\
    &E\left\{(p_{J-g+1}(\mathbf{X};\widetilde{\boldsymbol{\alpha}}_{J-g+1})-p_{J-g}(\mathbf{X};\widetilde{\boldsymbol{\alpha}}_{J-g}))\frac{p_z(\mathbf{X})\pi_Z(\bX)}{p_z(\mathbf{X};\widetilde{\balpha}_z)\pi_z(\bX;\widetilde{\bbeta}_z)}(m_z(\mathbf{X})-m_z(\mathbf{X};\widetilde{\boldsymbol{\gamma}}_z))\right\}\label{eq:trip_robust_decom1},\\
    &E\{m_z(\mathbf{X};\widetilde{\boldsymbol{\gamma}}_z)\psi_{S,J-g+1}\}=E\left\{m_z(\mathbf{X};\widetilde{\boldsymbol{\gamma}}_z)\left(\frac{\pi_z(\bX)(p_{J-g+1}(\bX)-p_{J-g+1}(\mathbf{X};\widetilde{\balpha}_{J-g+1}))}{\pi_z(\bX;\widetilde{\bbeta}_z)}+p_{J-g+1}(\mathbf{X};\widetilde{\balpha}_{J-g+1})\right)\right\}\label{eq:trip_robust_decom2},\\
    &E\{m_z(\mathbf{X};\widetilde{\boldsymbol{\gamma}}_z)\psi_{S,J-g}\}=E\left\{m_{z}(\mathbf{X};\widetilde{\boldsymbol{\gamma}}_z)\left(\frac{\pi_z(\bX)(p_{J-g}(\bX)-p_{J-g}(\mathbf{X};\widetilde{\balpha}_{J-g}))}{\pi_z(\bX;\widetilde{\bbeta}_z)}+p_{J-g}(\mathbf{X};\widetilde{\balpha}_{J-g})\right)\right\}\label{eq:trip_robust_decom3},\\
    &  E\{Y(z)\mathbf{1}(G=g)\}=\mu_g(z)\Pr(G=g)=E\{(p_{J-g+1}(\mathbf{X})-p_{J-g}(\mathbf{X}))m_z(\mathbf{X})\}  \label{eq:trip_robust_decom4},
\end{align}
where Equation \eqref{eq:trip_robust_decom4} follows from Equation \eqref{eq:alt_iden}. Since $p_z(\mathbf{X};\widetilde{\balpha}_z)$ and $\pi_z(\bX;\widetilde{\bbeta}_z)$ are uniformly bounded away from $0$ and $1$, we conclude from that $\eqref{eq:trip_robust_decom1}+\eqref{eq:trip_robust_decom2}-\eqref{eq:trip_robust_decom3}=\eqref{eq:trip_robust_decom4}$ whenever any two of the working models in $\{\pi_z(\bX;\bbeta_z),p_z(\bX;\balpha_z),m_z(\bX;\bgamma_z)\}$ are correctly specified. This conclude that $\mathbb{P}_n\{\widehat{\xi}_{zg}(\mathbf V)\}$ converges to $E\{Y(z)\mathbf{1}(G=g)\}$. Combining the above discussions, we obtain that
$$
\widehat\mu_g^{\text{DR}}(z) = \frac{\mathbb{P}_n\{\widehat{\xi}_{zg}(\mathbf V)\}}{\mathbb{P}_n\{\widehat{\psi}_{S,J-g+1}-\widehat{\psi}_{S,J-g}\}} = \frac{E\{Y(z)\mathbf{1}(G=g)\}}{E\{S(J-g+1)-S(J-g)\}} +o_p(1)= \mu_g(z)+o_p(1),
$$
whenever any two of the working models in $\{\pi_z(\bX;\bbeta_z),p_z(\bX;\balpha_z),m_z(\bX;\bgamma_z)\}$ are correctly specified. This concludes the triple robustness property. 

\begin{lemma}\label{lemma:psi_deriva}
Define $\boldsymbol{\zeta}=(\boldsymbol{\alpha}_{J-g+1}^\top,\boldsymbol{\alpha}_{J-g}^\top,\boldsymbol{\alpha}_{z}^\top,\bbeta_z^\top,\boldsymbol{\gamma}_z^\top)^\top$, which contains all nuisance parameters in the propensity score model, principal score model, and outcome mean model to construct $\widehat\mu_{g}^{\text{DR}}(z)$. Also let $\widetilde{\boldsymbol{\zeta}}=(\widetilde{\balpha}_{J-g+1}^\top,\widetilde{\balpha}_{J-g}^\top,\widetilde{\balpha}_{z}^\top,\widetilde{\bbeta}_z^\top,\widetilde{\bgamma}_z^\top)$ be the true value of $\boldsymbol{\zeta}$. 
Assume that expectation and derivative are exchangeable.
If the principal score and the outcome regression are both correctly specified, then 
$$
E\left\{\frac{\partial \xi_{zg}}{\partial \boldsymbol{\zeta}^\top}(\mathbf{V;\widetilde{\boldsymbol{\zeta}}})\right\}=\mathbf{0}.
$$
\end{lemma}

\begin{proof}
It follows from Equations \eqref{eq:trip_robust_decom1}-\eqref{eq:trip_robust_decom3}. 
\end{proof}
Consider a M-estimator $\widehat{\mu}_g(z)^\prime$ defined by the below estimating equation
\begin{equation}\label{eq:m-estimator}
    \mathbb{P}_n\{\xi_{zg}(\mathbf{V};\widehat{\mu}_g(z)^\prime,\widetilde{\boldsymbol{\zeta}})\}=0,
\end{equation}
where $\widetilde{\boldsymbol{\zeta}}$ is the convergent value of $\widehat{\boldsymbol{\zeta}}$. 
Recall that we use MLE or GEE to obtain $\widetilde{\boldsymbol{\zeta}}$, 
which implies that $\sqrt{n}(\widehat{\boldsymbol{\zeta}}-\widetilde{\boldsymbol{\zeta}})$ is a tight sequence, i.e., $\sqrt{n}(\widehat{\boldsymbol{\zeta}}-\widetilde{\boldsymbol{\zeta}})=O_p(1)$. By construction, our doubly robust estimator $\widehat{\mu}_g^{\text{DR}}(z)$ is defined by the estimating equation
\begin{equation}\label{eq:m-estimator_ori}
    \mathbb{P}_n\{\xi_{zg}(\mathbf{V};\widehat{\mu}_g^{\text{DR}}(z),\widehat{\boldsymbol{\zeta}})\}=0,
\end{equation}
where the only difference between \eqref{eq:m-estimator} and \eqref{eq:m-estimator_ori} is that the truth and the plug-in estimator of $\widetilde{\boldsymbol{\zeta}}$ are used respectively. Applying the first-order Taylor's theorem to \eqref{eq:m-estimator_ori} with respect to $\widetilde{\boldsymbol{\zeta}}$ gives
\begin{equation}\label{eq:taylor_plugin}
     \mathbb{P}_n\{\xi_{zg}(\mathbf{V};\widehat{\mu}_g^{\text{DR}}(z),\widehat{\boldsymbol{\zeta}})\}=\mathbb{P}_n\{\xi_{zg}(\mathbf{V};\widehat{\mu}_g^{\text{DR}}(z),\widetilde{\boldsymbol{\zeta}})\}+ \mathbb{P}_n\left\{\frac{\partial \xi_{zg}(\mathbf{V};\widehat{\mu}_g^{\text{DR}}(z),\widehat{\boldsymbol{\zeta}}^\prime)}{\partial \boldsymbol{\zeta}^\top}\right\}(\widehat{\boldsymbol{\zeta}}-\widetilde{\boldsymbol{\zeta}}),
\end{equation}
where $\widehat{\boldsymbol{\zeta}}^\prime$ lies between $\widehat{\boldsymbol{\zeta}}$ and $\widetilde{\boldsymbol{\zeta}}$. Similarly, applying the first-order Taylor's theorem to $\mathbb{P}_n\{\xi_{zg}(\mathbf{V};\widehat{\mu}_g^{\text{DR}}(z),\widetilde{\boldsymbol{\zeta}})\}$ with respect to $\widehat{\mu}_g(z)^\prime$ yields 
\begin{equation}\label{eq:taylor_plugin2}
     \mathbb{P}_n\{\xi_{zg}(\mathbf{V};\widehat{\mu}_g^{\text{DR}}(z),\widetilde{\boldsymbol{\zeta}})\}=\mathbb{P}_n\left\{\frac{\partial \xi_{zg}(\mathbf{V};\widehat{\mu}_g(z)^\ast,\widetilde{\boldsymbol{\zeta}})}{\partial (\mu_g(z))}\right\}(\widehat{\mu}_g^{\text{DR}}(z)-\widehat{\mu}_g(z)^\prime),
\end{equation}
where $\widehat{\mu}_g(z)^\ast$ lies between $\widehat{\mu}_g^{\text{DR}}(z)$ and $\widehat{\mu}_g(z)^\prime$ and $\mathbb{P}_n\{\xi_{zg}(\mathbf{V};\widehat{\mu}_g(z)^\prime,\widetilde{\boldsymbol{\zeta}})\}=0$ by construction. Combining \eqref{eq:taylor_plugin} and \eqref{eq:taylor_plugin2} gives
$$
\sqrt{n}(\widehat{\mu}_g^{\text{DR}}(z)-\widehat{\mu}_g(z)^\prime)=\frac{\mathbb{P}_n\Big\{\displaystyle\frac{\partial \xi_{zg}(\mathbf{V};\widehat{\mu}_g^{\text{DR}}(z),\widehat{\boldsymbol{\zeta}}^\prime)}{\partial \boldsymbol{\zeta}^\top}\Big\}\times \sqrt{n}(\widehat{\boldsymbol{\zeta}}-\widetilde{\boldsymbol{\zeta}})}{-\mathbb{P}_n\Big\{\displaystyle\frac{\partial \xi_{zg}(\mathbf{V};\widehat{\mu}_g(z)^\ast,\widetilde{\boldsymbol{\zeta}})}{\partial (\mu_g(z))}\Big\}}.
$$
Notice that $\displaystyle\frac{\partial \xi_{zg}(\mathbf{V};\widehat{\mu}_g(z)^\ast,\widetilde{\boldsymbol{\zeta}})}{\partial (\mu_g(z))}=-(\psi_{S,J-g+1}-\psi_{S,J-g})$, $-\mathbb{P}_n\Big\{\displaystyle\frac{\partial \xi_{zg}(\mathbf{V};\widehat{\mu}_g(z)^\ast,\widetilde{\boldsymbol{\zeta}})}{\partial (\mu_g(z))}\Big\}=e_g+o_p(1)$. Then, based on Lemma \ref{lemma:psi_deriva} and consistency of $\widehat{\mu}_g^{\text{DR}}(z)$ and $\widehat{\boldsymbol{\zeta}}$, we obtain
$$\mathbb{P}_n\left\{\frac{\partial \xi_{zg}(\mathbf{V};\widehat{\mu}^\text{DR}_g(z),\widehat{\boldsymbol{\zeta}}^\prime)}{\partial \boldsymbol{\zeta}^\top}\right\}=o_p(1).$$
Eventually, $\sqrt{n}(\widehat{\mu}_g^{\text{DR}}(z)-\widehat{\mu}_g(z)^\prime)=o_p(1)O_p(1)=o_p(1)$, which further implies that the influence functions of $\widehat{\mu}_g^{\text{DR}}(z)$ and $\widehat{\mu}_g(z)^\prime$ are identical.  By Equation (3.6) in \cite{tsiatis2006semiparametric}, the influence function of M-estimator $\widehat{\mu}_g(z)^\prime$ is $\Psi_{zg}(\mathbf{V})$, which completes the proof. 

\subsection{Characterizations of the robust sandwich variance estimators}\label{ss:variance}
In this section, we present the remaining robust sandwich variance estimators. We write out the forms of joint estimating equations and the remaining procedures are the same as the one given in the main manuscript. 
\subsubsection{Propensity score and principal score weighting estimator}\label{ss:var_proof_psw}

Define $\btheta^{\text{tp+ps}}=(\mu_g(z),\mu_g(z^\prime),\balpha_{J-g+1}^\top,\balpha_{J-g}^\top,\balpha_z^\top,\balpha_{z^\prime}^\top,\bbeta^\top,p_{J-g+1},p_{J-g})^\top$. Then, $\widehat{\btheta}^{\text{tp+ps}}$ can be seen as the solution of the following the joint estimating equations $\mathbb{P}_n\{\Phi(\mathbf{V};\btheta^{\text{tp+ps}})\}=\mathbf{0}$ with
\begin{equation}\label{eq:joint_estimating_eq_psw2}
     \Phi(\mathbf{V};\btheta^{\text{tp+ps}})= \begin{pmatrix}
    \displaystyle \frac{p_{J-g+1}(\mathbf{X};\balpha_{J-g+1})-p_{J-g}(\mathbf{X};\balpha_{J-g})}{p_z(\mathbf{X};\balpha_{z})}\frac{\mathbf{1}(Z=z)S}{{p}_{J-g+1}-{p}_{J-g}}Y-\pi_z(\bX;\bbeta_z){\mu}_g(z)\\
    \displaystyle \frac{p_{J-g+1}(\mathbf{X};\balpha_{J-g+1})-p_{J-g}(\mathbf{X};\balpha_{J-g})}{p_{z^\prime}(\mathbf{X};\balpha_{z^\prime})}\frac{\mathbf{1}(Z=z^\prime)S}{{p}_{J-g+1}-{p}_{J-g}}Y-\pi_z(\bX;\bbeta_z){\mu}_g(z^\prime)\\
\kappa_{J-g+1}(S,Z,\mathbf{X};{\balpha}_{J-g+1})\\
\kappa_{J-g}(S,Z,\mathbf{X};{\balpha}_{J-g})\\
\kappa_{z}(S,Z,\mathbf{X};{\balpha}_{z})\\
\kappa_{z^\prime}(S,Z,\mathbf{X};{\balpha}_{z^\prime})\\
\iota(Z,\bX;\bbeta)\\
    {p}_{J-g+1}(\bX;\balpha_{J-g+1})-{p}_{J-g+1}\\
  {p}_{J-g}(\bX;\balpha_{J-g+1})-{p}_{J-g}  
    \end{pmatrix}.
\end{equation}
Remove the third row in $\Phi(\mathbf{V};\btheta^{\text{tp+ps}})$ when $J-g+1=z\text{ or }z^\prime$, and remove the fourth and last row when $g=J$. 

\subsubsection{Estimator based on propensity score and outcome regression}\label{ss:var_or}

Define $\btheta^{\text{tp+or}}=(\mu_g(z),\mu_g(z^\prime),\bbeta,\bgamma_{z}^\top,\bgamma_{z^\prime}^\top,p_{J-g+1},p_{J-g})^\top$. Then $\widehat{\btheta}^{\text{tp+or}}$ can be viewed as the  solution of the following joint estimating equations $\mathbb{P}_n\{\Phi(\mathbf{V};\btheta^{\text{tp+or}})\}=\mathbf{0}$ with
\begin{equation}\label{eq:joint_estimating_eq_or2}
     \Phi(\mathbf{V};\btheta^{\text{tp+or}})= \begin{pmatrix}
       \displaystyle\left\{\frac{S\mathbf{1}(Z=J-g+1)}{\pi_{J-g+1}(\bX;\bbeta_{J-g+1})}-\frac{S\mathbf{1}(Z=J-g)}{\pi_{J-g}(\bX;\bbeta_{J-g})}\right\}m_{z}(\mathbf{X};
     {\bgamma}_{z})-({p}_{J-g+1}-{p}_{J-g}){\mu}_g(z)\\
       \displaystyle\left\{\frac{S\mathbf{1}(Z=J-g+1)}{\pi_{J-g+1}(\bX;\bbeta_{J-g+1})}-\frac{S\mathbf{1}(Z=J-g)}{\pi_{J-g}(\bX;\bbeta_{J-g})}\right\}m_{z^\prime}(\mathbf{X};
     {\bgamma}_{z^\prime})-({p}_{J-g+1}-{p}_{J-g}){\mu}_g(z^\prime)\\
     \iota(Z,\bX;\bbeta)\\
\tau_z(\mathbf{V};{\bgamma}_{z})\\
\tau_{z^\prime}(\mathbf{V};{\bgamma}_{z^\prime})\\
    S\mathbf{1}(Z=J-g+1)/\pi_{J-g+1}(\bX;\bbeta_{J-g+1})-{p}_{J-g+1}\\
  S\mathbf{1}(Z=J-g)/\pi_{J-g}(\bX;\bbeta_{J-g})-{p}_{J-g}
    \end{pmatrix}.
\end{equation}
Remove the last row in $\Phi(\mathbf{V};\btheta^{\text{tp+or}})$ when $g=J$.

\subsubsection{Estimator based on principal score and outcome regression}\label{ss:var_ps+or}    

Define $\btheta^{\text{ps+or}}=(\mu_g(z),\mu_g(z^\prime),\balpha_{J-g+1}^\top,\balpha_{J-g}^\top,\bgamma_{z}^\top,\bgamma_{z^\prime}^\top,p_{J-g+1},p_{J-g})^\top$. Then $\widehat{\btheta}^{\text{ps+or}}$ can be viewed as the  solution of the following joint estimating equations $\mathbb{P}_n\{\Phi(\mathbf{V};\btheta^{\text{ps+or}})\}=\mathbf{0}$ with
\begin{equation}\label{eq:joint_estimating_eq_or2}
     \Phi(\mathbf{V};\btheta^{\text{ps+or}})= \begin{pmatrix}
       \displaystyle\left\{p_{J-g+1}(\bX;\balpha_{J-g+1})-p_{J-g}(\bX;\balpha_{J-g})\right\}m_{z}(\mathbf{X};
     {\bgamma}_{z})-({p}_{J-g+1}-{p}_{J-g}){\mu}_g(z)\\
       \displaystyle\left\{p_{J-g+1}(\bX;\balpha_{J-g+1})-p_{J-g}(\bX;\balpha_{J-g})\right\}m_{z^\prime}(\mathbf{X};
     {\bgamma}_{z^\prime})-({p}_{J-g+1}-{p}_{J-g}){\mu}_g(z^\prime)\\
    \kappa_{J-g+1}(S,Z,\mathbf{X};{\balpha}_{J-g+1})\\
\kappa_{J-g}(S,Z,\mathbf{X};{\balpha}_{J-g})\\
\tau_z(\mathbf{V};{\bgamma}_{z})\\
\tau_{z^\prime}(\mathbf{V};{\bgamma}_{z^\prime})\\
    {p}_{J-g+1}(\bX;\balpha_{J-g+1})-{p}_{J-g+1}\\
  {p}_{J-g}(\bX;\balpha_{J-g+1})-{p}_{J-g}  
    \end{pmatrix}.
\end{equation}
Remove the last row in $\Phi(\mathbf{V};\btheta^{\text{ps+or}})$ when $g=J$. 

\section{Proof of the results without principal ignorability}\label{sec:proof_without_PI}
\subsection{Proof of the identification formulas}\label{sec:noPI;ss:iden}
Observe that 
\begin{align*}
    m_z(\mathbf{X})&=\sum_{\widetilde{g}\geq J+1-z}E\{Y|Z=z,S=1,G=\widetilde{g},\mathbf{X}\}\Pr(G=\widetilde{g}|Z=z,S=1,\mathbf{X})\quad\text{(LOTE)}\\
    &=\sum_{\widetilde{g}\geq J+1-z} E\{Y(z)|G=\widetilde{g},\mathbf{X}\}\Pr(G=\widetilde{g}|Z=z,S=1,\mathbf{X})\quad\text{(SUTVA and monotonicity)}\\
    &=\sum_{\widetilde{g}\geq J+1-z} E\{Y(z)|G=\widetilde{g},\mathbf{X}\}\frac{\Pr(G=\widetilde{g},Z=z|\mathbf{X})}{\Pr(Z=z,S=1|\mathbf{X})}\\
    &=\sum_{\widetilde{g}\geq J+1-z} E\{Y(z)|G=\widetilde{g},\mathbf{X}\}\frac{\Pr(G=\widetilde{g}|\mathbf{X})}{\Pr(S(z)=1|\mathbf{X})}\quad\text{(SUTVA and treatment ignorability)}\\
    &=\sum_{\widetilde{g}\geq J+1-z}E\{Y(z)|G=\widetilde{g},\mathbf{X}\}\frac{e_{\widetilde{g}}(\mathbf{X})}{\sum_{g^\prime\geq J+1-z}e_{g^\prime}(\mathbf{X})}\quad \text{(LOTE and monotonicity)}\\
    &=\{\Omega_{zg}(\mathbf{X})\}^{-1}E\{Y(z)|G=g,\mathbf{X}\},
\end{align*}
which implies $\mu_g(z)=E\{E\{Y(z)|G=g,\mathbf{X}\}|G=g\}=E\{\Omega_{zg}(\mathbf{X})m_z(\mathbf{X})|G=g\}$. We conclude from the proof in Section \ref{ss:proof_iden}.

\subsection{Derivation of the EIF}\label{sec:noPI;ss:EIF}
We inherit all the preliminaries in the proof of Theorem \ref{thm:eif} in Section \ref{sec:main;ss:eif}. We first show the following lemma.
\begin{lemma}\label{lemma:sen_eif_lemma1}
We have that
$$\dot{\Omega}_{zg,\boldsymbol{\theta}}|_{\boldsymbol{\theta}=\boldsymbol{\theta}_0}=E\{\eta_{zg}(\mathbf{V})S(S|Z,\mathbf{X})|\mathbf{X}\},$$
where $$
\eta_{zg}(\mathbf{V})=\frac{\Omega_{zg}(\mathbf{X})\psi_{S,z}}{p_z(\mathbf{X})}-\frac{\Omega_{zg}^2(\mathbf{X})\sum_{\widetilde{g}\geq J+1-z}\delta_{z\widetilde{g}}(\mathbf{X})(\psi_{S,J-\widetilde{g}+1}-\psi_{S,J-\widetilde{g}})}{\delta_{zg}(\mathbf{X})p_z(\mathbf{X})}.
$$
\end{lemma}

\begin{proof}
Define $N=\delta_{zg}(\mathbf{X})p_z(\mathbf{X})$ and $D=N/\Omega_{zg}(\mathbf{X})$. By Equation \eqref{eq:implication of lemma 4}, we conclude that
   $$
   \dot{N}_{\boldsymbol{\theta}}(\mathbf{X})|_{\boldsymbol{\theta}=\boldsymbol{\theta}_0}=E\Big\{\delta_{zg}(\mathbf{X})(\psi_{S,z}-p_z(\mathbf{X}))S(S|Z,\mathbf{X})|\mathbf{X}\Big\}.
   $$
Similarly, one can show 
   $$\dot{D}_{\boldsymbol{\theta}}(\mathbf{X})|_{\boldsymbol{\theta}=\boldsymbol{\theta}_0}=E\left\{\left[\sum_{\widetilde{g}\geq J+1-z}\delta_{z\widetilde{g}}(\mathbf{X})\{(\psi_{S,J-\widetilde{g}+1}-\psi_{S,J-\widetilde{g}})-(p_{J-\widetilde{g}+1}(\mathbf{X})-p_{J-\widetilde{g}}(\mathbf{X}))\}\right]S(S|Z,\mathbf{X})|\mathbf{X}\right\}.
   $$
We then conclude $\dot{\Omega}_{zg,\boldsymbol{\theta}}|_{\boldsymbol{\theta}=\boldsymbol{\theta}_0}=E\{\eta_{zg}(\mathbf{V})S(S|Z,\mathbf{X})|\mathbf{X}\}$ based on the quotient rule of influence function.
\end{proof}
By the identification formula without principal ignorability, we have $\mu_g(z)={N^{\text{PI}}}/{D^{\text{PI}}}$,
where $N^{\text{PI}}=E\left\{(p_{J-g+1}(\mathbf{X})-p_{J-g}(\mathbf{X}))\Omega_{zg}(\mathbf{X})m_z(\mathbf{X})\right\}$ and $D^{\text{PI}}=p_{J-g+1}-p_{J-g}$. For the denominator $D^{\text{PI}}$, we have already showed that 
$$\Psi^{\text{PI}}_D(\mathbf{V})=\Psi_D(\mathbf{V})=(\psi_{S,J-g+1}-\psi_{S,J-g})-D^{\text{PI}},$$
because $D=D^{\text{PI}}$. It is left to derive the EIF of the numerator $N^{\text{PI}}$, denoted by $\Psi^{\text{PI}}_N(\mathbf{V})$. By the chain rule, 
\begin{align}
    \dot{N}^{\text{PI}}_{\btheta}|_{\boldsymbol{\theta}=\boldsymbol{\theta}_0}=&E\{(p_{J-g+1}(\mathbf{X})-p_{J-g}(\mathbf{X}))\Omega_{zg}(\mathbf{X}) m_z (\mathbf{X})S(\mathbf{X})\}\label{eq:eifpicompo1}\\
    &+E\{(\dot{p}_{J-g+1,\btheta}(\mathbf{X})|_{\boldsymbol{\theta}=\boldsymbol{\theta}_0}-\dot{p}_{J-g,\btheta}(\mathbf{X}))|_{\boldsymbol{\theta}=\boldsymbol{\theta}_0}\Omega_{zg}(\mathbf{X}) m_z (\mathbf{X})\}\label{eq:eifpicompo2}\\
    &+ E\{(p_{J-g+1}(\mathbf{X})-p_{J-g}(\mathbf{X}))\dot{\Omega}_{zg,\btheta}(\mathbf{X})|_{\boldsymbol{\theta}=\boldsymbol{\theta}_0} m_z (\mathbf{X})\}\label{eq:eifpicompo3}\\
    &+ E\{(p_{J-g+1}(\mathbf{X})-p_{J-g}(\mathbf{X}))\Omega_{zg}(\mathbf{X})\dot{m}_{z,\btheta}(\mathbf{X})|_{\boldsymbol{\theta}=\boldsymbol{\theta}_0}\}\label{eq:eifpicompo4}.
\end{align}
Because $E\{NS(\mathbf{X})\}=0$, \eqref{eq:eifpicompo1} can be mean-centered as
$$
E\{[(p_{J-g+1}(\mathbf{X})-p_{J-g}(\mathbf{X}))\Omega_{zg}(\mathbf{X}) m_z (\mathbf{X})-N]S(\mathbf{X})\}.
$$ 
Similar to the proof of Theorem \ref{thm:eif}, \eqref{eq:eifpicompo2} and \eqref{eq:eifpicompo4} can be written as 
\begin{align*}
   &E\{(\psi_{S,J-g+1}-\psi_{S,J-g}-p_{J-g+1}(\mathbf{X})+p_{J-g}(\mathbf{X}))\Omega_{zg}(\mathbf{X}) m_z (\mathbf{X})S(S|Z,\mathbf{X})\},\\
   &E\left \{(p_{J-g+1}(\mathbf{X})-p_{J-g}(\mathbf{X}))\frac{\psi_{YS,z}- m_z (\mathbf{X})\psi_{S,z}}{p_z(\mathbf{X})}\Omega_{zg}(\mathbf{X})S(Y|S,Z,\mathbf{X})\right \},
\end{align*}
respectively. By Lemma \ref{lemma:sen_eif_lemma1}, \eqref{eq:eifpicompo3} reduces to
$$E\{\eta_{zg}(\mathbf{V})(p_{J-g+1}(\mathbf{X})-p_{J-g}(\mathbf{X})) m_z (\mathbf{X})S(S|Z,\mathbf{X})\}.$$
Moreover, it is straightforward to verify that
\begin{align*}
    &(p_{J-g+1}(\mathbf{X})-p_{J-g}(\mathbf{X}))\Omega_{zg}(\mathbf{X}) m_z (\mathbf{X})-N\in \mathcal{F}_1,\\
    &(\psi_{S,J-g+1}-\psi_{S,J-g}-p_{J-g+1}(\mathbf{X})+p_{J-g}(\mathbf{X}))\Omega_{zg}(\mathbf{X}) m_z (\mathbf{X})\in \mathcal{F}_3,\\
    &\eta_{zg}(\mathbf{V})(p_{J-g+1}(\mathbf{X})-p_{J-g}(\mathbf{X})) m_z (\mathbf{X})\in \mathcal{F}_3,\\
    &(p_{J-g+1}(\mathbf{X})-p_{J-g}(\mathbf{X}))\frac{\psi_{YS,z}- m_z (\mathbf{X})\psi_{S,z}}{p_z(\mathbf{X})}\Omega_{zg}(\mathbf{X})\in \mathcal{F}_4,
\end{align*}
which implies that
$$\Psi^{\text{PI}}_N(\mathbf{V})=(p_{J-g+1}(\mathbf{X})-p_{J-g}(\mathbf{X}))\frac{\psi_{YS,z}\Omega_{zg}(\mathbf{X})}{p_z(\mathbf{X})}-N+(\psi_{S,J-g+1}-\psi_{S,J-g})\Omega_{zg}(\mathbf{X}) m_z (\mathbf{X})$$
$$-(p_{J-g+1}(\mathbf{X})-p_{J-g}(\mathbf{X})) m_z (\mathbf{X})\Omega_{zg}^2(\mathbf{X})\frac{\sum_{\widetilde{g}\geq J+1-z}\delta_{z\widetilde{g}}(\mathbf{X})(\psi_{S,J-\widetilde{g}+1}-\psi_{S,J-\widetilde{g}})}{\delta_{zg}(\mathbf{X})p_z(\mathbf{X})}.$$
We then conclude the proof by the quotient rule of influence function.

\subsection{Proof of the robustness and efficiency properties}\label{sec:noPI;ss:efficiency}
We first show that $\widehat{\mu}^{\text{BC-PI}}_g(z)$ is doubly robust, i.e., it is consistent if either the propensity score model or the outcome mean model is correctly specified, provided that the principal score model is correctly specified. The proof in Section \ref{sec:noPI;ss:iden} implies 
$$\mu_g(z)=\frac{E\{\Omega_{zg}(\mX)m_z(\mX)\mathbf{1}(G=g)\}}{\Pr(G=g)}=\frac{E\{\Omega_{zg}(\mX)m_z(\mX)e_g(\mX)\}}{e_g},$$
where the last equality is due to the LOTE. It is clear that $\mathbb{P}_n\{\widehat{\psi}_{J-g+1}-\widehat{\psi}_{J-g}\}$ converges in probability to $e_g$ if either the propensity score or the principal score model is correctly specified, as shown in Section \ref{subsec:proof_dr}. It is left to show that $\mathbb{P}_n\{\widehat{\Xi}^{\text{PI}}\}$ converges in probability to $E\{\Omega_{zg}(\mX)m_z(\mX)e_g(\mX)\}$ in a doubly robust sense. By construction, $\mathbb{P}_n\{\widehat{\Xi}^{\text{PI}}\}$ converges in probability to
\begin{align*}
    &E\Bigg\{\frac{\delta_{zg}(\mX)(p_{J-g+1}(\mX;\widetilde{\balpha}_{J-g+1})-p_{J-g}(\mX;\widetilde{\balpha}_{J-g}))}{\sum_{g^\prime\geq J+1-z}\delta_{zg^\prime}(\mX)\{p_{J-g^\prime+1}(\mX;\widetilde{\balpha}_{J-g^\prime+1})-p_{J-g^\prime}(\mX;\widetilde{\balpha}_{J-g^\prime})\}}\times\\
    &\left\{\frac{\pi_z(\bX)[m_z(\bX)p_z(\bX)-m_z(\bX;\widetilde{\bgamma}_z)p_z(\bX;\widetilde{\balpha}_z)]}{\pi_z(\bX;\widetilde{\bbeta}_z)}+m_z(\bX;\widetilde{\bgamma}_z)p_z(\bX;\widetilde{\balpha}_z)-\frac{p_z(\mX;\widetilde{\balpha}_z)m_z(\mX;\widetilde{\bgamma}_z)\sum_{g^\prime\geq J+1-z}\delta_{zg^\prime}(\mX)\{p_{J-g^\prime+1}(\mX)-p_{J-g^\prime}(\mX)\}}{\sum_{g^\prime\geq J+1-z}\delta_{zg^\prime}(\mX)\{p_{J-g^\prime+1}(\mX;\widetilde{\balpha}_{J-g^\prime+1})-p_{J-g^\prime}(\mX;\widetilde{\balpha}_{J-g^\prime})\}}\right\}+\\
    &\frac{\delta_{zg}(\mX)p_z(\mX;\widetilde{\balpha}_z)m_z(\mX;\widetilde{\bgamma}_z)\left[\frac{\pi_{J-g+1}(\bX)(p_{J-g+1}(\mX)-p_{J-g+1}(\mX;\widetilde{\balpha}_{J-g+1}))}{\pi_{J-g+1}(\bX;\widetilde{\bbeta}_{J-g+1})}+p_{J-g+1}(\mX;\widetilde{\balpha}_{J-g+1})-\frac{\pi_{J-g}(\bX)(p_{J-g}(\mX)-p_{J-g}(\mX;\widetilde{\balpha}_{J-g}))}{\pi_{J-g}(\bX;\widetilde{\bbeta}_{J-g})}-p_{J-g}(\mX;\widetilde{\balpha}_{J-g})\right]}{\sum_{g^\prime\geq J+1-z}\delta_{zg^\prime}(\mX)\{p_{J-g^\prime+1}(\mX;\widetilde{\balpha}_{J-g^\prime+1})-p_{J-g^\prime}(\mX;\widetilde{\balpha}_{J-g^\prime})\}}\Bigg\}.
\end{align*}
If the principal score model is correctly specified so that $p_z(\mX,\widetilde {\balpha}_z)=p_z(\mX)$ for all $z$, the above can be simplified to 
\begin{align*}
        &\quad E\Bigg\{\frac{\delta_{zg}(\mX)p_z(\mX)(p_{J-g+1}(\mX)-p_{J-g}(\mX))}{\sum_{g^\prime\geq J+1-z}\delta_{zg^\prime}(\mX)\{p_{J-g^\prime+1}(\mX)-p_{J-g^\prime}(\mX)\}}\left\{m_z(\mX)-m_z(\mX;\widetilde{\bgamma}_z)\right\}p_z(\bX)\frac{\pi_z(\bX)}{\pi_z(\bX;\widetilde{\bbeta}_z)}+\\
    &\quad \frac{\delta_{zg}(\mX)p_z(\mX)m_z(\mX;\widetilde{\bgamma}_z)(p_{J-g+1}(\mX)-p_{J-g}(\mX))}{\sum_{g^\prime\geq J+1-z}\delta_{zg^\prime}(\mX)\{p_{J-g^\prime+1}(\mX)-p_{J-g^\prime}(\mX)\}}\Bigg\}\\
    &=E\left\{\frac{\delta_{zg}(\mX)p_z(\mX)(p_{J-g+1}(\mX)-p_{J-g}(\mX))m_z(\mX)}{\sum_{g^\prime\geq J+1-z}\delta_{zg^\prime}(\mX)\{p_{J-g^\prime+1}(\mX)-p_{J-g^\prime}(\mX)\}}\right\}\\
    &=E\{\Omega_{zg}(\mX)m_z(\mX)e_g(\mX)\},
\end{align*}
where the first inequality holds if either $\pi_z(\bX)=\pi_z(\bX;\widetilde{\bbeta}_z)$ or $m_z(\bX)=m_z(\bX;\widetilde{\bgamma}_z)$. This concludes that $\widehat{\mu}^{\text{BC-PI}}_g(z)$ converges to $\mu_g(z)$ if either the propensity score model or the outcome mean model is correctly specified, provided that the principal score model is correct. The proof of the semiparametric efficiency in Theorem \ref{thm:triply robustness} applies as long as Lemma \ref{lemma:psi_deriva} holds with $\xi_{zg}^{\text{PI}}=\Psi_{zg}^{\text{PI}}(p_{J-g+1}-p_{J-g})$. One can check the validity of Lemma \ref{lemma:psi_deriva} similarly. Then $\widehat{\mu}^{\text{BC-PI}}_g(z)$ achieves the semiparametric variance lower bound when all models are correctly specified.

\section{Proof of the results without monotonicity}\label{sec:proof_without_mono}

\subsection{Identification formulas for the principal score without monotonicity}\label{sec:noMono;ss:ps}
The observed stratum $S=1|Z=z$ is a mixture of $\mathcal{G}_{z}$, which indicates 
\begin{align*}
    p_z(\mathbf{X})=&\sum_{g=J-z+1}^J\Pr(G=g|Z=z,\mathbf{X})+\sum_{\mathfrak{g}\in\mathcal{G}_{z}\backslash\mathcal{Q}}\Pr(G=\mathfrak{g}|Z=z,\mathbf{X})\\
    =&\sum_{g=J-z+1}^{J}e_g(\mathbf{X})+e_r(\mathbf{X})\sum_{\mathfrak{g}\in\mathcal{G}_{z}\backslash\mathcal{Q}}\rho_{\mathfrak{g}}(\mathbf{X})\quad\text{(Treatment ignorability)}\\
    =&\sum_{g=J-z+1}^{J}e_g(\mathbf{X})+e_r(\mathbf{X})q_z(\mathbf{X}).
\end{align*}
We first consider $r\geq1$. To solve the above system of equations (for $z\in\mathcal{J}$), we first consider two of them with $z=J-r+1$ and $z=J-r$:
\begin{align*}
    &p_{J-r+1}(\mathbf{X})=\sum_{g=r}^J e_g(\mathbf{X})+e_r(\mathbf{X})q_{J-r+1}(\mathbf{X}),\\
    &p_{J-r}(\mathbf{X})=\sum_{g=r+1}^J e_g(\mathbf{X})+e_r(\mathbf{X})q_{J-r}(\mathbf{X}),
\end{align*}
which implies
\begin{align*}
    &p_{J-r+1}(\mathbf{X})-p_{J-r}(\mathbf{X})=e_r(\mathbf{X})(1+q_{J-r+1}(\mathbf{X})-q_{J-r}(\mathbf{X})).
\end{align*}
Thus, $e_r(\mathbf{X})$ is obtained. Eventually, subtracting $p_{J-g}(\mathbf{X})$ from $p_{J-g+1}(\mathbf{X})$ yields
\begin{equation*}
    p_{J-g+1}(\mathbf{X})-p_{J-g}(\mathbf{X})=e_g(\mathbf{X})+e_r(\mathbf{X})(q_{J-g+1}(\mathbf{X})-q_{J-g}(\mathbf{X})).
\end{equation*}
For $r=0$, setting $z=J$ implies
\begin{align*}
    p_J(\mathbf{X})=&1-e_0(\mathbf{X})-\sum_{\mathfrak{g}\in\mathcal{G\backslash\mathcal{Q}}} e_{\mathfrak{g}}(\mathbf{X})+e_0(\mathbf{X})q_J(\mathbf{X})\\
    =&1-e_0(\mathbf{X})-e_0(\mathbf{X})\sum_{\mathfrak{g}\in\mathcal{G\backslash\mathcal{Q}}} \rho_{\mathfrak{g}}(\mathbf{X})+e_0(\mathbf{X})q_J(\mathbf{X})\\
    =&1-e_0(\mathbf{X})-e_0(\mathbf{X})q_{J+1}(\mathbf{X})+e_0(\mathbf{X})q_J(\mathbf{X})\\
    =&1-e_0(\mathbf{X})(1+q_{J+1}(\mathbf{X})-q_J(\mathbf{X})).
\end{align*}
Thus, 
\begin{align*}
    e_0(\mathbf{X})=&\frac{1-p_J(\mathbf{X})}{1+q_{J+1}(\mathbf{X})-q_J(\mathbf{X})}\\
    =&\frac{p_{J+1}(\mathbf{X})-p_J(\mathbf{X})}{1+q_{J+1}(\mathbf{X})-q_J(\mathbf{X})}.
\end{align*}


\subsection{Proof of the identification formulas}\label{sec:noMono;ss:iden}
We prove the identification formulas based on the moment conditions under violation of monotonicity. 
For all $g\in\mathcal{G}_{z}$, we have that
\begin{align*}
    E\{Y(z)|{G}=\mathfrak{g},\mathbf{X}\}&=E\{Y(z)|G\in\mathcal{G}_{z},\mathbf{X}\}\quad\text{(by Assumption \ref{assump:stronger PI for sens_mono})}\\
    &=E\{Y(z)|S=1,Z=z,\mathbf{X}\}=m_z(\mathbf{X})\quad\text{(by treatment ignorability and SUTVA)},
\end{align*}
which implies
\begin{align}
    E\{Y(z)|G=\mathfrak{g}\}&=E\{E\{Y(z)|G=\mathfrak{g},\mathbf{X}\}|G=\mathfrak{g}\}\quad\text{(by LOTE)}\nonumber\\
    &=E\{m_z(\mathbf{X})|G=\mathfrak{g}\}\nonumber\\
    &=E\left\{\frac{e_{\mathfrak{g}}(\mathbf{X})}{e_{\mathfrak{g}}}m_z(\mathbf{X})\right\}\quad\text{(by Lemma \ref{iden:lemma2})}\label{eq:5.2eqref1},
\end{align}
which, combined with the identification formulas for the principal score, yields the identification formula based on principal score and outcome regression.
The identification formula based on propensity score and outcome regression follows from Lemma \ref{iden:lemma3} and Equation \eqref{eq:5.2eqref1}. By LOTE, we further have
\begin{align*}
    &E\left\{\frac{e_{\mathfrak{g}}(\mathbf{X})}{e_{\mathfrak{g}}}\frac{m_z(\mathbf{X})\Pr(S=1,Z=z|\mathbf{X})}{p_z(\mathbf{X})\pi_{z}(\bX)}\right\}=E\left\{\frac{e_{\mathfrak{g}}(\mathbf{X})}{e_{\mathfrak{g}}}m_z(\mathbf{X})\right\},
\end{align*}
which shows the identification formula based on weighting. 

\subsection{Derivation of the EIF}\label{sec:noMono;ss:EIF}
We inherit all the preliminaries in the proof of Theorem \ref{thm:eif} in Section \ref{sec:main;ss:eif}. By Equation \eqref{eq:5.2eqref1}, 
$$\mu_{\mathfrak{g}}(z)=\frac{N^{\text{MO}}}{D^{\text{MO}}},$$
where $N^{\text{MO}}\equiv E\{e_{\mathfrak{g}}(\mathbf{X})m_z(\mathbf{X})\}$ and $D^{\text{MO}}\equiv E\{e_{\mathfrak{g}}(\mathbf{X})\}$. The identification formulas for the principal score without monotonicity imply that $e_{\mathfrak{g}}(\mX)$ is a summation of the building blocks $p_z(\bX)h(\bX)=E\{S h(\mX)|Z=z,\mX\}$ ($h(\mX)$ depends on the sensitivity parameters). By Lemma \ref{eif:lemma2} with $F(Y,S,\mathbf{X})=S\ h(\mathbf{X})$, the EIF for each building block is $h(\mathbf{X})\psi_{S,z}-E\{p_z(\mathbf{X})h(\mathbf{X})\}$ by noting that
\begin{align*}
    \dot{\mu}_{z,Sh(\mathbf{X}),\btheta}|_{\btheta=\btheta_0}&=E\{(\psi_{Sh(\mathbf{X}),z}-E\{p_z(\mathbf{X})h(\mathbf{X})\})S(\mathbf{V})\}\\
    &=E\{(h(\mathbf{X})\psi_{S,z}-E\{p_z(\mathbf{X})h(\mathbf{X})\})S(\mathbf{V})\},
\end{align*}
which implies the influence function for $D^{\text{MO}}$ is given by $\Psi^{\text{MO}}_{D}=\psi_{\mathfrak{g}}^\ast-e_{\mathfrak{g}}$ by linearity of expectation. By the chain rule, 
$$\dot{N}^\text{MO}_{\boldsymbol{\theta}}|_{\boldsymbol{\theta}=\boldsymbol{\theta}_0}=E\{e_{\mathfrak{g}}(\mathbf{X})m_z(\mathbf{X})S(\mathbf{X})\}+E\{\dot{e}_{\mathfrak{g},\btheta}(\mathbf{X})|_{\boldsymbol{\theta}=\boldsymbol{\theta}_0}m_z(\mathbf{X})\}+E\{e_{\mathfrak{g}}(\mathbf{X})\dot{m}_{z,\btheta}(\mathbf{X})|_{\boldsymbol{\theta}=\boldsymbol{\theta}_0}\}.$$
Because $E\{N^{\text{MO}}S(\mathbf{X})\}=0$, 
$$E\{e_{\mathfrak{g}}(\mathbf{X})m_z(\mathbf{X})S(\mathbf{X})\}=E\{(e_{\mathfrak{g}}(\mathbf{X})m_z(\mathbf{X})-N^{\text{MO}})S(\mathbf{X})\}.$$
Furthermore, applying Lemma \ref{eif:lemma1} with $F(Y,S,\mathbf{X})=S h(\mathbf{X})$ implies that
\begin{align*}
    \dot{\mu}_{z,Sh(\mathbf{X}),\btheta}(\mathbf{X})|_{\btheta=\btheta_0}&=E\{(\psi_{Sh(\mathbf{X}),z}-p_z(\mathbf{X})h(\mathbf{X}))S(Y,S|Z,\mathbf{X})|\mathbf{X}\}\\
    &=E\{(h(\mathbf{X})\psi_{S,z}-p_z(\mathbf{X})h(\mathbf{X}))S(Y,S|Z,\mathbf{X})|\mathbf{X}\}\\
    &=E\left\{(h(\mathbf{X})\psi_{S,z}-p_z(\mathbf{X})h(\mathbf{X}))S(S|Z,\mathbf{X})|\mathbf{X}\right\},
\end{align*}
where the last equality holds due to the fact that
\begin{equation*}
    E\{(h(\mathbf{X})\psi_{S,z}-p_z(\mathbf{X})h(\mathbf{X}))S(Y|S,Z,\mathbf{X})|\mathbf{X}\}=0.
\end{equation*}
Thus,
\begin{equation}\label{eq:deri_e_bar_g}
   \dot{e}_{\mathfrak{g},\btheta}(\mathbf{X})|_{\boldsymbol{\theta}=\boldsymbol{\theta}_0}=E\{(\psi^\ast_{\mathfrak{g}}-e_{\mathfrak{g}}(\mathbf{X}))S(S|Z,\mathbf{X})|\mathbf{X}\}.
\end{equation}
One can verify that Lemma \ref{eif:lemma3} still holds without monotonicity, which implies
\begin{align*}
    E\{e_{\mathfrak{g}}(\mathbf{X})\dot{m}_{z,\btheta}(\mathbf{X})|_{\boldsymbol{\theta}=\boldsymbol{\theta}_0}\}&=E\left\{E\Big \{e_{\mathfrak{g}}(\mathbf{X})\frac{\psi_{YS,z}-m_z(\mathbf{X})\psi_{S,z}}{p_z(\mathbf{X})}S(Y|S,Z,\mathbf{X})\Big|\mathbf{X}\Big \}\right\}\\
    &=E\left \{e_{\mathfrak{g}}(\mathbf{X})\frac{\psi_{YS,z}-m_z(\mathbf{X})\psi_{S,z}}{p_z(\mathbf{X})}S(Y|S,Z,\mathbf{X})\right \}.
\end{align*}
It is straightforward to verify that
\begin{align*}
    &e_{\mathfrak{g}}(\mathbf{X})m_z(\mathbf{X})-N^{\text{MO}}\in \mathcal{F}_1,\\
    &m_z(\mathbf{X})(\psi_{\mathfrak{g}}^\ast-e_{\mathfrak{g}}(\mathbf{X}))\in\mathcal{F}_3,\\
    &e_{\mathfrak{g}}(\mathbf{X})\frac{\psi_{YS,z}-m_z(\mathbf{X})\psi_{S,z}}{p_z(\mathbf{X})}\in \mathcal{F}_4,
\end{align*}
which implies that the influence function for $N^{\text{MO}}$, $\Psi^{\text{MO}}_{N}(\mathbf{V})$, is given by 
$$\Psi^{\text{MO}}_{N}(\mathbf{V})=e_{\mathfrak{g}}(\mathbf{X})m_z(\mathbf{X})-N^{\text{MO}}+m_z(\mathbf{X})(\psi_{\mathfrak{g}}^\ast-e_{\mathfrak{g}}(\mathbf{X}))+e_{\mathfrak{g}}(\mathbf{X})\displaystyle\frac{\psi_{YS,z}-m_z(\mathbf{X})\psi_{S,z}}{p_z(\mathbf{X})}.$$
We then conclude the EIF based on the quotient rule of influence function. 

\subsection{Double robustness and semiparametric efficiency}
The proof is similar to the proof of Theorem \ref{thm:triply robustness}. $\mathbb{P}_n\{\widehat{\psi}_{\mathfrak{g}}^\ast\}$ converges in probability to $e_{\mathfrak{g}}$ if either the propensity score model or the principal score model is correctly specified. This result follows from standard arguments for the doubly robust estimator used to estimate the average treatment effect of the intermediate outcome. It is left to show that the numerator of $\widehat{\mu}_{\mathfrak{g}}^\text{BC-MO}(z)$, ${ \mathbb{P}_n\left\{\displaystyle {\widehat{e}_{\mathfrak{g}}(\mathbf{X})S\mathbf{1}(Z=z)}(Y-\widehat{m}_{z}(\mathbf{X}))/(\widehat{p}_z(\mathbf{X})\widehat{\pi}_z(\bX))+\widehat{m}_{z}(\mathbf{X})\widehat{{\psi}}^\ast_{\mathfrak{g}}\right\}}$, converges in probability to $E\{Y(z)\mathbf{1}(G=\mathfrak{g})\}$ whenever any two of the working models in $\{\pi_z(\bX;\bbeta_z),p_z(\bX;\balpha_z),m_z(\bX;\bgamma_z)\}$ are correctly specified. By Equation \eqref{eq:5.2eqref1}, 
\begin{equation}\label{eq:trip_robust_problimi}
  E\{Y(z)\mathbf{1}(G=\mathfrak{g})\}=\mu_{\mathfrak{g}}(z)\Pr(G=\mathfrak{g})=E\{e_{\mathfrak{g}}(\mathbf{X})m_z(\mathbf{X})\}. 
\end{equation}
The probability limit for ${ \mathbb{P}_n\left\{\displaystyle {\widehat{e}_{\mathfrak{g}}(\mathbf{X})S\mathbf{1}(Z=z)}(Y-\widehat{m}_{z}(\mathbf{X}))/(\widehat{p}_z(\mathbf{X})\widehat{\pi}_z(\bX))+\widehat{m}_{z}(\mathbf{X})\widehat{{\psi}}^\ast_{\mathfrak{g}}\right\}}$ is given by
\begin{align*}
    &\quad { \mathbb{P}_n\left\{\displaystyle \frac{\widehat{e}_{\mathfrak{g}}(\mathbf{X})S\mathbf{1}(Z=z)}{\widehat{p}_z(\mathbf{X})\widehat{\pi}_z(\bX)}(Y-\widehat{m}_{z}(\mathbf{X}))+\widehat{m}_{z}(\mathbf{X})\widehat{{\psi}}^\ast_{\mathfrak{g}}\right\}}\\
    &= E\left\{\frac{{e}_{\mathfrak{g}}(\mathbf{X};\widetilde{\balpha})S\mathbf{1}(Z=z)}{p_z(\mathbf{X};\widetilde{\balpha}_z)\pi_z(\bX;\widetilde{\bbeta}_z)}(Y-m_z(\mathbf{X};\widetilde{\bgamma}_z))+m_z(\mathbf{X};\widetilde{\bgamma}_z)e_{\mathfrak{g}}(\mathbf{X};\widetilde{\balpha})\right\}+o_p(1)\\
    &=E\left\{\frac{{e}_{\mathfrak{g}}(\mathbf{X};\widetilde{\balpha})p_z(\mathbf{X})\pi_z(\bX)}{p_z(\mathbf{X};\widetilde{\balpha}_z)\pi_z(\bX;\widetilde{\bbeta}_z)}(m_z(\mathbf{X})-m_z(\mathbf{X};\widetilde{\bgamma}_z))+m_z(\mathbf{X};\widetilde{\bgamma}_z)e_{\mathfrak{g}}(\mathbf{X};\widetilde{\balpha})\right\}+o_p(1)\quad\text{(LOTE)},
\end{align*}
where $\widetilde{\balpha}$ is the probability limit for a vector of all the model parameters specified for estimating $e_{\mathfrak{g}}(\mathbf{X})$. The triple robustness follows from the above immediately. The proof of semiparametric efficiency when all models are correctly specified follows from the proof of Theorem \ref{thm:triply robustness} because one can verify that Lemma \ref{lemma:psi_deriva} holds with $\xi_{z\mathfrak{g}}^{\text{MO}}=\Psi_{z\mathfrak{g}}^{\text{MO}}e_{\mathfrak{g}}$.

\section{Supplementary Material for the simulation study}\label{ss:additional_sim}

\subsection{Specification of the outcome mean model $m_z(\mathbf{X})$}

We show that the outcome model $m_z(\mathbf{X})$ is a linear function of $\mathbf{X}$. By LOTE, we have that
\begin{align*}
    &\quad E\{Y(z)|S=1,Z=z,\mathbf{X}\}\\
    &=E\{E\{Y(z)|S=1,Z=z,\mathbf{X},G\}|S=1,Z=z,\mathbf{X}\}\\
    &=\sum_{g\geq J-z+1}\Pr(G=g|\mathbf{X},Z=z,S=1)E\{Y(z)|S=1,Z=z,\mathbf{X},G=g\}\\
    &=\sum_{g\geq J-z+1}\Pr(G=g|\mathbf{X},Z=z,S=1)E\{Y(z)|Z=z,\mathbf{X},G=g\}\quad\text{(Monotonicity)}\\
    &=\sum_{g\geq J-z+1}\Pr(G=g|\mathbf{X},Z=z,S=1)E\{Y(z)|\mathbf{X},G=g\}\quad\text{(Treatment ignorability)}\\
    &=E\{Y(z)|\mathbf{X},G\in U_z\}\quad\text{(Principal ignorability)}.
\end{align*}
The arguments above also apply in the case of extended principal ignorability when the monotonicity assumption is violated.

\section{Supplementary material tables and figures}\label{sec:tables}
We attach supplementary material tables and figures below.

\begin{figure}
    \centering
    \includegraphics[scale=0.4]{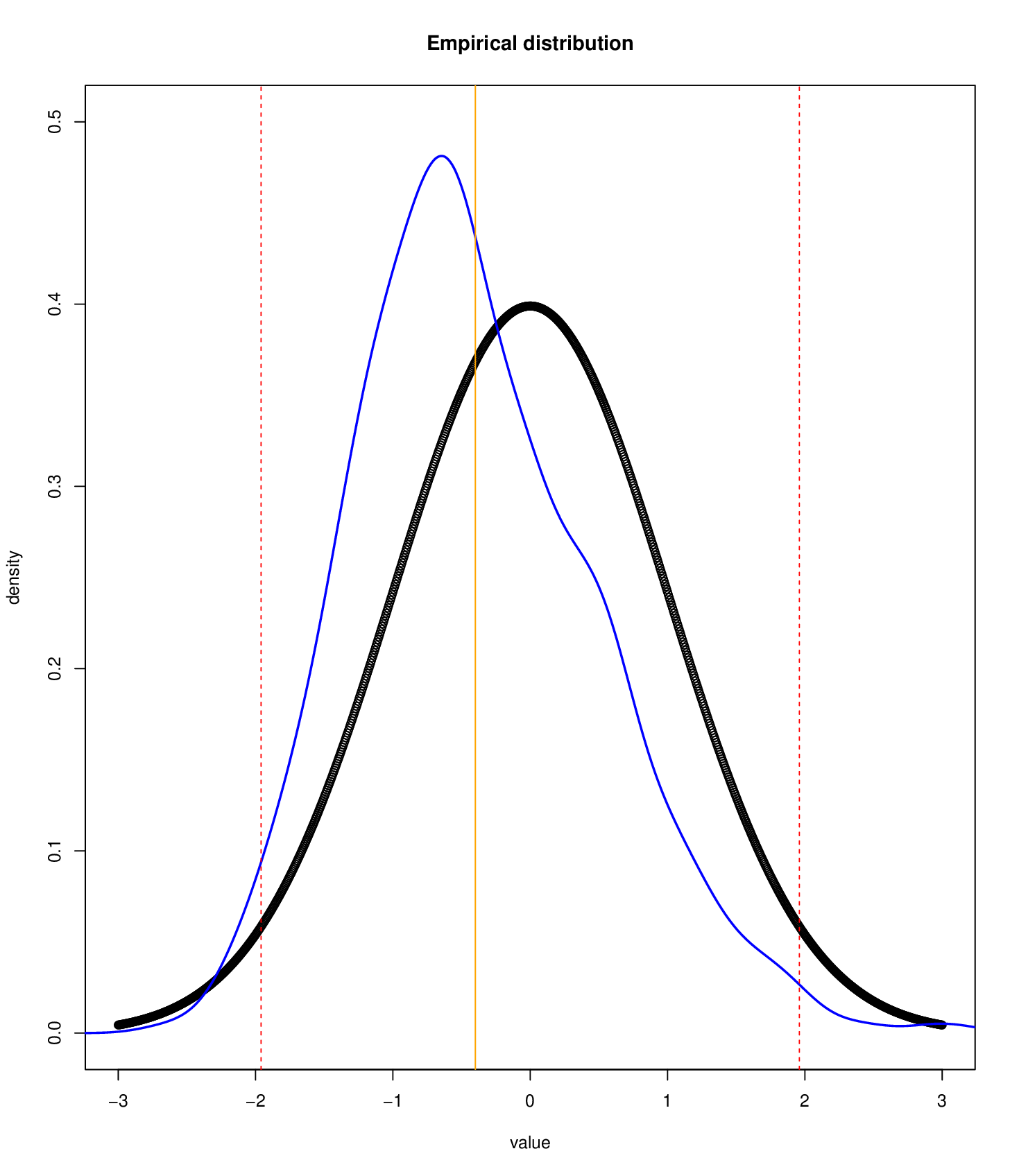}
    \caption{A comparison between the empirical distribution of the standardized (with respect to the truth and robust sandwich variance estimate) principal score weighting estimator (blue curve) and the standard normal distribution (black curve) when the sample size is small ($n=500$) and the principal score model is incorrectly specified. The orange vertical line indicates the mean of empirical distribution and the red dashed vertical line indicates the normal CI margins $[-1.96,1.96]$. The blue curve is expected to be a mean-shift from the black curve if the asymptotic normal approximation is accurate. The empirical coverage probability is the area under the blue curve bounded by two red dashed lines.}
    \label{fig:sim_n500_psw_abnormal_explain}
\end{figure}

\begin{figure}
    \centering
    \includegraphics[scale=0.6]{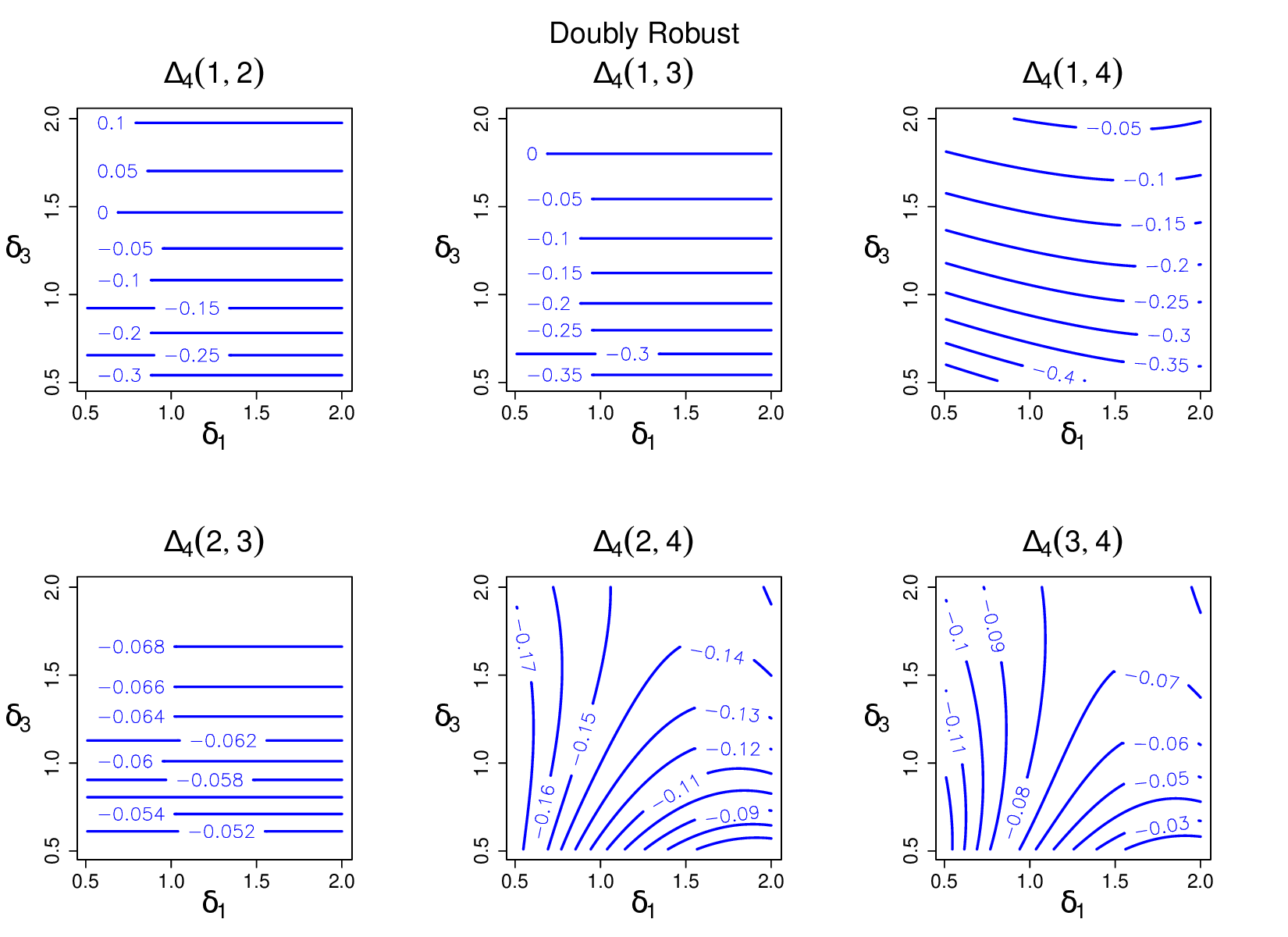}
    \caption{The contour plots for the point estimates of SACEs within the stratum $g=4$ in NTP study using the bias-corrected doubly robust estimator given equal conditional mean potential outcomes between the stratum $g=2$ and the stratum $g=4$, i.e., $\delta_2=1$, and the ratios of conditional mean potential outcome for the stratum $g=1$ or $g=3$ with respect to the stratum $g=4$ varying from half to twice, i.e., $\delta_1,\delta_3\in[0.50,2.00]$. As explained in Section \ref{s:SA;subsec:PI}, the bias-corrected doubly robust estimator is in fact singly robust as it requires correct specification of the principal score model. However, we retain the ``doubly robust'' in the estimator name to differentiate it from the simple weighting and regression estimators.}
    \label{fig:sa_pi_delta2}
\end{figure}

\begin{figure}
    \centering
    \includegraphics[scale=0.6]{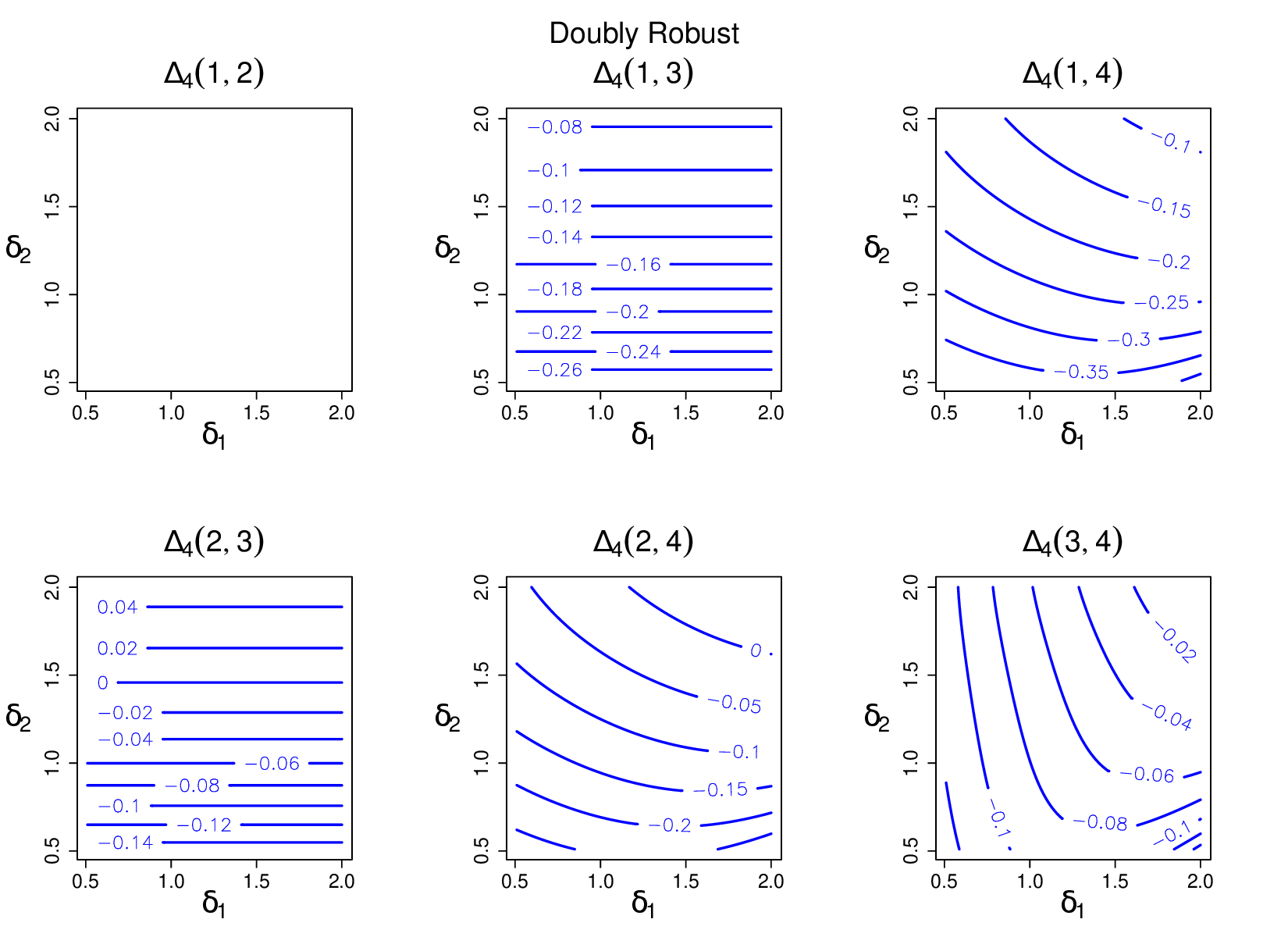}
    \caption{The contour plots for the point estimates of SACEs within the stratum $g=4$ in NTP study using the bias-corrected doubly robust estimator given equal conditional mean potential outcomes between the stratum $g=3$ and the stratum $g=4$, i.e., $\delta_3=1$, and the ratios of conditional mean potential outcome for the stratum $g=1$ or $g=2$ with respect to the stratum $g=4$ varying from half to twice, i.e., $\delta_1,\delta_2\in[0.50,2.00]$. As explained in Section \ref{s:SA;subsec:PI}, the bias-corrected doubly robust estimator is in fact singly robust as it requires correct specification of the principal score model. However, we retain the ``doubly robust'' in the estimator name to differentiate it from the simple weighting and regression estimators.}
    \label{fig:sa_pi_delta3}
\end{figure}

\begin{figure}
    \centering
    \includegraphics[scale=0.48]{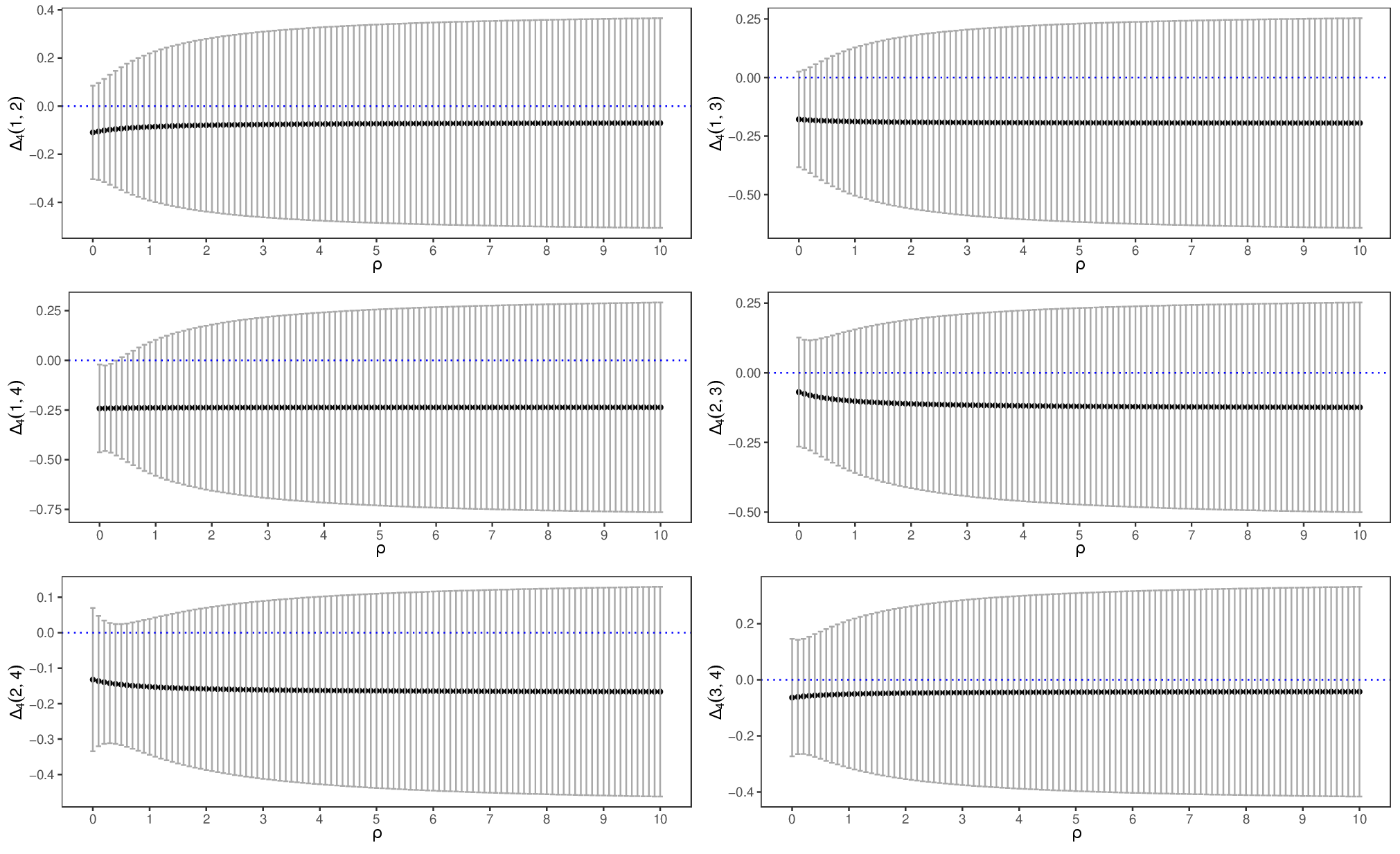}
    \caption{The point estimates and the associated 95\% Wald confidence intervals for the bias-corrected principal score weighting estimator of $\Delta_4(z,z^\prime)$ when the monotonicity is violated with sensitivity parameters $\rho\in[0,10]$. Here, the parameter $\rho$ measures the magnitude of deviation from the monotonicity assumption. The blue dotted line indicates the null.}
    \label{fig:sa_mono_psw}
\end{figure}

\begin{figure}
    \centering
    \includegraphics[scale=0.48]{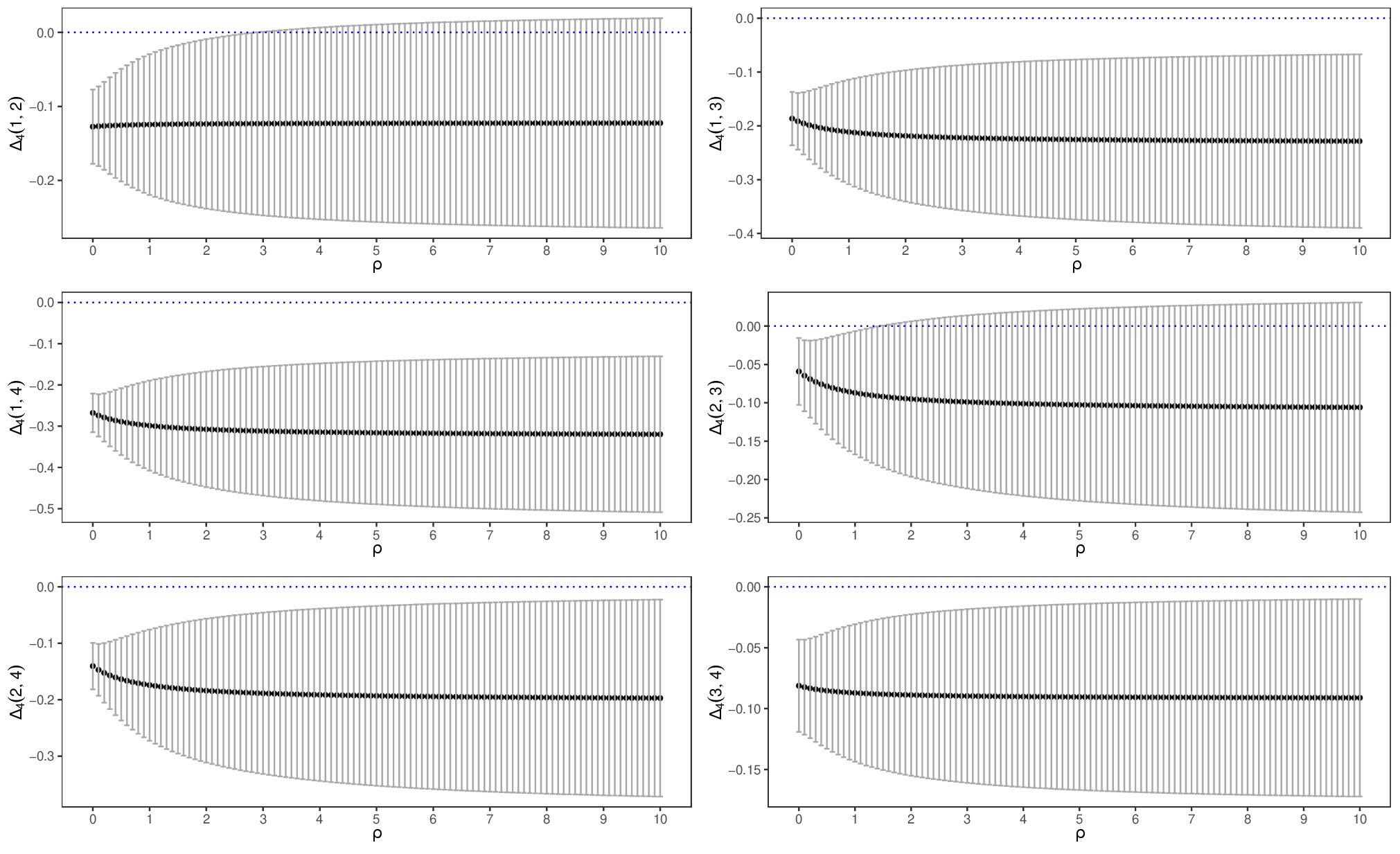}
    \caption{The point estimates and the associated 95\% Wald confidence intervals for the bias-corrected outcome regression estimator of $\Delta_4(z,z^\prime)$ when the monotonicity is violated with sensitivity parameters $\rho\in[0,10]$. Here, the parameter $\rho$ measures the magnitude of deviation from the monotonicity assumption. The blue dotted line indicates the null.}
    \label{fig:sa_mono_or}
\end{figure}

\begin{figure}
    \centering
    \includegraphics[scale=0.48]{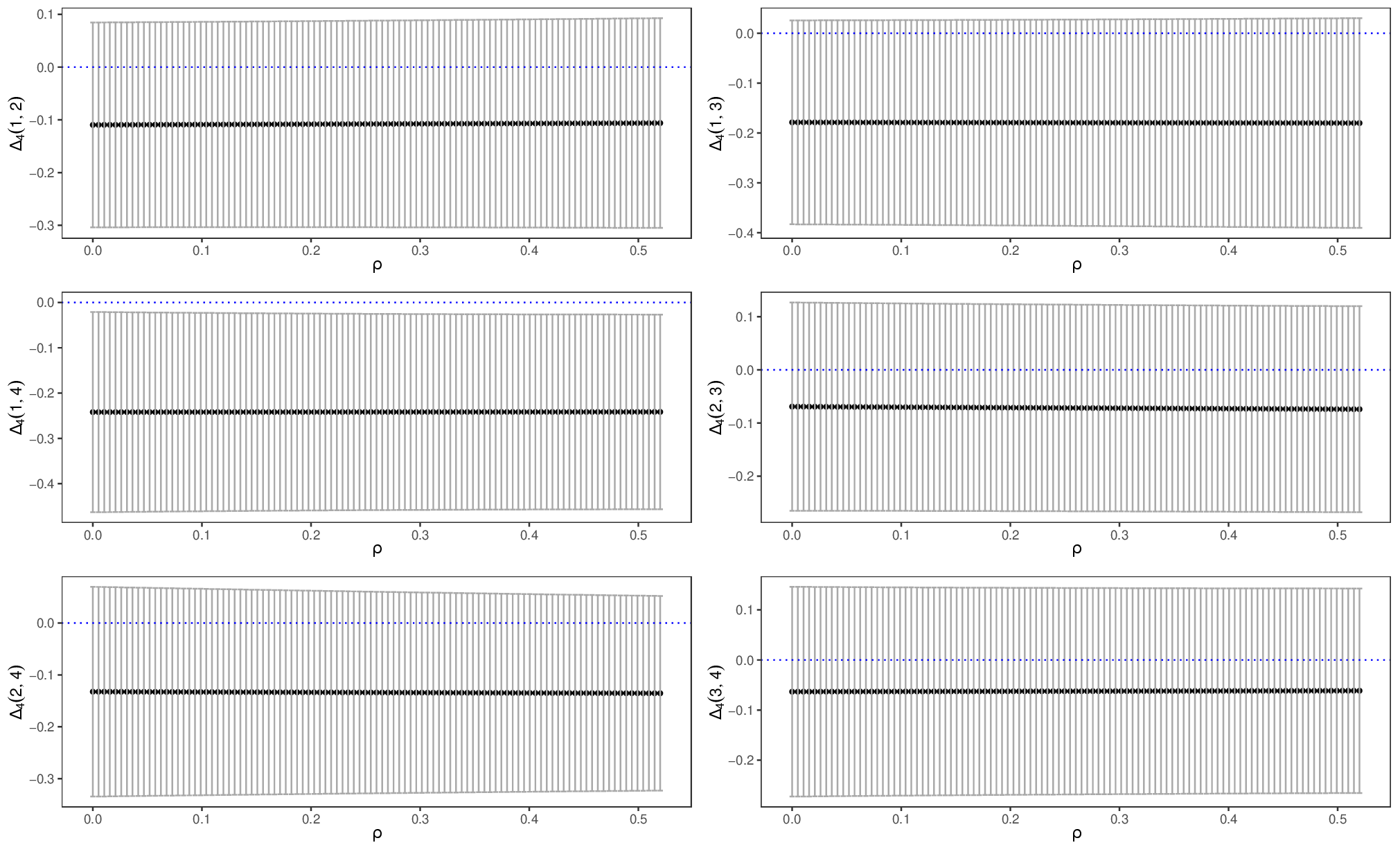}
    \caption{The point estimates and the associated 95\% Wald confidence intervals for the bias-corrected principal score weighting estimator of $\Delta_4(z,z^\prime)$ when the monotonicity is violated only between adjacent strata with sensitivity parameters $\rho\in[0,0.52]$. Here, the parameter $\rho$ measures the magnitude of deviation from the monotonicity assumption. The blue dotted line indicates the null.}
    \label{fig:sa_mono_psw_partial}
\end{figure}

\begin{figure}
    \centering
    \includegraphics[scale=0.48]{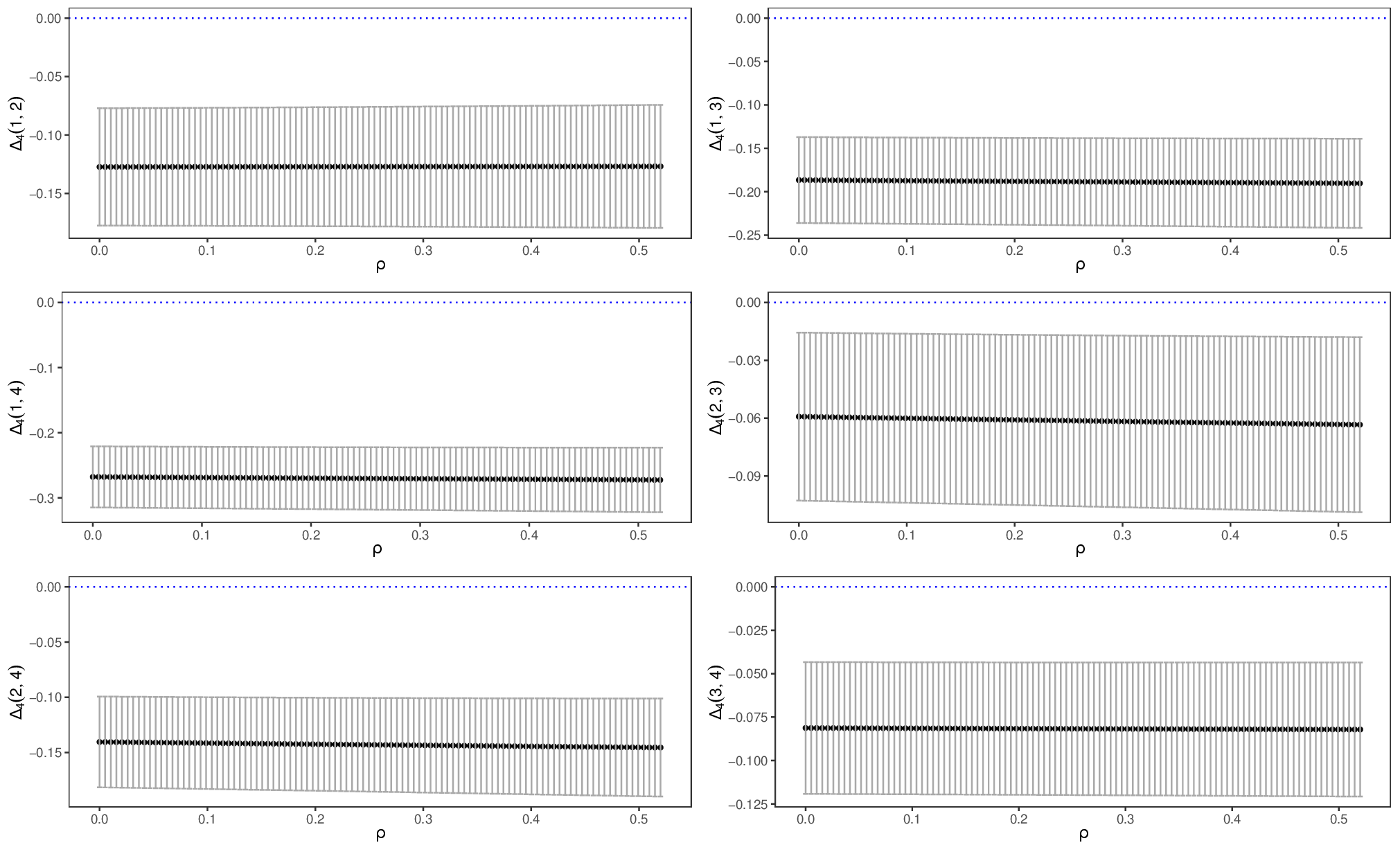}
    \caption{The point estimates and the associated 95\% Wald confidence intervals for the bias-corrected outcome regression estimator of $\Delta_4(z,z^\prime)$ when the monotonicity is violated only between adjacent strata with sensitivity parameters $\rho\in[0,0.52]$. Here, the parameter $\rho$ measures the magnitude of deviation from the monotonicity assumption. The blue dotted line indicates the null.}
    \label{fig:sa_mono_or_partial}
\end{figure}

\begin{figure}
    \centering
    \includegraphics[scale=0.48]{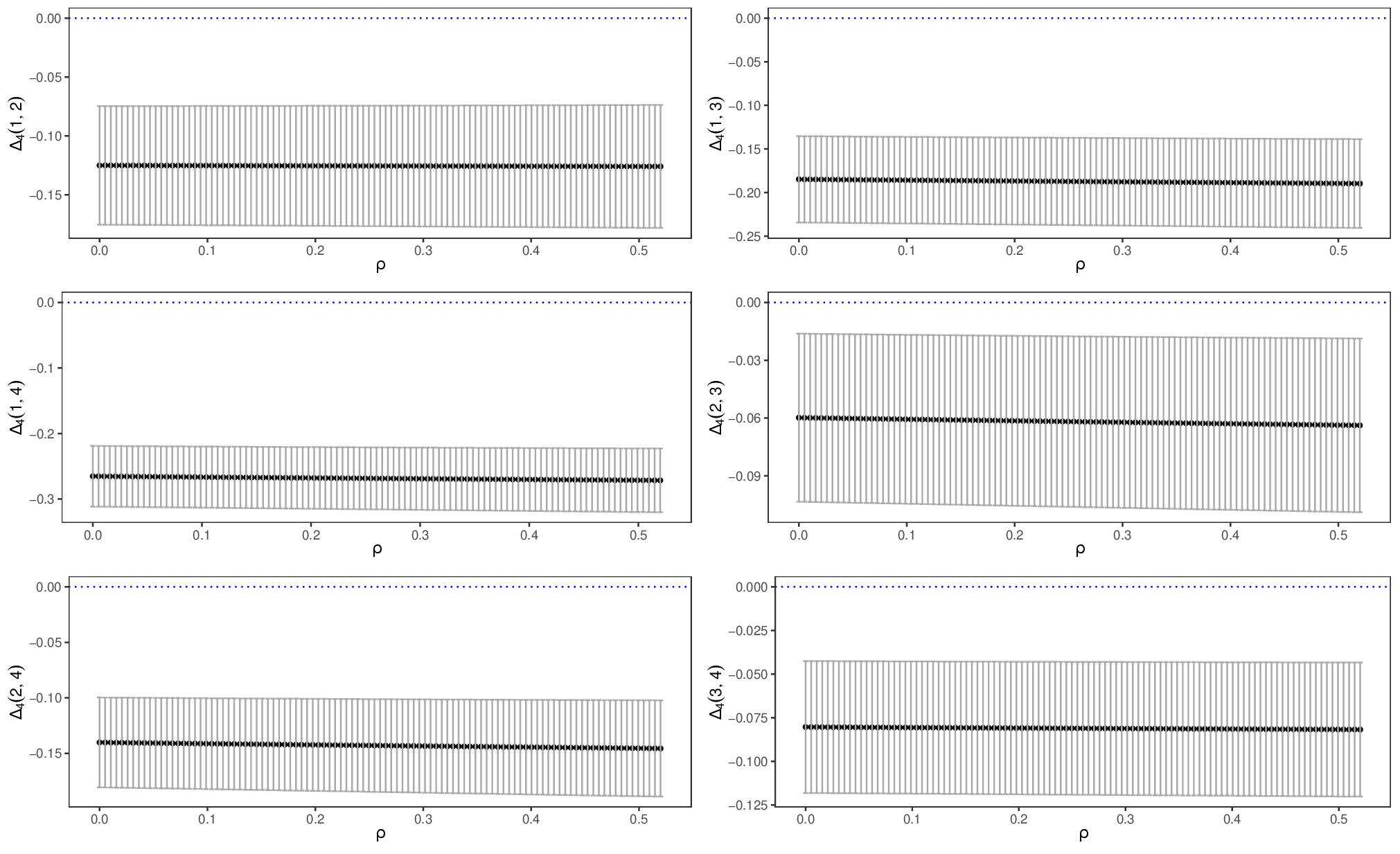}
    \caption{The point estimates and the associated 95\% Wald confidence intervals for the bias-corrected doubly robust estimator of $\Delta_4(z,z^\prime)$ when the monotonicity is violated only between adjacent strata with sensitivity parameters $\rho\in[0,0.52]$. Here, the parameter $\rho$ measures the magnitude of deviation from the monotonicity assumption. The blue dotted line indicates the null.}
    \label{fig:sa_mono_dr_partial}
\end{figure}

\clearpage

\begin{table}[htbp]
\caption{Bias, Monte Carlo standard deviations (`MCSD'), average empirical standard errors (`AESE') based on robust sandwich variance estimators, and empirical coverage (`CP') using AESE for all possible contrasts $\Delta_g(z,z^\prime)$, based on the principal score weighting estimator with bias-correction (`PSW-BC'), outcome regression estimator with bias-correction (`OR-BC'), and doubly robust estimator with bias-correction (`DR-BC'). The data-generating process assumes that principal ignorability is violated while monotonicity holds, and that the covariate-dependent sensitivity parameter is misspecified by fixing it at its mean value. The associated working models for each estimator are assumed to be correctly specified, or compatible with the true data-generating process. 
}
\label{tab:sim_PI_mis}
\begin{adjustbox}{width=1\textwidth}
\begin{tabular}{llll rrr ccc ccc ccc}
    \toprule
            &&   &   &\multicolumn{3}{c}{BIAS} & \multicolumn{3}{c}{CP} &\multicolumn{3}{c}{MCSD} &\multicolumn{3}{c}{AESE}\\ 
           \cmidrule(lr){5-7}\cmidrule(lr){8-10} \cmidrule(lr){11-13} \cmidrule(lr){14-16}
       $n$&$g$&$z$&$z^\prime$              & PSW-BC   &OR-BC    &DR-BC    & PSW-BC  &OR-BC   &DR-BC & PSW-BC  &OR-BC   &DR-BC& PSW-BC  &OR-BC   &DR-BC  \\ \hline
        500 & 2 & 2 & 3 & -0.06 &  0.05 &  0.04 & 95.3 & 95.6 & 99.4 & 0.99 & 0.39 & 0.38 & 1.00 & 0.41 & 0.36\\ 
        ~ & 3 & 1 & 2 & -0.03 &  0.01 &  0.01 & 95.0 & 94.3 & 95.0 & 0.79 & 0.22 & 0.21 & 0.78 & 0.22 & 0.21\\ 
        ~ & ~ & ~ & 3 & -0.05 &  0.00 &  0.00 & 95.3 & 95.0 & 94.6 & 0.76 & 0.29 & 0.26 & 0.77 & 0.29 & 0.27\\ 
        ~ & ~ & 2 & 3  & -0.01 & -0.03 & -0.03 & 95.5 & 94.9 & 95.0 & 0.72 & 0.19 & 0.18 & 0.74 & 0.19 & 0.18 \\ 
        \hline
        2000 & 2 & 2 & 3 &  0.02 &  0.06 &  0.06 & 95.3 & 94.4 & 91.5 & 0.43 & 0.17 & 0.13 & 0.43 & 0.17 & 0.12 \\ 
        ~ & 3 & 1 & 2 &  0.04 &  0.05 &  0.05 & 94.8 & 92.2 & 91.2 & 0.38 & 0.10 & 0.10 & 0.37 & 0.11 & 0.10 \\ 
        ~ & ~ & ~ & 3 &  0.00 &  0.01 &  0.01 & 95.1 & 94.0 & 94.6 & 0.37 & 0.15 & 0.13 & 0.37 & 0.15 & 0.13\\ 
        ~ & ~ & 2 & 3 & -0.03 & -0.04 & -0.04 & 94.6 & 91.5 & 90.9 & 0.34 & 0.09 & 0.08 & 0.35 & 0.09 & 0.08\\ 
        \bottomrule
    \end{tabular}
\end{adjustbox}
\end{table}

\begin{table}[htbp]
\caption{Bias, Monte Carlo standard deviations (`MCSD'), average empirical standard errors (`AESE') based on robust sandwich variance estimators, and empirical coverage (`CP') using AESE for all possible contrasts $\Delta_g(z,z^\prime)$, based on the principal score weighting estimator with bias-correction (`PSW-BC'), outcome regression estimator with bias-correction (`OR-BC'), and doubly robust estimator with bias-correction (`DR-BC'). The data-generating process assumes violation of principal ignorability with sensitivity functions $\delta_1=\delta_2=0.5$, while monotonicity is maintained. The associated working models for each estimator are assumed to be correctly specified, or compatible with the true data-generating process. 
}
\label{tab:sim_PI_delta1=delta2=0.5}
\begin{adjustbox}{width=1\textwidth}
\begin{tabular}{llll rrr ccc ccc ccc}
    \toprule
            &&   &   &\multicolumn{3}{c}{BIAS} & \multicolumn{3}{c}{CP} &\multicolumn{3}{c}{MCSD} &\multicolumn{3}{c}{AESE}\\ 
           \cmidrule(lr){5-7}\cmidrule(lr){8-10} \cmidrule(lr){11-13} \cmidrule(lr){14-16}
       $n$&$g$&$z$&$z^\prime$              & PSW-BC   &OR-BC    &DR-BC    & PSW-BC  &OR-BC   &DR-BC & PSW-BC  &OR-BC   &DR-BC& PSW-BC  &OR-BC   &DR-BC  \\ \hline
        500 & 2 & 2 & 3 & -0.07 &  0.00 &  0.04 & 95.0 & 95.1 & 94.2 & 0.48 & 0.25 & 0.21 & 0.49 & 0.26 & 0.22 \\
        ~ & 3 & 1 & 2 & -0.03 &  0.01 &  0.14 & 95.1 & 95.0 & 93.7 & 0.84 & 0.39 & 0.42 & 0.80 & 0.29 & 0.42 \\
        ~ & ~ & ~ & 3 & -0.06 &  0.00 &  0.08 & 94.9 & 95.7 & 95.4 & 0.80 & 0.38 & 0.38 & 0.79 & 0.34 & 0.39 \\
        ~ & ~ & 2 & 3  & -0.05 & -0.01 & -0.06 & 95.3 & 94.3 & 94.0 & 0.79 & 0.36 & 0.35 & 0.79 & 0.34 & 0.36 \\
        \hline
        2000 & 2 & 2 & 3 &  0.00 &  0.01 &  0.01 & 94.8 & 94.8 & 94.5 & 0.20 & 0.11 & 0.10 & 0.18 & 0.12 & 0.10 \\
        ~ & 3 & 1 & 2 & -0.01 &  0.01 &  0.02 & 96.1 & 94.8 & 95.4 & 0.41 & 0.20 & 0.19 & 0.41 & 0.20 & 0.19 \\ 
        ~ & ~ & ~ & 3 &  0.00 &  0.00 &  0.00 & 95.2 & 96.2 & 95.6 & 0.38 & 0.19 & 0.18 & 0.36 & 0.20 & 0.19 \\
        ~ & ~ & 2 & 3 & -0.01 &  0.00 &  0.00 & 95.7 & 95.6 & 95.5 & 0.38 & 0.18 & 0.17 & 0.35 & 0.16 & 0.15 \\
        \bottomrule
    \end{tabular}
\end{adjustbox}
\end{table}

\begin{table}[htbp]
\caption{Bias, Monte Carlo standard deviations (`MCSD'), average empirical standard errors (`AESE') based on robust sandwich variance estimators, and empirical coverage (`CP') using AESE for all possible contrasts $\Delta_g(z,z^\prime)$, based on the principal score weighting estimator with bias-correction (`PSW-BC'), outcome regression estimator with bias-correction (`OR-BC'), and doubly robust estimator with bias-correction (`DR-BC'). The data-generating process assumes violation of principal ignorability with sensitivity functions $\delta_1=\delta_2=2$, while monotonicity is maintained. The associated working models for each estimator are assumed to be correctly specified, or compatible with the true data-generating process.
}
\label{tab:sim_PI_delta1=delta2=2}
\begin{adjustbox}{width=1\textwidth}
\begin{tabular}{llll rrr ccc ccc ccc}
    \toprule
            &&   &   &\multicolumn{3}{c}{BIAS} & \multicolumn{3}{c}{CP} &\multicolumn{3}{c}{MCSD} &\multicolumn{3}{c}{AESE}\\ 
           \cmidrule(lr){5-7}\cmidrule(lr){8-10} \cmidrule(lr){11-13} \cmidrule(lr){14-16}
       $n$&$g$&$z$&$z^\prime$              & PSW-BC   &OR-BC    &DR-BC    & PSW-BC  &OR-BC   &DR-BC & PSW-BC  &OR-BC   &DR-BC& PSW-BC  &OR-BC   &DR-BC  \\ \hline
        500 & 2 & 2 & 3 & -0.22 &  0.04 & -0.20 & 95.3 & 96.7 & 94.9 & 1.79 & 1.10 & 0.70 & 1.89 & 0.99 & 0.74 \\
        ~ & 3 & 1 & 2 & -0.03 & -0.02 & -0.15 & 94.4 & 95.4 & 94.5 & 0.84 & 0.40 & 0.41 & 0.80 & 0.29 & 0.41 \\
        ~ & ~ & ~ & 3 & -0.03 & -0.01 & -0.08 & 94.9 & 94.7 & 94.1 & 0.79 & 0.36 & 0.33 & 0.78 & 0.32 & 0.34 \\
        ~ & ~ & 2 & 3  &  0.03 &  0.01 &  0.06 & 94.8 & 94.8 & 94.1 & 0.80 & 0.35 & 0.39 & 0.77 & 0.31 & 0.39 \\
        \hline
        2000 & 2 & 2 & 3 & -0.07 & -0.02 & -0.03 & 94.6 & 94.6 & 94.2 & 0.76 & 0.39 & 0.32 & 0.94 & 0.39 & 0.34 \\
        ~ & 3 & 1 & 2 & -0.07 & -0.05 & -0.05 & 93.7 & 94.6 & 94.2 & 0.41 & 0.19 & 0.19 & 0.42 & 0.20 & 0.20 \\ 
        ~ & ~ & ~ & 3 & -0.03 & -0.02 & -0.02 & 95.4 & 95.3 & 94.7 & 0.38 & 0.17 & 0.16 & 0.46 & 0.22 & 0.21 \\
        ~ & ~ & 2 & 3 &  0.03 &  0.04 &  0.04 & 95.1 & 95.0 & 94.4 & 0.39 & 0.19 & 0.19 & 0.42 & 0.17 & 0.17 \\
        \bottomrule
    \end{tabular}
\end{adjustbox}
\end{table}

\begin{table}[htbp]
\caption{Bias, Monte Carlo standard deviations (`MCSD'), average empirical standard errors (`AESE') based on robust sandwich variance estimators, and empirical coverage (`CP') using AESE for all possible contrasts $\Delta_g(z,z^\prime)$, based on the principal score weighting estimator with bias-correction (`PSW-BC'), outcome regression estimator with bias-correction (`OR-BC'), and doubly robust estimator with bias-correction (`DR-BC'). The data-generating process assumes a mild violation of monotonicity with sensitivity parameter $\rho=0.2$, while principal ignorability is maintained. The working models for each estimator are assumed to be correctly specified, or compatible with the true data-generating process.
}
\label{tab:sim_mono_rho=0.2}
\begin{adjustbox}{width=1\textwidth}
\begin{tabular}{llll rrr ccc ccc ccc}
    \toprule
            &&   &   &\multicolumn{3}{c}{BIAS} & \multicolumn{3}{c}{CP} &\multicolumn{3}{c}{MCSD} &\multicolumn{3}{c}{AESE}\\ 
           \cmidrule(lr){5-7}\cmidrule(lr){8-10} \cmidrule(lr){11-13} \cmidrule(lr){14-16}
       $n$&$g$&$z$&$z^\prime$              & PSW-BC   &OR-BC    &DR-BC    & PSW-BC  &OR-BC   &DR-BC & PSW-BC  &OR-BC   &DR-BC& PSW-BC  &OR-BC   &DR-BC  \\ \hline
        500 & 2 & 2 & 3 & -0.04 &  0.00 &  0.02 & 96.4 & 97.5 & 94.6 & 0.95 & 0.62 & 0.31 & 1.08 & 0.62 & 0.32 \\ 
        ~ & 3 & 1 & 2 & -0.07 &  0.00 &  0.00 & 94.7 & 94.5 & 93.9 & 0.87 & 0.22 & 0.21 & 0.90 & 0.22 & 0.21 \\
        ~ & ~ & ~ & 3 &  0.00 &  0.00 &  0.00 & 94.9 & 95.1 & 95.1 & 0.88 & 0.35 & 0.32 & 0.91 & 0.33 & 0.30 \\
        ~ & ~ & 2 & 3  &  0.07 &  0.00 & -0.01 & 94.9 & 94.4 & 94.8 & 0.84 & 0.21 & 0.19 & 0.88 & 0.21 & 0.19 \\
        \hline
        2000 & 2 & 2 & 3 & -0.02 & -0.01 & -0.01 & 95.4 & 94.8 & 95.3 & 0.35 & 0.19 & 0.13 & 0.35 & 0.19 & 0.14 \\
        ~ & 3 & 1 & 2 &  0.00 &  0.00 &  0.00 & 95.3 & 95.4 & 95.1 & 0.40 & 0.11 & 0.10 & 0.42 & 0.11 & 0.10 \\ 
        ~ & ~ & ~ & 3 & -0.03 &  0.00 &  0.00 & 95.1 & 95.3 & 95.8 & 0.45 & 0.17 & 0.15 & 0.44 & 0.17 & 0.15 \\
        ~ & ~ & 2 & 3 &  0.03 &  0.00 &  0.00 & 94.1 & 94.9 & 95.3 & 0.42 & 0.10 & 0.09 & 0.41 & 0.10 & 0.09 \\
        \bottomrule
    \end{tabular}
\end{adjustbox}
\end{table}

\begin{table}[htbp]
\caption{Bias, Monte Carlo standard deviations (`MCSD'), average empirical standard errors (`AESE') based on robust sandwich variance estimators, and empirical coverage (`CP') using AESE for all possible contrasts $\Delta_g(z,z^\prime)$, based on the principal score weighting estimator with bias-correction (`PSW-BC'), outcome regression estimator with bias-correction (`OR-BC'), and doubly robust estimator with bias-correction (`DR-BC'). The data-generating process assumes a severe violation of monotonicity with sensitivity parameter $\rho=5$, while principal ignorability is maintained. The working models for each estimator are assumed to be correctly specified, or compatible with the true data-generating process.
}
\label{tab:sim_mono_rho=5}
\begin{adjustbox}{width=1\textwidth}
\begin{tabular}{llll rrr ccc ccc ccc}
    \toprule
            &&   &   &\multicolumn{3}{c}{BIAS} & \multicolumn{3}{c}{CP} &\multicolumn{3}{c}{MCSD} &\multicolumn{3}{c}{AESE}\\ 
           \cmidrule(lr){5-7}\cmidrule(lr){8-10} \cmidrule(lr){11-13} \cmidrule(lr){14-16}
       $n$&$g$&$z$&$z^\prime$              & PSW-BC   &OR-BC    &DR-BC    & PSW-BC  &OR-BC   &DR-BC & PSW-BC  &OR-BC   &DR-BC& PSW-BC  &OR-BC   &DR-BC  \\ \hline
        500 & 2 & 2 & 3 & $-0.11$ &  0.01  &  0.00  & 95.2 & 95.6 & 94.7 & 0.91 & 0.43 & 0.26 & 0.93 & 0.43 & 0.27 \\
        ~ & 3 & 1 & 2 &  0.05   &  0.01  &  0.00  & 95.2 & 94.8 & 95.4 & 1.22 & 0.41 & 0.32 & 1.29 & 0.42 & 0.33 \\
        ~ & ~ & ~ & 3 &  0.26   &  0.06  &  0.01  & 97.1 & 96.6 & 94.1 & 2.08 & 0.82 & 0.51 & 1.93 & 0.82 & 0.52 \\
        ~ & ~ & 2 & 3  &  0.24   &  0.00  &  0.00  & 95.0 & 94.7 & 94.9 & 1.31 & 0.38 & 0.31 & 1.34 & 0.39 & 0.30 \\
        \hline
        2000 & 2 & 2 & 3 &  0.00   &  0.01  &  0.01  & 95.4 & 95.3 & 94.8 & 0.31 & 0.17 & 0.12 & 0.33 & 0.17 & 0.13 \\
        ~ & 3 & 1 & 2 &  0.00   & $-0.01$ & $-0.02$ & 94.9 & 95.6 & 94.6 & 0.47 & 0.17 & 0.14 & 0.47 & 0.18 & 0.14 \\
        ~ & ~ & ~ & 3 &  0.03   & $-0.01$ & $-0.01$ & 94.4 & 95.1 & 95.0 & 0.68 & 0.31 & 0.24 & 0.69 & 0.31 & 0.24 \\
        ~ & ~ & 2 & 3 &  0.06   &  0.00  &  0.00  & 95.6 & 94.4 & 95.2 & 0.52 & 0.17 & 0.13 & 0.52 & 0.17 & 0.13 \\
        \bottomrule
    \end{tabular}
\end{adjustbox}
\end{table}

\bibliographystyle{chicago}

\bibliography{ref}

\begin{thebibliography}{}

\bibitem[\protect\citeauthoryear{Bang and Robins}{Bang and Robins}{2005}]{bang2005doubly}
Bang, H. and J.~M. Robins (2005).
\newblock Doubly robust estimation in missing data and causal inference models.
\newblock {\em Biometrics\/}~{\em 61\/}(4), 962--973.

\bibitem[\protect\citeauthoryear{Bickel, Klaassen, Bickel, Ritov, Klaassen, Wellner, and Ritov}{Bickel et~al.}{1993}]{bickel1993efficient}
Bickel, P.~J., C.~A. Klaassen, P.~J. Bickel, Y.~Ritov, J.~Klaassen, J.~A. Wellner, and Y.~Ritov (1993).
\newblock {\em Efficient and Adaptive Estimation for Semiparametric Models}.
\newblock New York: Springer.

\bibitem[\protect\citeauthoryear{Cheng, Guo, Liu, Wruck, Li, and Li}{Cheng et~al.}{2023}]{cheng2023multiply}
Cheng, C., Y.~Guo, B.~Liu, L.~Wruck, F.~Li, and F.~Li (2023).
\newblock Multiply robust estimation for causal survival analysis with treatment noncompliance.
\newblock {\em arXiv preprint arXiv:2305.13443\/}.

\bibitem[\protect\citeauthoryear{Cheng and Li}{Cheng and Li}{2025}]{cheng2025identification}
Cheng, C. and F.~Li (2025).
\newblock Identification and multiply robust estimation in causal mediation analysis across principal strata.
\newblock {\em Journal of the Royal Statistical Society Series B: Statistical Methodology\/}, qkaf037.

\bibitem[\protect\citeauthoryear{Chernozhukov, Chetverikov, Demirer, Duflo, Hansen, Newey, and Robins}{Chernozhukov et~al.}{2018}]{Chernozhukov2018}
Chernozhukov, V., D.~Chetverikov, M.~Demirer, E.~Duflo, C.~Hansen, W.~Newey, and J.~Robins (2018).
\newblock {Double/debiased machine learning for treatment and structural parameters}.
\newblock {\em The Econometrics Journal\/}~{\em 21\/}(1), 1--68.

\bibitem[\protect\citeauthoryear{Ding, Geng, Yan, and Zhou}{Ding et~al.}{2011}]{ding2011identifiability}
Ding, P., Z.~Geng, W.~Yan, and X.-H. Zhou (2011).
\newblock Identifiability and estimation of causal effects by principal stratification with outcomes truncated by death.
\newblock {\em Journal of the American Statistical Association\/}~{\em 106\/}(496), 1578--1591.

\bibitem[\protect\citeauthoryear{Ding and Lu}{Ding and Lu}{2016}]{DingandLu2016}
Ding, P. and J.~Lu (2016, 06).
\newblock {Principal stratification analysis using principal scores}.
\newblock {\em Journal of the Royal Statistical Society Series B: Statistical Methodology\/}~{\em 79\/}(3), 757--777.

\bibitem[\protect\citeauthoryear{Elliott, Joffe, and Chen}{Elliott et~al.}{2006}]{elliott2006potential}
Elliott, M.~R., M.~M. Joffe, and Z.~Chen (2006).
\newblock A potential outcomes approach to developmental toxicity analyses.
\newblock {\em Biometrics\/}~{\em 62\/}(2), 352--360.

\bibitem[\protect\citeauthoryear{{European Medicines Agency}}{{European Medicines Agency}}{2020}]{ema2020}
{European Medicines Agency} (2020).
\newblock {ICH E9 (R1)} addendum on estimands and sensitivity analysis in clinical trials to the guideline on statistical principles for clinical trials.
\newblock \url{https://www.ema.europa.eu/en/documents/scientific-guideline/ich-e9-r1-addendum-estimands-sensitivity-analysis-clinical-trials-guideline-statistical-principles_en.pdf}.

\bibitem[\protect\citeauthoryear{Frangakis and Rubin}{Frangakis and Rubin}{2002}]{frangakis2002principal}
Frangakis, C.~E. and D.~B. Rubin (2002).
\newblock Principal stratification in causal inference.
\newblock {\em Biometrics\/}~{\em 58\/}(1), 21--29.

\bibitem[\protect\citeauthoryear{Jiang, Yang, and Ding}{Jiang et~al.}{2022}]{JiangJRSSB2022}
Jiang, Z., S.~Yang, and P.~Ding (2022).
\newblock {Multiply robust estimation of causal effects under principal ignorability}.
\newblock {\em Journal of the Royal Statistical Society Series B: Statistical Methodology\/}~{\em 84\/}(4), 1423--1445.

\bibitem[\protect\citeauthoryear{Juszczak, Altman, Hopewell, and Schulz}{Juszczak et~al.}{2019}]{juszczak2019reporting}
Juszczak, E., D.~G. Altman, S.~Hopewell, and K.~Schulz (2019).
\newblock Reporting of multi-arm parallel-group randomized trials: extension of the consort 2010 statement.
\newblock {\em Journal of the American Medical Association\/}~{\em 321\/}(16), 1610--1620.

\bibitem[\protect\citeauthoryear{Kennedy}{Kennedy}{2023}]{kennedy2023semiparametric}
Kennedy, E.~H. (2023).
\newblock Semiparametric doubly robust targeted double machine learning: a review.

\bibitem[\protect\citeauthoryear{Li and Li}{Li and Li}{2019}]{li2019propensity}
Li, F. and F.~Li (2019).
\newblock Propensity score weighting for causal inference with multiple treatments.
\newblock {\em The Annals of Applied Statistics\/}~{\em 13\/}(4), 2389--2415.

\bibitem[\protect\citeauthoryear{Luo, Li, and He}{Luo et~al.}{2023}]{luo2023causal}
Luo, S., W.~Li, and Y.~He (2023).
\newblock Causal inference with outcomes truncated by death in multiarm studies.
\newblock {\em Biometrics\/}~{\em 79\/}(1), 502--513.

\bibitem[\protect\citeauthoryear{{National Toxicology Program}}{{National Toxicology Program}}{2017}]{National_Toxicology_Program2017-qs}
{National Toxicology Program} (2017, December).
\newblock Toxicology and carcinogenesis studies of antimony trioxide in wistar han [{Crl:WI} (han)] rats and {B6C3F1/N} mice (inhalation studies).
\newblock {\em National Toxicology Program Technical Report Series\/}~(590).

\bibitem[\protect\citeauthoryear{Rosenbaum and Rubin}{Rosenbaum and Rubin}{1983}]{rosenbaum1983central}
Rosenbaum, P.~R. and D.~B. Rubin (1983).
\newblock The central role of the propensity score in observational studies for causal effects.
\newblock {\em Biometrika\/}~{\em 70\/}(1), 41--55.

\bibitem[\protect\citeauthoryear{Rubin}{Rubin}{2006}]{rubin2006causal}
Rubin, D.~B. (2006).
\newblock Causal inference through potential outcomes and principal stratification: application to studies with ``censoring'' due to death.
\newblock {\em Statistical Science\/}~{\em 21\/}(3), 299--309.

\bibitem[\protect\citeauthoryear{Tsiatis}{Tsiatis}{2006}]{tsiatis2006semiparametric}
Tsiatis, A.~A. (2006).
\newblock {\em Semiparametric Theory and Missing Data}.
\newblock New York: Springer.

\bibitem[\protect\citeauthoryear{Van~der Vaart}{Van~der Vaart}{2000}]{van2000asymptotic}
Van~der Vaart, A.~W. (2000).
\newblock {\em Asymptotic Statistics}, Volume~3.
\newblock Cambridge University Press.

\bibitem[\protect\citeauthoryear{Wang, Richardson, and Zhou}{Wang et~al.}{2017}]{wang2017causal}
Wang, L., T.~S. Richardson, and X.-H. Zhou (2017).
\newblock Causal analysis of ordinal treatments and binary outcomes under truncation by death.
\newblock {\em Journal of the Royal Statistical Society Series B: Statistical Methodology\/}~{\em 79\/}(3), 719--735.

\end{thebibliography}
\end{document}